\documentclass[hyper,12pt,letterpaper]{JHEP3}

\usepackage{latexsym,amsmath,amsfonts,amssymb}
\usepackage[dvips]{graphicx}
\usepackage{bbm}
\usepackage{bm}
\usepackage{subfigure}
\usepackage{url}

\renewenvironment{thebibliography}[1]{%
\begin{oldthebibliography}{#1}%
\setlength\itemsep{-1mm}%
}%
{%
\end{oldthebibliography}%
}

\newcommand{\re}{\,\mathbb{R}\mbox{e}\,}

\newcommand{\vol}{\mathrm{vol}}

\newcommand{\parfrac}[2]{\frac{\partial #1}{\partial #2}}

\newcommand{\ud}[2]{^{#1}_{\phantom{#1}#2}}

\newcommand{\eg}{\textit{e.g.}}

\newcommand{\ie}{\textit{i.e.}}

\newcommand{\rep}[1]{\mathbf{#1}}

\newcommand{\kq}{/\!/}
\newcommand{\trans}{\mathsf{T}}

\numberwithin{equation}{section}

\newcommand{\nn}{\nonumber}
\newcommand{\mat}[1]{\begin{pmatrix} #1 \end{pmatrix}}
\newcommand{\be}{\begin{equation}} \newcommand{\ee}{\end{equation}}
\newcommand{\bea}{\begin{equation} \begin{aligned}} \newcommand{\eea}{\end{aligned} \end{equation}}
\newcommand{\tabs}{\rule[-1ex]{0pt}{3.5ex}}

\newcommand{\cC}{\mathcal{C}}
\newcommand{\cD}{\mathcal{D}}

\newcommand{\cF}{\mathcal{F}}
\newcommand{\cG}{\mathcal{G}}

\newcommand{\cI}{\mathcal{I}}

\newcommand{\cL}{\mathcal{L}}

\newcommand{\cN}{\mathcal{N}}
\newcommand{\cO}{\mathcal{O}}
\newcommand{\cP}{\mathcal{P}}
\newcommand{\cQ}{\mathcal{Q}}

\newcommand{\bC}{\mathbb{C}}

\newcommand{\bP}{\mathbb{P}}

\newcommand{\bR}{\mathbb{R}}
\newcommand{\bZ}{\mathbb{Z}}

\newcommand{\fg}{\mathfrak{g}}
\newcommand{\fh}{\mathfrak{h}}
\newcommand{\fu}{\mathfrak{u}}

\newcommand{\CP}{\bC\bP}
\newcommand{\CPt}{{\bC\bP^2}}
\newcommand{\aDs}{\ensuremath{\overline{\text{D6}}}}

\DeclareMathOperator{\Tr}{Tr}
\DeclareMathOperator{\tr}{tr}

\DeclareMathOperator{\sign}{sign}

\title{Quantum moduli space of Chern-Simons quivers, wrapped D6-branes and AdS$_4$/CFT$_3$}

\let\CC\spadesuit
\let\AA\blacklozenge
\let\BB\clubsuit

\author{Francesco Benini$^\CC$, Cyril Closset$^\AA$, Stefano Cremonesi$^\BB$\\

$^\CC$ Department of Physics, Princeton University, \\
Princeton, NJ 08544, USA\\
$^\AA$ Department of Particle Physics and Astrophysics \\
Weizmann Institute of Science, Rehovot 76100, Israel. \\
$^\BB$ Raymond and Beverly Sackler School of Physics and Astronomy \\
Tel-Aviv University, Ramat-Aviv 69978, Israel \\

\email{fbenini@princeton.edu}, \email{closset@weizmann.ac.il}, \email{stefano@tau.ac.il}
}

\abstract{%
We study the quantum moduli space of $\cN=2$ Chern-Simons quivers with generic ranks and CS levels, proving along the way exact formulas for the charges of bare monopole operators.
We then derive $\cN=2$ Chern-Simons quiver theories dual to $AdS_4\times Y^{p,q}(\CPt)$ M-theory backgrounds, for the whole family of Sasaki-Einstein seven-manifolds and for any value of the torsion $G_4$ flux. The derivation of the gauge theories relies on the reduction to type IIA string theory, in which M2-branes become D2-branes while the conical geometry maps to RR flux and D6-branes wrapped on compact four-cycles. M5-branes on torsion cycles map to flux and wrapped D4-branes. The moduli space of the quiver is shown to contain the corresponding CY$_4$ cone and all its crepant resolutions.
}

\preprint{PUTP-2373 \\WIS/04/11-MAY-DPPA \\TAUP-2928/11}

\keywords{Chern-Simons Theories, AdS-CFT Correspondence, M-Theory, Solitons Monopoles and Instantons}

\begin{document}

%
%
%
%
%


\section{Introduction}
\label{sec: intro}

According to the AdS/CFT correspondence, M-theory on $AdS_4\times Y_7$, with $Y_7$ a seven-dimensional manifold preserving some supersymmetry (SUSY), is dual to a three-di\-men\-sio\-nal superconformal field theory (3d SCFT) which describes the low energy dynamics of M2-branes on the cone $C(Y_7)$. Important progress
\cite{Bagger:2006sk, Gustavsson:2007vu, Bagger:2007jr, Aharony:2008ug, Benna:2008zy, Imamura:2008nn, Aharony:2008gk, Jafferis:2008qz, Hanany:2008cd, Hanany:2008fj, Klebanov:2008vq, Aganagic:2009zk, Martelli:2009ga}
has been made in the last few years to write down such field theories for simple geometries, especially in the wake of the ABJM proposal \cite{Aharony:2008ug}. Nevertheless how to provide an explicit description of the SCFT for a generic $Y_7$ remains an open problem. The cases which are best understood so far are those for which the M-theory setup has a dual type IIB brane description \`a la Hanany-Witten  \cite{Hanany:1996ie}, from which one can read off the low energy theory \cite{Aharony:2008ug, Jafferis:2008qz, Imamura:2008ji, Gaiotto:2009tk}.

In this paper we tackle the problem of deriving the field theory for $C(Y_7)$ a
Calabi-Yau (CY) four-fold, preserving $\cN=2$ SUSY in 3d. Notice that $\cN=2$ is the highest amount of SUSY for which the field theory can have matter in non-real representations, which we loosely call \emph{chiral} like in 4d.

We follow the logic first articulated in \cite{Aganagic:2009zk} by Aganagic, who suggested to Kaluza-Klein (KK) reduce a CY$_4$ M-theory background to type IIA on a wisely chosen circle $S^1$, such that the type IIA background is a CY$_3$ fibered over $\bR$ and with Ramond-Ramond (RR) fluxes. In the reduction M2-brane probes become D2-brane probes, and such configuration allows in principle to extract the low energy field theory thanks to the knowledge of D-brane theories at CY$_3$ singularities. It is exciting to discover how the 4d and 3d cases are linked in this way.

In practice the method is hampered by two difficulties: 1) knowing the field theory dual to D-branes at the tip of the CY$_3$; 2) being able to interpret whichever extra singularity, besides the D2-branes, the reduction to type IIA brings about.
In particular in \cite{Aganagic:2009zk} the simplest case in which only D2-branes are present was considered. We will argue that this is only possible if the CY$_3$ does not have compact exceptional%
\footnote{An exceptional cycle is one which does not exist in the base of the CY cone but that appears upon partial resolution of the conical singularity.}
 4-cycles, the reason being that 4-cycles do not enjoy flop transitions. Then the CY$_3$ is a generalized conifold, a dual Hanany-Witten brane construction exists, and the
$\cN=2$ field theory one obtains is non-chiral.

The immediate generalization, which is the subject of this paper, is to allow D6-branes in type IIA. They arise whenever the M-theory circle shrinks on a codimension-four submanifold in the eight-dimensional cone. Examples with non-compact D6-branes have already been considered preserving both $\cN=3$ \cite{Hohenegger:2009as, Gaiotto:2009tk, Hikida:2009tp} and $\cN=2$ \cite{Benini:2009qs, Jafferis:2009th} supersymmetry. Their effect is to add to the quiver chiral multiplets in (anti)fundamental representations and (non-Abelian) flavor symmetries; most $\cN=2$ instances are chiral theories. In this paper we observe that whenever the CY$_3$ has exceptional 4-cycles, a necessary condition for the type IIA background to be geometric is that the singularity hides some compact D6-branes wrapped on those 4-cycles. The wrapped D6-branes contribute dynamical, as opposed to external, three-dimensional gauge bosons: they are fractional D2-branes that unbalance the gauge ranks of the quiver. Going back to M-theory, the construction allows us to describe four-folds with exceptional 6-cycles.

Compact D6-branes are similar to the fractional branes added by ABJ \cite{Aharony:2008gk} to the ABJM theory \cite{Aharony:2008ug}, as both modify some ranks of the quiver gauge theory, yet they are very different in other respects. The fractional M2-branes of ABJ descend to---in the 4d language---``non-anomalous'' fractional D2-branes, that is D4-branes wrapping non-exceptional 2-cycles in the CY$_3$.
The ABJ example is $\bC^4/\bZ_k$: upon IIA reduction it gives the conifold, which has such a 2-cycle. These branes are related to ``non-anomalous'' baryonic symmetries, enjoy Seiberg-like dualities, and do not give quantum corrections to monopole operators.
On the other hand compact D6-branes on exceptional 4-cycles and D4-branes on exceptional 2-cycles are ``\emph{anomalous}'' fractional D2-branes: the quivers they give rise to would be anomalous in 4d, although they are perfectly well-defined in 3d. The addition of such branes changes gauge ranks and CS levels, induces quantum corrections to the charges of monopole operators and deform some relations in the chiral ring, making a classical analysis of the field theory inadequate.

To make our life easier, and to circumvent problem 1) above, we limit ourselves to toric CY$_4$ geometries in M-theory, which give toric CY$_3$ manifolds in type IIA upon suitable reduction. We do this to have full control on the geometry and on D-branes therein, however we believe that our construction is valid more generally.
In this paper we discuss in details the simplest example: a family of conical toric CY$_4$ geometries, cones over the so-called $Y^{p,q}(\CPt)$ \cite{Gauntlett:2004hh, Martelli:2008rt} (or simply%
\footnote{In the literature the name $Y^{p,q}$ usually refers to some five-dimensional Sasaki-Einstein spaces. We hope not to create confusion.}
$Y^{p,q}$ in the following) Sasaki-Einstein seven-manifolds which are $S^3/\bZ_p$ bundles over $\CPt$ (a notable member is $Y^{2,3} = M^{3,2}$). The type IIA reduction gives $\bC^3/\bZ_3$ with $p$ D6-branes wrapping the exceptional $\CPt$. Metrics for these Sasaki-Einstein spaces are known \cite{Gauntlett:2004hh, Martelli:2008rt}, but we will not need them. A broader discussion of more general geometries is left for a companion paper \cite{to:appear}.

Field theories for M2-branes probing a subclass of $C(Y^{p,q})$ geometries have been proposed by Martelli and Sparks (MS) based on Chern-Simons (CS) quivers with equal ranks \cite{Martelli:2008si} (see also \cite{Hanany:2008cd}). The MS field theories correctly reproduce the $C(Y^{p,q})$ geometries as a branch of their moduli space in the parameter range $q \in [p,2p]$, to be contrasted with the wider range $q \in [0,3p]$ for which the metrics are known. Moreover the partial resolutions of the $\bC^2/\bZ_p$ fiber
are not present in their field theories. We will clarify the reason for these puzzling properties.

The reduction from M-theory to type IIA was also considered in \cite{Martelli:2008rt}, but performed only on the $AdS_4\times Y^{p,q}$ near horizon background.
In this case the degeneration locus of the M-theory circle action coincides with the tip of the cone, which is not part of the near horizon geometry after the backreaction of the stack of M2-branes placed there is taken into account: the D6-branes have disappeared.
Resolving the CY$_4$ singularity in M-theory blows up the 4-cycle wrapped by the D6-branes in the CY$_3$, making them visible even when the backreaction of regular branes is taken into account.
On the other hand, a careful analysis of charges allows us to take into account the D6-branes even in the conformal near-horizon geometry.

We derive field theories for M2-branes probing $C(Y^{p,q})$ in the full parameter range $q \in [0,3p]$: they have quiver diagrams as in \cite{Martelli:2008si} but unequal ranks for ``anomalous'' groups and different CS levels.
This has some consequences. First, a classical analysis of the field theory is not adequate, for instance to find its moduli space. The chiral ring is generated by chiral fields appearing in the Lagrangian, plus some monopole operators. The monopoles acquire global charges at one-loop (in fact we prove, with localization techniques, that the charges are one-loop exact), and satisfy quantum F-term relations not directly ensuing from the superpotential. We collect the relevant field theory tools at the beginning of the paper. Second, the field theories that we propose
contain in their moduli space all toric crepant resolutions of the CY$_4$.

The geometry $Y^{p,q}$ has an interesting homology group $H_3(Y^{p,q},\bZ)$ which is a finite Abelian group of order $q(3p-q)$ \cite{Martelli:2008rt}: M5-branes can be wrapped on its elements, giving rise to torsion $G_4$-flux in M-theory, and $B_2\wedge F_2$ flux in type IIA. We find the full family of superconformal theories dual to these different $AdS_4\times Y^{p,q}$ backgrounds, generalizing the torsionless case.
The moduli spaces of the field theories dual to M-theory vacua with torsion flux generically do not account for all the partial resolutions of the conical geometry. This field-theoretic result qualitatively agrees with the observation that torsion $G_4$-fluxes may obstruct partial resolutions \cite{Benishti:2009ky}.
Among this family, we find two quiver CS theories (related by parity times charge conjugation) with equal ranks and levels as in \cite{Martelli:2008si}, if (and only if) $q\in [p,2p]$. The reasons for the restricted parameter space and the lack of some partial resolutions in the original proposal of \cite{Martelli:2008si} are now clear,
and tied to the presence of torsion $G_4$-flux in the gravity dual. Similar puzzles are generic with the chiral CS quivers appearing in the literature: we expect similar phenomena to take place.

The paper is organized as follows. In section \ref{sec:_VMS_monopoles}, which can be read independently of the rest of the paper, we present exact formul\ae{} for the
charges of chiral monopole operators in $\cN=2$ CS-matter theories, and we use them to compute the quantum chiral ring of CS quiver gauge theories.
We also present a more conventional one-loop computation of the moduli space, following known results.
In section \ref{sec:M_to_IIA} we discuss M-theory on the cone over $Y^{p,q}$, and its type IIA reduction to the resolved orbifold $\bC^3/\bZ_3$ fibered over $\bR$. We discuss the D-brane Page charges present in type IIA, including the effect of the Freed-Witten anomaly, and explain how type IIA reproduces the finite group $H^4(Y^{p,q})$.
In section \ref{sec: frac on C3Z3 orbifold} we review some useful facts about fractional branes on $\bC^3/\bZ_3$. In section \ref{sec: field theory} we use that information to derive the field theories describing M2-branes on $C(Y^{p,q})$. We check our proposals by computing the moduli space,
which contains all partial resolutions of the CY$_4$ singularity.
Several useful computations are relegated to the appendices. Moreover, in appendix \ref{app:flavored_ABJM} we analyze partial resolutions of the moduli spaces of the flavored ABJM theories of \cite{Benini:2009qs, Jafferis:2009th}.

\section{Monopole operators and moduli space of $\cN=2$ quivers}
\label{sec:_VMS_monopoles}

A striking difference between 3d $\cN=2$ SCFTs and their $\cN=1$ four-dimensional cousins is the existence in three dimensions of local monopole operators. These can be seen as the dimensional reduction of 4d 't Hooft line operators along the line \cite{Kapustin:2005py, Kapustin:2006pk}.
They carry global charges under the topological (or magnetic) currents $J = \ast\Tr  F$ associated to 3d photons. A subset of monopole operators transforms in short representations of the SUSY algebra: chiral multiplets.
Therefore topological charges grade the chiral ring of 3d theories.

Monopole operators in 3d CFTs have been studied in the pioneering works \cite{Borokhov:2002ib, Borokhov:2002cg, Borokhov:2003yu}, and among the more recent ones we emphasize \cite{Kim:2009wb, Benna:2009xd, Bashkirov:2010kz, Samtleben:2010eu}. The crucial point for us is that the charges of monopole operators can receive corrections at one-loop. It was recently shown, using localization techniques, that the one-loop result is exact for BPS monopole operators in theories with at least $\cN = 3$ superconformal symmetry  \cite{Bashkirov:2010kz}. That analysis can be extended to $\cN=2$ superconformal theories, as we show in appendix \ref{app: charges of monopoles}.
The main difficulty is to determine the exact superconformal R-charge: within the set of R-symmetries of the theory, only one is in the supermultiplet of the stress tensor and determines the dimension of chiral primaries.%
\footnote{A method to determine the superconformal R-symmetry has recently been proposed in \cite{Jafferis:2010un}.}
However for our purposes it suffices to consider any R-symmetry, with the superconformal R-symmetry being some linear combination of it with the other Abelian symmetries of the theory:
\be
R_\text{superconformal} = R + \sum_n \alpha_n Q_n + \sum_i \beta_i H_i^{(T)} \;,
\ee
where $Q_n$ denote Abelian flavor symmetries and $H_i^{(T)}$ topological symmetries.
We assume that the IR R-symmetry is some combination of the UV symmetries.%
\footnote{A theory with such a property can be called a ``good theory'', similarly to the discussion of $\cN=4$ quivers of Gaiotto and Witten \cite{Gaiotto:2008ak}.}

\subsection{Charges of half-BPS monopoles}
\label{subsec: charges monopoles}

Classically and in radial quantization, a half-BPS monopole operator in a $\cN=2$ superconformal theory is a configuration
\be
A_\mu dx^\mu = H \, B_i dx^i \;,\qquad\qquad \sigma = \frac{H}{2r} \;,\qquad\qquad \phi = 0
\ee
with all other fields vanishing. Here $H$ is the magnetic flux in the algebra $\fg$ of the gauge group $\cG$, $B_i$ is the Dirac monopole configuration of magnetic charge one, $\sigma$ is the adjoint real scalar field in the $\cN=2$ vector multiplet $(A_\mu, \lambda, \sigma)$, $r$ is the radius of the 2-sphere, and $\phi$ are complex scalars in chiral multiplets.
$H$ can always be gauge rotated to the Cartan subalgebra $\fh$ of $\fg$. Classically and in the absence of CS terms, monopoles are only charged under the topological symmetries.

Monopoles can acquire charges, both gauge and global, from CS terms. In the following we will distinguish between gauge, R-, flavor and topological symmetries, as well as manifest and hidden: gauge, R- and flavor symmetries are manifest in the Lagrangian, and commute with each other; topological Abelian symmetries are not manifest---the only thing we see are the currents $J=*\Tr F$. Hidden symmetries arise at the fixed point (possibly as enhancement of topological symmetries), but are not symmetries of the UV Lagrangian.

Consider a monopole with flux $H=n$ in an Abelian factor. Then a CS term $\frac k{4\pi} \int A\wedge F$ induces electric charge $nk$.
Mixed Abelian CS terms $\sum_{ij} \frac{k_{ij}}{4\pi} \int A_i \wedge dA_j$ between dynamical and external (global) gauge fields induce charges under manifest global symmetries. Let $A_f$ be an external gauge field associated to an Abelian symmetry, then $\frac{k_{g,f}}{2\pi} \int A_f \wedge F$ (where $k_{g,f}$ is a gauge-flavor CS term) induces global charge $nk_{g,f}$. This argument immediately suggests that a monopole cannot transform under simple (in the sense of simple group) manifest global symmetries, because we cannot write a mixed CS term with a simple group. Notice that we are talking about \emph{bare} monopole operators: in the terminology of \cite{Borokhov:2002ib, Borokhov:2002cg, Borokhov:2003yu} they are the Fock vacuum in the fermionic Fock space of zero-modes, and so cannot form non-unidimensional representations. On the contrary the gauge-invariant monopoles of \cite{Borokhov:2002ib, Borokhov:2002cg, Borokhov:2003yu} are obtained by multiplication by fundamental fields, so that they can transform under simple manifest global symmetries.

In \cite{Kapustin:2006pk} it was shown that monopoles with flux in a generic gauge group $\cG$, in the presence of a CS term at level $k$, transform in a non-trivial representation of the gauge group. The flux is specified by a homeomorphism $\eta:U(1) \to \cG$ or equivalently by $H \equiv \eta\big( 1 \in \fu(1) \big) \in \fg$ (constrained by Dirac quantization), up to conjugation. The monopole action transforms as
\be
\delta S = k \Tr (H \, \delta A) \Big|_\text{monopole}
\ee
under a gauge transformation $\delta A$, so that the monopole transforms in a representation whose highest weight is $k\Tr(H \,\cdot\,)$.%
\footnote{In the case of a $U(N)$ gauge group that we will consider below we have $\fg\cong \fg^*$ and we can therefore write the weights simply as $kH$.}

Charges of monopole operators receive quantum corrections due to zero-modes, and for $\cN=2$ superconformal theories the one-loop answer is exact. In appendix \ref{app: charges of monopoles} we discuss the formul\ae{} for the quantum correction $\delta q$ to any Abelian charge, in case of generic gauge group $\cG$ and matter representations $R_\Phi$:
\be
\delta q = - \frac12 \sum_{\text{fermions } f} \,\, \sum_{\rho \in R_f} \big| \rho(H) \big| \, q_f
\ee
where $f$ are all fermions in the theory and $\rho$ are the weights of the representation.
It is easy to see that there are no quantum corrections to topological charges and that the charge under any Cartan of a simple manifest global symmetry group is zero, confirming that monopoles do not transform under simple manifest global symmetries.

Let us specialize here to quiver theories, where matter is restricted to the adjoint, bifundamental or (anti)fundamental representation.
We consider an $\cN =2$ CS quiver with gauge group $\cG= \prod_{i=1}^G U(N_i)$,  CS levels%
\footnote{In general $U(1)$ and $SU(N)$ in $U(N)$ can have different CS levels. We will discuss this possibility in section \ref{sec:_off_diag_CS} in relation to anomalies.}
$k_i$, bifundamental chiral superfields $X_{ij}$, fundamentals $q_{im}$ and antifundamentals $\tilde q_{mi}$. Let $F_i^+$ ($F_i^-$) be the number of (anti)fundamentals of the group $U(N_i)$.
A monopole operator is characterized by its magnetic charges (GNO charges \cite{Goddard:1976qe, Kapustin:2005py})
\be
\label{generic monopole op}
H_i=(n_{i,1}, \cdots, n_{i, N_i}) \;, \qquad\qquad i=1, \cdots, G
\ee
in the Cartan subalgebra $\fh$. Under any R-symmetry $U(1)_R$ (normalized such that gaugini have R-charge $1$), the monopole of (\ref{generic monopole op}) acquires a charge
\bea
R &= - \frac12 \sum_{X_{ij}} ( R[X_{ij}] - 1 ) \sum_{k=1}^{N_i}\sum_{l=1}^{N_j} |n_{i,k}-n_{j,l}| \,\, -\frac12 \sum_{i=1}^G \sum_{k=1}^{N_i}\sum_{l=1}^{N_i} |n_{i,k}-n_{i,l}| \\
&\quad - \frac12 \sum_i \bigg( \sum_{m=1}^{F^+_i} (R[q_{im}] - 1) + \sum_{m=1}^{F^-_i} (R[\tilde{q}_{mi}] - 1) \bigg) \sum_{k=1}^{N_i}|n_{i,k}| \;.
\eea
The second line, due to flavors, is the correction studied in \cite{Benini:2009qs, Jafferis:2009th}.
Since any $U(1)$ flavor symmetry is the difference of two R-symmetries, under any non-R symmetry $Q$ we have the induced charge
\bea
Q &= -\frac12 \sum_{X_{ij}} Q[X_{ij}] \sum_{k=1}^{N_i}\sum_{l=1}^{N_j} |n_{i,k}-n_{j,l}| \\
&\quad - \frac12 \sum_i \bigg( \sum_{m=1}^{F^+_i} Q[q_{im}] + \sum_{m=1}^{F^-_i} Q[\tilde{q}_{mi}] \bigg) \sum_{k=1}^{N_i}|n_{i,k}| \;.
\eea
In this paper we will not consider flavors, so the second lines can be neglected.

Let us consider now gauge symmetries. In \cite{Kim:2009wb, Imamura:2011su} it was shown (see appendix \ref{app: charges of monopoles}) that in doing localization, the integrand in the path-integral picks up a phase
\be
e^{ib_0(a)} \;,\qquad\text{with}\qquad b_0(a) = - \frac12 \sum_\Phi \sum_{\rho \in R_\Phi} |\rho(H)| \, \rho(a)
\ee
where $a \in \fh$ is the Cartan gauge field and the sum is over all chiral multiplets.
Since the function $b_0(a)$ is linear, it is easy to work out the variation of the action with respect to $A$, from which we infer that the monopole transform in a representation whose highest weight is
\be
\label{highest weight general}
w = k \Tr(H \,\cdot\,) - \frac12 \sum_\Phi \sum_{\rho \in R_\Phi} |\rho(H)| \, \rho \;.
\ee
In the special case of a quiver theory, the induced electric charges $g_{i,k}$ under each $U(1)_{i,k}$ ($k_i = 1,\cdots, N_i$) in Cartan subgroup of $\prod_i U(N_i)$ are
\be
\label{full gauge charge under U1}
g_{i,k} = k_i \, n_{i,k}   \; + \, \delta g_{i, k } \;.
\ee
Here $\delta g$ receives contributions from bifundamental and (anti)fundamental matter, but not from adjoint matter nor from gaugini:
\be
\delta g_{i, k } = - \frac12 \sum_{X_{ij}} \sum_{l=1}^{N_j} |n_{i,k}-n_{j,l}| \; + \frac12 \sum_{X_{ji}} \sum_{l=1}^{N_j} |n_{i,k}-n_{j,l}|\,\, -\,\,\frac12 \left(F_i^+-F_i^- \right) |n_{i,k}| \;.
\ee
We see that only ``chiral matter'' (matter in non-real representations) can induce gauge charges. The $U(1)$ charges (\ref{full gauge charge under U1}) define a weight for the gauge group $U(N_i)$,
\be
\label{full hw}
w_i = (g_{i,1}, g_{i,2}, \cdots, g_{i,N_i}) \;,
\ee
which is the highest weight (\ref{highest weight general}) of the representation under which the monopole transforms.
Whenever $\delta g_{i,k}$ are half-integer, the parity anomaly \cite{Niemi:1983rq,Redlich:1983kn, Redlich:1983dv} forces $k_i$ to be half-integer as well so that $g_{i,k}$ are always integers.

Finally, consider hidden symmetries whose Cartan currents are visible as topological currents $J = *\Tr F$: monopoles are by definition charged under them. Full representations of a hidden symmetry can be formed by different monopoles (as opposed to different states of a single monopole), therefore monopoles can transform under simple hidden global symmetries.%
\footnote{For $\cN\geq 4$ SUSY theories these hidden symmetries have been studied in \cite{Gaiotto:2008ak, Bashkirov:2010kz}.}
Notice that the distinction between manifest and hidden global symmetries is unphysical, but so is the distinction between monopoles and matter fields.

\subsection{Toric quivers and diagonal monopoles}\label{subsec:_toric_quiver_and_diag_mono}

The knowledge of the quantum charges of monopole operators allows us in principle to work out all holomorphic gauge invariant operators of the theory. What we are interested in, however, is the chiral ring, defined through relations between the holomorphic operators:
\be
A(\text{quiver}) = \frac{ \bC \left[\cO_1, \cO_2, \cdots\right] }{\cI} \;.
\ee
In 4d quiver SCFTs, the chiral ring is generated by chiral fields in the Lagrangian and the ideal $\cI$ can be read from the classical F-terms: $\cI= (\partial W)$; more precisely $\cI$ consists of all gauge invariant relations that follow from $\partial W=0$, together with the so-called syzygies \cite{Benvenuti:2006qr}. In 3d there are two differences: 1) the chiral ring is
generated by chiral monopole operators besides chiral fundamental fields; 2) there might exist relations between monopole operators which are not easily derived from the superpotential. For this reason the chiral ring of a 3d quiver is much more complicated to analyze than that of its 4d parent. Since we do not know of honest field theory methods to compute the quantum chiral ring relations in general, we will keep the strategy followed in \cite{Benini:2009qs, Jafferis:2009th}: an educated guess of the most important chiral ring relations between so-called diagonal monopole operators, based on their global charges.

Consider a quiver with ranks $N_i = \tilde{N} + M_i$, $\tilde{N}=\textrm{min}\{N_i\}$, and no flavors. We focus on diagonal monopole operators $T_H$, which turn on fluxes
\be
T_H\;: \qquad n_{i,k} \equiv n_{0,k} \quad \forall i \;, \quad \text{if } k\leq \tilde N \;, \qquad\ \text{and} \quad n_{i, k}=0 \;, \quad \text{if } k > \tilde N \;.
\ee
We will be mainly interested in the simplest diagonal monopole operators
\be
\label{def T, Tt}
T, \,\tilde{T} \;: \qquad H=(\pm 1, 0, \cdots, 0) \;.
\ee
In the case of equal ranks, $M_i=0$, the R-charges of $T_H$ are
\be
\label{R charge quiver with Ni eq N}
R[T_H] = -\frac12 \Bigg( G + \sum_{\text{fields } X} (R[X]-1) \Bigg) \sum_{k=1}^{\tilde{N}}\sum_{l=1}^{\tilde{N}} |n_{0,k}-n_{0,l}| \;.
\ee
The quantity in parenthesis automatically vanishes for \emph{toric quivers}, also known as brane
tilings (see \cite{Kennaway:2007tq} for a review). We will restrict the following analysis to such theories. A brane tiling is a bipartite graph, where each gauge group is represented by a face $F_i$ ($i=1, \cdots, G$), each bifundamental field by an edge $X_{ij}$ between two faces, and each superpotential term $W_{\alpha}$ by a black/white vertex. We have the further constraint that the graph tiles a torus, which implies $G + V-E =0$, where $V$ and $E$ are the number of vertices and edges respectively. We have $\sum_X R[X] = \frac12 \sum_{W_\alpha} \sum_{X \in W_\alpha} R[X] = \frac12 \sum_{W_\alpha} 2 = V$, so that the quantity in parenthesis equals $G + V - E = 0$. Similarly one can show that diagonal monopoles do not receive any quantum correction at all. The chiral ring is generated by chiral fields in the Lagrangian as well as $T$, $\tilde T$, subject to the classical relation $T\tilde T = 1$ \cite{Benini:2009qs}. Hence the classical analysis of the moduli space, as in \cite{Martelli:2008si}, gives the correct result.

If the ranks $N_i= \tilde{N}+M_i$ are unequal, the extra contribution to the R-charge of diagonal monopole operators is
\be
\label{deltaR general}
\delta R[T_H] = - \frac12 \Bigg( \sum_{\text{fields } X_{ij}} (R[X_{ij}] - 1)(M_i+M_j) + \sum_{i=1}^G 2M_i \Bigg) \sum_{k=1}^{\tilde{N}} |n_{0,k}| \;.
\ee
For a toric quiver we have
\be
\sum_{\text{fields } X_{ij}} M_i = \sum_{\text{fields } X_{ij}} M_j = \frac 12 \sum_{W_\alpha} \sum_{X_{ij}\in W_{\alpha}} M_i = \frac12 \sum_{W_\alpha} \sum_{F_i \in W_{\alpha}} M_i = \frac12 \sum_{F_i} M_i \sum_{W_\alpha \in F_i} 1 \;,
\ee
where $F_i \in W_\alpha$ means the faces that touch the vertex $W_\alpha$, and reciprocally for $W_\alpha \in F_i$. Denoting the number of edges (or equivalently vertices) around a face $F_i$ by $E_i$, we can reshuffle (\ref{deltaR general}) into
\be
\label{R-charge_monopole}
R[T_H] = \frac12 \Bigg( \sum_i ( E_i-2 ) M_i \, -\sum_{\text{fields } X_{ij}} R[X_{ij}](M_i+M_j)   \Bigg) \sum_{k=1}^{\tilde{N}} |n_{0,k}|  \;.
\ee
Remark that $E_i-2 >0$ (unless there are double bonds \cite{Davey:2009sr} in the brane tiling, a situation we will not consider); importantly, $E_i-2$ is even.

Similarly we can compute the electric charge (\ref{full gauge charge under U1}) under each $U(1)_{i,k}$ in the Cartan of the gauge group:
\be\label{monopole_electric charge}
g_{i,k}[T_H] = k_i\, n_{0,k} \,  - \frac12 \Bigg( \sum_{X_{ij} } M_j \, - \sum_{X_{ji}} M_j  \Bigg)\, |n_{0,k}| \;,
\ee
for $k=1, \cdots, \tilde{N}$, while $g_{i, k} =0$ for $k> \tilde{N}$. Monopoles transform in a representation whose highest weight is $w=(g_{i,1}, \cdots ,g_{i, N_i })$ and for the simplest monopoles $T$, $\tilde{T}$ (\ref{def T, Tt}) we have
\be
w_i(T) =(g_i, 0, \cdots, 0) \;, \qquad\qquad w_i(\tilde{T}) =(\tilde{g}_i, 0, \cdots, 0) \;,
\ee
with
\be
\label{charges Qi of T Tt in CFT}
g_i = k_i -\frac12 \Bigg( \sum_{X_{ij}} M_j \, - \sum_{X_{ji}} M_j  \Bigg)  \;,\qquad \tilde{g}_i = -k_i -\frac12 \Bigg( \sum_{X_{ij}} M_j \, - \sum_{X_{ji}} M_j  \Bigg) \;.
\ee
Thus the monopole operators $T$, $\tilde{T}$ transform in symmetric representations of the $U(N_i)$ gauge groups.

The quantum numbers of $T$, $\tilde{T}$ strongly constrain the possibilities for chiral ring relations involving $T\tilde{T}$. Generically we can have the gauge invariant relation (there could be several ways to contract gauge indices)
\be
\label{general_chiral_ring_rel_TTt}
T \tilde{T} \, \prod_X X_{ij}^{(M_i+M_j)} \,  = \, \prod_{\alpha} (W_{\alpha})^{m_{\alpha}} \;,
\ee
with $ \sum_{\alpha} m_{\alpha} = \tfrac{1}{2}\sum_i (E_i-2)M_i $. Here $W_{\alpha}$ stands for the superpotential terms. Superpotential terms have the property of having R-charge 2, vanishing charges under non-R $U(1)$ symmetries, and of being gauge invariant.  This is why they appear in the previous relation. Note that all superpotential terms of a toric quiver are equivalent in the chiral ring, therefore there is no ambiguity related to a choice of superpotential term. (\ref{general_chiral_ring_rel_TTt}) can sometimes be simplified if a gauge invariant operator can be factored out.

The gauge invariant relations involving both $T$ and $\tilde{T}$ do not all necessarily take the form \eqref{general_chiral_ring_rel_TTt}. If a field appears in both sides of \eqref{general_chiral_ring_rel_TTt}, we can replace it by another (elementary or composite) field charged under the same representation of the gauge group, if the latter exists. So doing we may find gauge invariant relations which transform covariantly under the global symmetries of the quiver gauge theory.


\subsection{Moduli spaces of toric quivers from monopole operators}
\label{subsec:_chiral_ring_for_CY4}

Our purpose is to apply the method explained above to compute the Coulomb branch of the moduli space of CS quivers, and eventually compare it with some CY$_4$ used in the M-theory background. In particular if we have a 3d CS quiver conjectured to describe the low energy dynamics of M2-branes on a CY$_4$, we expect the moduli space of the theory to contain $\tilde N$ symmetrized copies of it.%
\footnote{$\tilde N$ need not be equal to the total number $N$ of M2-branes, but only to the number of mobile M2-branes. We will see in examples that generically $\tilde N \leq N$.}

One way of characterizing such geometric branch of the Coulomb moduli space is to compute the chiral ring of the quiver, and compare it with the coordinate ring of the CY$_4$. Let us consider the pseudo-Abelian case, $\tilde N=1$. We collect the gauge invariant operators constructed out of $\{ X_{ij}, T, \tilde{T}\}$ with at most one power of $T$ or $\tilde{T}$, which we denote $\cO^{(0)}$, $\cO^{(\pm 1)}$ according to their magnetic charge, and construct
\be
\label{geometric_moduli_space}
A(\text{quiver})_{\mathrm{geom}} \equiv  \frac{\bC \left[\cO^{(0)}, \cO^{(+1)}, \cO^{(- 1)} \right]}{\big( \partial W,\;  T\tilde{T} \cP_1(X)-\cP_2(X) \big)} \;,
\ee
where $T\tilde{T} \cP_1(X) =\cP_2(X)$ are the quantum F-term relations proposed in (\ref{general_chiral_ring_rel_TTt}). We will find in examples that indeed $A(\text{quiver})_{\mathrm{geom}} = A(\text{CY}_4)$. In the case of flavored quivers with equal ranks it was possible \cite{Benini:2009qs} to give a general proof of that, due to manipulations of the brane tiling techniques of \cite{Imamura:2008qs, Hanany:2008fj}. In the present case of quivers with unequal ranks it seems that such easy techniques are not available.

We will talk about the geometry in later sections, but we already anticipated in the introduction that to derive the field theories we KK reduce the CY$_4$ along a wisely-chosen circle, such that we obtain a CY$_3$ fibered along $\bR$, and then exploit the quiver that describes D-branes at the tip of the CY$_3$. Indeed in the field theory we can consider the subring
\be
\label{CY3 subring}
A(\text{CY}_3) = \frac{\bC \left[\cO^{(0)}\right]}{(\partial W)}
\ee
which is precisely the coordinate ring of the aforementioned CY$_3$.


\subsection{Moduli spaces from a semi-classical computation}
\label{subsec:_VMS_from_1loop}

We can approach the computation of the Coulomb branch of the moduli space in a more conventional way, by performing a semi-classical calculation that includes one-loop effects in the effective theory on the Coulomb branch. This approach follows the work of  \cite{deBoer:1997kr, Aharony:1997bx, Dorey:1999rb, Tong:2000ky} on 3d $\cN=2$ Yang-Mills (YM) and Yang-Mills-Chern-Simons (YM-CS) theories, and gives results that perfectly match with the quantum chiral ring presented above. The advantage of the computation with monopoles is that it is one-loop exact;%
\footnote{There can be non-perturbative corrections to the superpotential on the moduli space, though.}
the disadvantage is that it knows only the complex structure of the moduli space and not its K\"ahler structure. On the other hand the semi-classical computation probes the K\"ahler structure and so it is particularly suited to analyze partial resolutions (to be discussed in a specific example in section \ref{subsec: mod space resolutions}), even though it does not capture possible non-perturbative corrections to the metric on the moduli space.

Consider a 3d $\cN=2$ YM-CS quiver as in the previous sections, with gauge group $\prod_{i=1}^G  U(\tilde{N}+M_i)$, YM coupling constants $e_i$, and bifundamental fields $X_{ij}$. The classical YM-CS scalar potential reads (see \eg{} \cite{Martelli:2008si} for a nice account):
\be
V = \sum_{i=1}^{G} \, \frac{e^2_i}2 \Big( \cD_i - \sigma_i \frac{k_i}{2\pi} - \xi_i^\text{bare} \Big)^2 + \sum_{\text{fields } X_{ij}} \big| \sigma_i X_{ij}- X_{ij}\sigma_j \big|^2 + \sum_{\text{fields } X_{ij}} \Big| \parfrac{W}{X_{ij}} \Big|^2 \;,
\ee
where $\sigma_i$ are the Hermitian scalar fields in 3d $\cN=2$ vector multiplets, $\cD_i$ are the 4d D-terms
\be
\cD_i \equiv \sum_{X_{ij}} X_{ij} X_{ij}^{\dagger} - \sum_{X_{ji}}  X_{ji}^{\dagger} X_{ji} \;,
\ee
$W$ is the superpotential and $\xi_i^\text{bare}$ are possible bare Fayet-Iliopoulos (FI) terms, that for the moment we set to zero. Vanishing of $V$ would lead to the equations
\be
\label{VMS equ: Dterm}
\cD_i = \sigma_i \frac{k_i}{2\pi} \;, \qquad\qquad \sigma_i X_{ij}- X_{ij}\sigma_j =0 \qquad \forall \,i,\,j \;,\qquad\qquad \partial_{X_{ij}} W=0 \;.
\ee
Notice that the first set of equations become pure constraints in a purely CS theory (formally in the limit $e_i \rightarrow \infty$).

The semi-classical analysis of the moduli space goes as follows. First we choose a background for the Hermitian scalars in vector multiplets, diagonalized via gauge transformations:
\be
\label{generic_VEVs_for_sigma}
\sigma_i = \mathrm{diag}(\sigma_{n_i}^{(i)}) \;, \qquad \quad n_i=1, \cdots, \tilde{N}+M_i \;.
\ee
These VEVs partially break the gauge group (generically to the maximal torus) and give an effective real mass to most of the components of bifundamentals $X_{ij}$,
\be
\delta M[(X_{ij})\ud{n_i}{n_j}] = \sigma^{(i)}_{n_i} - \sigma^{(j)}_{n_j},
\ee
freezing the massive fields to vanishing VEV.
Integrating out the massive chiral multiplets (which include fermions) generates CS interactions at one-loop, as reviewed in appendix \ref{app:_shift_of_CS_levels}. We thus compute effective field-dependent CS levels $k_i^\text{eff}$ and FI parameters $\xi_i^\text{eff}$ for the unbroken gauge group, which depend on the VEVs \eqref{generic_VEVs_for_sigma}.
Finally, we look for SUSY vacua solving D-term (which contain the effective FI parameters) and F-term equations of the effective theory, and modding out by the unbroken gauge group.

The moduli space contains more directions: all photons in the effective theory which are not coupled to matter can be dualized to real periodic pseudoscalars---the \emph{dual photons}---which, because of the CS couplings, shift under some gauge transformations. One can use those gauge transformations to gauge fix the dual photons, as in \cite{Aharony:2008ug, Martelli:2008si} but here in the effective theory. As a result the space of solutions of F-term and $\sigma$-dependent D-term equations has to be modded out by a subgroup of the unbroken gauge group (possibly including a residual discrete gauge symmetry that depends on $k_i^{eff}$).

Let us consider the \emph{geometric branch} of the moduli space, defined by
\be
\label{geometric_branch_sigma}
\sigma_i = \mathrm{diag}(\sigma_1, \cdots, \sigma_{\tilde{N}}, 0, \cdots, 0) \qquad \forall\, i=1,\cdots,G \;,
\ee
in which all real scalars are equal. To begin with, we fix to zero the extra $M_i$ components of $\sigma_i$, that in the following we will call $\tilde \sigma_m$ ($m = 1,\cdots, \max(M_i)$). The gauge group is generically broken as
\be
\prod_{i=1}^G \, U(\tilde{N}+M_i) \quad\to\quad \left(U(1)^G\right)^{\tilde{N}} \, \prod_{i=1}^G \, U(M_i)
\ee
and the allowed bifundamental VEVs generically are
\be
\label{bifund_VEV_CFT}
X_{ij} = \mat{ X^\text{diag}_{ij} & 0 \\ 0 & 0 } \;,
\ee
where $X^\text{diag}_{ij}$ are $\tilde{N}\times \tilde{N}$ diagonal matrices. Depending on the quiver and the ranks, additional diagonal entries of $X_{ij}$ might acquire VEV, but we consider here VEVs which are always allowed by \eqref{geometric_branch_sigma}. In the CFTs of the next sections, diagonal \eqref{bifund_VEV_CFT} are the most general VEVs allowed by generic \eqref{geometric_branch_sigma}.

Each of the $\tilde{N}$ $U(1)^G$ factors represents a copy of the Abelian quiver, under which only the corresponding eigenvalues of $X^\text{diag}_{ij}$ are charged. Between each copy and the remaining $\prod_i U(M_i)$ gauge groups there can be chiral fields that get massive on the Coulomb branch and should be integrated out, shifting the CS levels.
On the other hand, for generic \eqref{geometric_branch_sigma} the $\tilde N$ Abelian quivers decouple.
Since permutations of eigenvalues are a residual gauge symmetry, the geometric branch of the CFT is an $\tilde{N}$-symmetric product of a $U(1)^G$ quiver moduli space with D- and F-term equations
\be
\label{geometric_branch_CFT_abelian}
\xi_{i}^\text{eff}(\sigma) = \sum_{X_{ij}} |x_{ij}|^2 - \sum_{X_{ji}}  |x_{ji}|^2 \;, \qquad\qquad  \partial_{x_{ij}} W(x)=0 \;,
\ee
where $x_{ij} \in \bC$. The equations for $\prod_i U(M_i)$ are trivially solved. The effective CS terms are $k_i^\text{eff} = k_i + \frac12 \sign(\sigma) \big( \sum_{X_{ij}} M_j - \sum _{X_{ji}} M_j \big)$, while the effective FI parameters are
\be
\label{xi_eff_geometric_branch}
\xi_i^\text{eff}(\sigma)= \frac{k_i}{2\pi} \sigma  + \frac1{4\pi} \bigg(\sum_{X_{ij}} M_j - \sum_{X_{ji}} M_j \bigg) |\sigma| \;,
\ee
which is a particular case of formula (\ref{general formula for xi eff}) and includes the classical and one-loop contribution.
At fixed $\sigma$ \eqref{geometric_branch_CFT_abelian} describes the CY$_3$ associated to the quiver, as is well known, with resolution parameters corresponding to $\xi_i^\text{eff}(\sigma)$. Together they describe a CY$_3$ fibered over a line $\bR \cong \{\sigma\}$.

We can rewrite (\ref{xi_eff_geometric_branch}) more compactly using the adjacency matrix $A_{ij}$ of the quiver: the integer $A_{ij}=-A_{ji}$ is the net number of arrows from node $i$ to node $j$. We have
\be
\label{xi_eff_geometric_branch bis}
\xi_i^\text{eff}(\sigma)= \frac{k_i}{2\pi} \sigma + \frac1{4\pi} |\sigma| \sum_{j=1}^G A_{ij}\, N_j \;.
\ee
Note that since the quivers we consider  have as many incoming as outgoing arrows at each node,  $\sum_j A_{ij} N_j = \sum_j A_{ij} M_j$ for $N_j = \tilde{N}+M_j$.

Each Abelian quiver provides a dual photon as well: the diagonal $U(1)$ vector in $U(1)^G$ can be dualized to $\varphi$ as in \cite{Aharony:1997bx, Aharony:2008ug, Martelli:2008si}, and together with the CY$_3$ bundle over $\bR$ they make the moduli space a four-fold (in fact a CY$_4$). The dual photon shifts under the topological symmetry associated to the diagonal magnetic flux (while matter fields are invariant), and this topological symmetry maps to a $U(1)_M$ isometry of the four-fold. The main difference with respect to \cite{Martelli:2008si} is that the effective theory is one-loop corrected with respect to the ``bare quiver'', and the regions at $\sigma \gtrless 0$ are not continuously related. Indeed the quantized effective CS levels jump at $\sigma=0$, and this is possible only if the circle parametrized by the dual photon shrinks there.%
\footnote{To be more precise, since the dual photon is not gauge invariant, we should say that the global circle action---that shifts the dual photon and leaves the other fields invariant---has a fixed point.}

Finally let us briefly consider switching on bare FI parameters $\xi_i^\text{bare}$ and the extra eigenvalues $\tilde \sigma_m$ in
\be
\sigma_i = \mathrm{diag}(\sigma_1, \cdots, \sigma_{\tilde{N}}, \tilde\sigma_1, \cdots, \tilde\sigma_{M_i}) \qquad \forall\, i=1,\cdots,G \;.
\ee
The $\xi_i^\text{bare}$ simply enter in $V$ and therefore in (\ref{xi_eff_geometric_branch}); the $\tilde\sigma_m$ effectively provide real masses and so deform (\ref{xi_eff_geometric_branch}) further:
\be
\label{eff FI resolved}
\xi_i^\text{eff} = \xi_i^\text{bare} + \frac{k_1}{2\pi} \sigma + \frac1{4\pi} \bigg( \sum_{X_{ij}} \sum_{m=1}^{M_j} |\sigma - \tilde\sigma_m| - \sum_{X_{ji}} \sum_{m=1}^{M_j} |\sigma - \tilde\sigma_m| \bigg) \;.
\ee
This time $\prod_i U(M_i) \to U(1)^{\sum_i M_i}$ and their equations are non-trivial.
Depending on the CS levels and ranks of the quiver, it is not generically possible to solve the D-term equations for the massless fields. However when it is possible, they provide resolution parameters of the geometric moduli space, as we will see in section \ref{subsec: mod space resolutions}. It is clear from (\ref{eff FI resolved}) that every time $\sigma$ crosses one of the $\tilde\sigma_m$, a real mass changes sign, the effective CS levels jump and the effective field theory changes. Therefore at $\sigma = \tilde\sigma_m$ the dual photon degenerates. Finally, each of the resolutions modes $\tilde\sigma_m$ is also related to a dual photon: together they describe complexified K\"ahler classes of the four-fold.

To conclude, let us stress how the two approaches, semi-classical and via monopoles, agree: The SCFT chiral ring analysis of sections \ref{subsec:_toric_quiver_and_diag_mono}--\ref{subsec:_chiral_ring_for_CY4} reproduces the coordinate ring of a complex variety (it will turn out to be a CY$_4$ in our examples), which by (\ref{CY3 subring}) contains as a subvariety the CY$_3$ associated to the quiver.
On the other hand the semi-classical approach
reproduces the data of a IIA background: it yields a foliation of a CY$_3$ along $\bR$.
The full four-fold moduli space is achieved with the inclusion of the dual photon.


\section{From M-theory to type IIA}
\label{sec:M_to_IIA}

Consider M2-branes probing a CY$_4$ geometry, which is the cone over a Sasaki-Einstein seven-manifold $Y_7$.
In \cite{Aganagic:2009zk} Aganagic proposed an elegant method to derive the 3d low energy field theory on the M2-branes: one KK reduces M-theory along a wisely-chosen circle in the CY$_4$ in such a way that the resulting type IIA background is a CY$_3$ fibration along a real line $\bR = \{r_0\}$ with RR 2-form flux.%
\footnote{This is a topological statement: the IIA metric on a CY$_3$ fiber at fixed $r_0$ needs not be Ricci-flat.}
The analysis of \cite{Aganagic:2009zk} was restricted to a $U(1)_M$ group  acting freely outside the CY$_4$ singularity, but we will generalize it allowing fixed points of $U(1)_M$ below.
We will call the resulting type IIA manifold $X_7 \equiv \text{CY}_4/U(1)_M$, which can be represented as a fibration over $\bR$ of the CY$_3$ given by the K\"ahler quotient $\text{CY}_3 = \text{CY}_4 \kq U(1)_M$. If the CY$_4$ is a cone over a Sasaki-Einstein $Y_7$, $X_7$ can also be viewed as a cone over the six-manifold $M_6 = Y_7/U(1)_M$.
M2-branes are mapped to D2-branes, so the problem boils down to finding the field theory dual to transverse D-branes probing a CY$_3$ with 2-form fluxes. Much about this problem is known, mainly from the study of D3-branes on three-folds: the low energy worldvolume theory on the branes is a quiver gauge theory. We will focus on toric CY$_4$'s which descend to toric CY$_3$'s, because it is easy to analyze their geometry and the field theory dual to toric CY three-folds with D-branes is always known \cite{Hanany:2005ss}; however the same logic goes through the more general non-toric case.

In type IIA the CY$_3$ is fibered along $\bR = \{r_0\}$, with its K\"ahler moduli being linear functions of $r_0$. When the CY$_4$ is conical, the CY$_3$ fiber over $r_0 = 0$ is conical rather than resolved.
Since $U(1)_M$ is in general non-trivially fibered over its base, the reduction also introduces RR $F_2$ flux in IIA.
In addition (see appendix \ref{app: torsion flux}) a flat dynamically quantized NSNS potential $B_2$ may appear. D2-branes at the CY$_3$ singularity break up into mutually BPS states, called \emph{fractional D2-branes}, which in the large volume limit can be thought of as D4-branes wrapped on 2-cycles or D6-branes wrapped on 4-cycles, possibly with worldvolume fluxes $F$. It is easy to see from the Wess-Zumino (WZ) action that $F_2$ and $\cF = \hat B_2 + F$ (hat means pull-back) induce CS terms in the three-dimensional action. The 3d theory is then a quiver gauge theory with CS terms.

With $\cN=2$ supersymmetry (the one preserved by M2-branes on a CY$_4$) the CS term is contained in the superspace piece $k\int d^4\theta\, V \, \Sigma(V) \supset k \big( \Tr \omega_{CS}(A) + 2\Tr D\sigma \big)$, where $k$ is the CS level, $V$ the vector multiplet, $\Sigma$ the linear multiplet containing its field strength, $\omega_{CS}$ the CS form, $\sigma$ the real scalar in the vector multiplet and $D$ the D-term. As we vary $\sigma$, that we will see corresponds to $r_0$, we get an effective FI parameter linear in $\sigma$. Recalling that FI parameters in the gauge theory measure resolution parameters of the transverse CY$_3$ geometry, we see that whenever $F_2$ and $B_2$ are constant, the K\"ahler parameters of the CY$_3$ are linear functions of $r_0$. Supersymmetry relates the constant $F_2$ flux through a 2-cycle in the CY$_3$ to the first derivative with respect to $r_0$ of the K\"ahler parameter of the 2-cycle.

So far we have assumed as in \cite{Aganagic:2009zk} that the KK reduction of the M-theory CY$_4$ geometry is not singular: in that case the integrals of $F_2$ and $B_2$ on the CY$_3$ are constant along $r_0$. This need not be the case, and in fact generically the $U(1)_M$ action degenerates.
The simplest degeneration that we can allow are sets of fixed points for the $U(1)_M$ action. Such KK monopoles lead to D6-branes wrapping 4-cycles of the CY$_3$: if the wrapped 4-cycle is noncompact the D6-brane is visible also in $M_6$ (as studied in \cite{Benini:2009qs, Jafferis:2009th}); if instead it is a compact 4-cycle hidden at the CY$_3$ singularity over $r_0=0$ the D6-brane does not appear in $M_6$. However they become visible as soon as a partial resolution of the CY$_4$ blows up the exceptional 4-cycle in the CY$_3$. In any case it is simple to trace their presence in the reduction of the CY$_4$ to $X_7$ because the cohomology class of $F_2$ in the CY$_3$ jumps at their location on the base $\bR$. The type IIA background involves piecewise constant (in $r_0$) RR 2-form fluxes through 2-cycles, equal by supersymmetry to the first derivatives in $r_0$ of the piecewise linear K\"ahler parameters of the same 2-cycles. This behavior is reflected in the dual field theory: the D6-branes add matter fields (from D2-D6 strings) with real masses dependent on $\sigma$, leading to jumps in the effective CS levels (first derivatives of effective FI parameters) whenever a real mass changes sign.

In fact whenever the CY$_3$ has exceptional 4-cycles (this is the case if the CY$_4$ has exceptional 6-cycles), a necessary condition for the type IIA background to be in a geometric phase of the CY$_3$ sigma model is that compact D6-branes be present.
We make the latter requirement since we will rely on the DBI-WZ action of D-branes
to deduce their worldvolume field theory.
Holomorphic 2-cycles inside exceptional 4-cycles in the CY$_3$ cannot flop, therefore their K\"ahler parameters must remain non-negative to keep the CY$_3$ fiber in a geometric phase over the entire $\bR$ line. Since they are piecewise linear on $\bR$, a suitable number of D6-branes wrapping the exceptional 4-cycle are needed.

The discussion above was general, but exploiting toric geometry allows us to be very explicit. After performing an $SL(3,\bZ)$ transformation such that the toric $U(1)_M$ corresponds to the vertical direction in the 3d toric diagram of the CY$_4$, the 2d toric diagram of the CY$_3$ is the vertical projection of the latter.
Each $U(1)_M$ KK monopole (D6-brane) is identified by a pair of adjacent vertically aligned points in the 3d toric diagram of the CY$_4$ \cite{Benini:2009qs}, and wraps a toric divisor in the CY$_3$ associated to the point in the 2d toric diagram that the pair of points projects to.

In the following we will analyze these issues, thus generalizing \cite{Benini:2009qs} to compact D6-branes, in the simplest model: the family of CY$_4$ cones over the so-called $Y^{p,q}(\CPt)$ 7-manifolds \cite{Gauntlett:2004hh, Martelli:2008rt}. Notable members are $Y^{2,3} \equiv M^{3,2}= M^{1,1,1}$ and $Y^{p,0} = S^7/\bZ_{3p}$. The type IIA reduction gives $\bC^3/\bZ_3$, which has an exceptional $\CPt$ at the tip. A discussion of more general geometries is left for a companion paper \cite{to:appear}.

\subsection{The geometry of $Y^{p,q}(\CPt)$}

We consider, as a specific example, M2-branes in M-theory probing a family of toric CY four-folds which are cones over the Sasaki-Einstein (SE) seven-manifolds $Y^{p,q}(\CPt)$ introduced in \cite{Gauntlett:2004hh, Martelli:2008rt}. $Y^{p,q}$ are $S^3/\bZ_p$ bundles over $\CP^2$, parameterized by the integer $q$. The SE$_7$ metrics are known \cite{Gauntlett:2004hh, Martelli:2008rt}, but we will not use them. The manifolds are smooth for $0<q<3p$; if $\gcd(p,q) = k>1$, then $Y^{p,q} = Y^{p/k,q/k}/\bZ_k$ where the orbifold group acts freely. When $q=0,3p$, $Y^{p,q} = Y^{1,q/p}/\bZ_p$ but this time the resulting cone has a complex codimension-three line of orbifold singularities which locally looks like $\bC^3/\bZ_3$.
Replacing $q \rightarrow 3p-q$ gives an identical geometry \cite{Martelli:2008si}. It corresponds to a change of orientation of the M-theory circle, as we explain in section \ref{subsec: parities} below.

\begin{figure}[tn]
\begin{center}
\includegraphics[width=4cm]{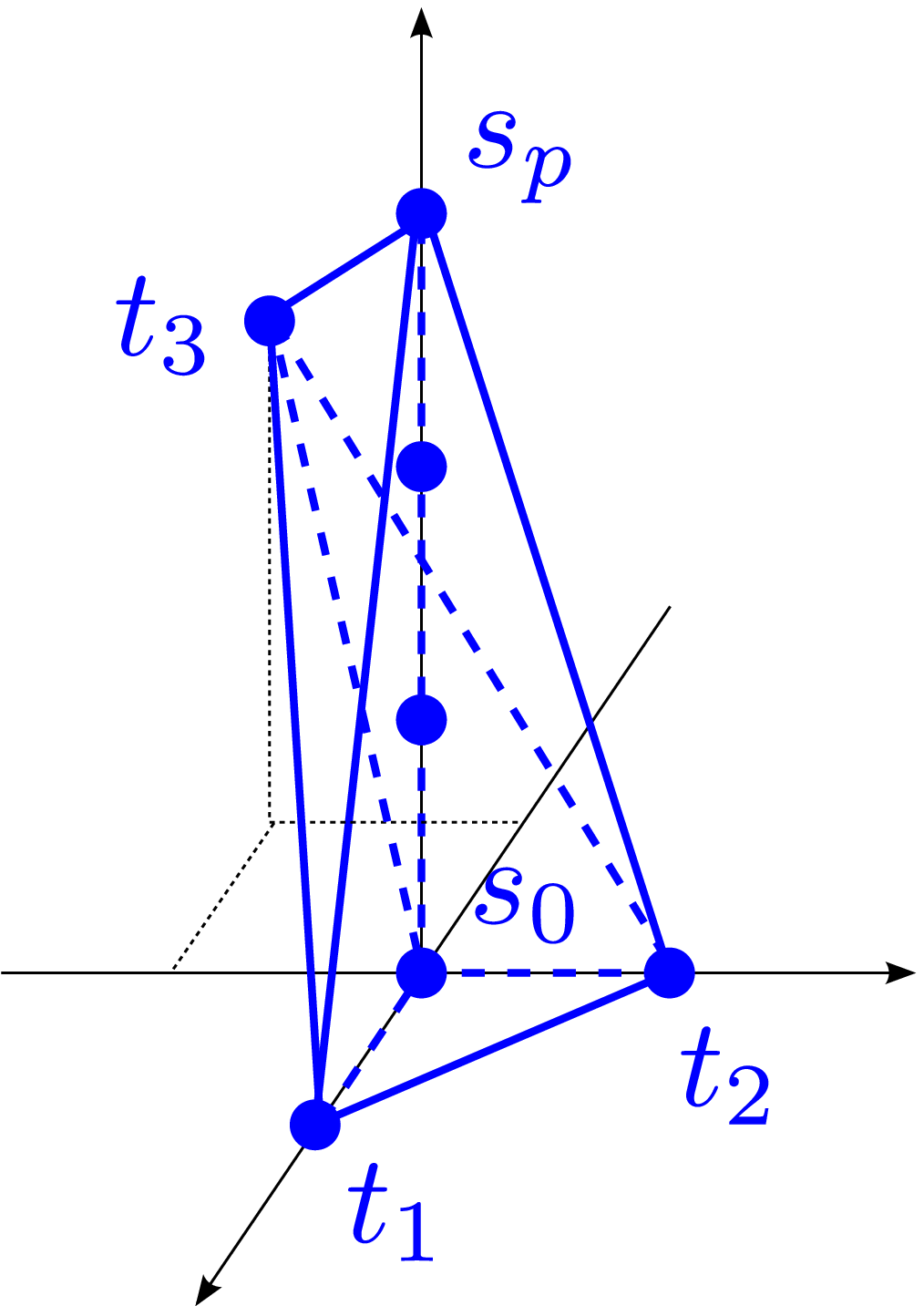}
\caption{Toric diagram of $C(Y^{p,q})$ (drawn for $Y^{3,2}$). \label{fig: toric diagram Ypq}}
\end{center}
\end{figure}

The cone over $Y^{p,q}$ is a toric CY$_4$ and its toric diagram has five external points \cite{Martelli:2008rt}
\be\label{toric vectors for Ypq}
t_1=(1,0,0) \qquad t_2 = (0,1,0) \qquad t_3 = (-1,-1,q) \qquad s_0 = (0,0,0) \qquad s_p = (0,0,p)
\ee
as in figure \ref{fig: toric diagram Ypq}. Recall that the 3d toric diagram is defined up to $SL(3,\bZ)$ transformations. We have already arranged the diagram such that the K\"ahler reduction along $U(1)_M$ we are interested in corresponds to a vertical projection.

In general a toric CY$_4$, including all its toric crepant resolutions%
\footnote{Remark that in complex dimension 2 crepant resolutions always exist and are unique, in dimension 3 always exist but need not be unique, in dimension 4 and bigger need not exist. The simplest example is $\bC^4/\bZ_k$, which does not have crepant resolutions. Whenever $\Lambda = \{1\}$, the toric CY can be completely smoothed out by toric crepant resolutions.}%
, can be realized as the moduli space of a supersymmetric Abelian gauged linear sigma model (GLSM), quotiented by a finite Abelian group $\Lambda$. Calling $w_A \in \bZ^3$ the points of the 3d toric diagram, the GLSM is obtained by assigning a complex field to each vector and assigning a $U(1)$ gauge symmetry to each linear relation among $v_A = (1,w_A) \in \bZ^4$. The Abelian group is then $\Lambda = \bZ^4 / \text{Span}_\bZ \{ v_A\}$. In our examples $\Lambda$ is trivial.

For the cones over $Y^{p,q}$, the GLSM can be written (excluding the last line, that will be relevant for the reduction to type IIA later):
\be
\label{full GLSM dP0}
\begin{array}{c|cccccccccc|c}
\text{CY}_4 & t_1 & t_2 & t_3 & s_0 & s_1 & s_2 & \dots & s_{p-2} & s_{p-1} & s_p & \text{FI} \\
\hline
\cC & 1 & 1 & 1 & q-3 & -q & 0 & \dots & 0 & 0 & 0 & \xi^c \\
\cC_2 & 0 & 0 & 0 & 1 & -2 & 1 & \dots & 0 & 0 & 0 & \xi_2 \\
\cC_3 & 0 & 0 & 0 & 0 & 1 & -2 & \dots & 0 & 0 & 0 & \xi_3 \\
\vdots & \vdots & \vdots & \vdots & \vdots & \vdots & \vdots & \ddots & \vdots & \vdots & \vdots & \vdots \\
\cC_{p-1} & 0 & 0 & 0 & 0 & 0 & 0 & \dots & -2 & 1 & 0 & \xi_{p-1} \\
\cC_p & 0 & 0 & 0 & 0 & 0 & 0 & \dots & 1 & -2 & 1 & \xi_p \\
\hline
U(1)_M & 0 & 0 & 0 & 1 & -1 & 0 & \dots & 0 & 0 & 0 & r_0
\end{array}
\ee
The first line denotes the fields. The second line describes their charges under a $U(1)$ subgroup, and the subset $\{t_1,t_2,t_3\}$ describes $\CPt$. The following $p-1$ lines describe the GLSM for a $\bC^2/\bZ_p$ singularity, fibered over $\CPt$. The last column includes FI parameters of the GLSM: they control resolutions of the geometry, and $\xi_{2,\cdots,p}$ should be non-negative to keep the GLSM in a geometric phase. By abuse of notation we have identified the $p$ $U(1)$ gauge groups with holomorphic 2-cycles $\cC_a$.

In section \ref{sec: field theory} it will be useful to have an algebraic description of the CY$_4$ as a non-complete intersection. From the GLSM we construct the following gauge invariants:
\bea
a_{(3)} &= t_{i_1} t_{i_2} t_{i_3} s_0 s_1\dots s_p \qquad\qquad & b_{(q)} &= t_{i_1} \dots t_{i_q} s_1 s_2^2 \dots s_p^p \\
&& c_{(3p-q)} &= t_{i_1} \dots t_{i_{3p-q}} s_0^p s_1^{p-1} \dots s_{p-1}
\eea
where the subscripts indicate the number of symmetrized $t_i$. They identically satisfy the relations
\be\label{algebraic_Ypq}
b_{(q)} \, c_{(3p-q)} = a_{(3)}^p \;,
\ee
together with total symmetrization of $SU(3)$ indices in their products.

The topology of $Y^{p,q}$ was discussed in \cite{Martelli:2008rt}.  The homology groups are
\be
\begin{array}{c|cccccccc}
Y^{p,q} & H_0 & H_1 & H_2 & H_3 & H_4 & H_5 & H_6 & H_7 \\
\hline
& \quad \bZ \quad & \quad \bZ_{\gcd(p,q)} \quad & \quad \bZ \quad & \quad \Gamma \quad & \quad 0 \quad & \quad \bZ \oplus \bZ_{\gcd(p,q)} \quad & \quad 0 \quad & \quad \bZ \quad
\end{array}
\ee
The most interesting one is the third homology group, which is torsion (we refer to our companion paper \cite{to:appear} for a detailed discussion):
\be
H_3(Y^{p,q},\bZ) \cong H^4(Y^{p,q},\bZ) \cong \Gamma  \;.
\ee
$\Gamma$ is a finite Abelian group of dimension%
\footnote{If $q=0$ or $3p$,  CY$_4 = (\bC^3/\bZ_3 \times \bC)/\bZ_p$ and the torsion group is $\Gamma = \bZ_p$.}
$q(3p-q)$:
\be
\label{torsion group}
\Gamma = \bZ^2 \,/\, \langle (3q,q) \,,\, (q,p) \rangle \;.
\ee

Given the $Y^{p,q}$ manifold with its SE metric, the $\cN=2$ supersymmetric AdS$_4$ solution of 11d supergravity is
\bea
\label{Mth metric and flux}
ds^2 &=  R^2 \Big( \frac14 \,ds^2(AdS_4) + ds^2(Y^{p, q}) \Big) \\
G_4 &= \frac38 R^3 \, d\vol{(AdS_4)} \;.
\eea
There are $N$ units of M2-brane charge on $Y^{p,q}$, where $N$ is related to the radius $R$ by:
\be
\frac1{(2\pi l_p)^6} \int_{Y_7} \ast G_4 = N\, = \,   \, \frac{6 R^6}{(2 \pi l_p)^6} \mathrm{Vol}(Y^{p,q})   \;.
\ee
Turning on some torsion $G_4$ flux (which can be represented as a flat $C_3$ connection, see appendix \ref{app: torsion flux}) on $Y^{p,q}$ does not affect the supergravity equations of motion, hence there is a distinct M-theory background for each element
\be
\label{torsion G4 in Ypq}
[G_4] \in \Gamma \;.
\ee

We will discuss in section \ref{sec: field theory} the family of field theories dual to such backgrounds, for each torsion flux (\ref{torsion G4 in Ypq}).
In terms of the CY$_4$, such a solution corresponds to $N$ M2-branes at the tip, together with M5-branes wrapped on the relevant torsion cycle in $H_3 \cong \Gamma$.

\subsection{The geometric reduction of $C(Y^{p,q})$ to type IIA}

We proceed to KK reduce M-theory on the CY$_4$ to type IIA on the seven-manifold $X_7 = \text{CY}_4 / U(1)_M$. The reduction is performed along an isometry circle $U(1)_M$, chosen such that the K\"ahler quotient $\text{CY}_4\kq U(1)_M$ is a CY$_3$. In terms of the GLSM description, this happens if the charges of the fields under $U(1)_M$ sum up to zero \cite{Aganagic:2009zk}.

The last line in (\ref{full GLSM dP0}) defines our choice of $U(1)_M$ symmetry acting on the M-theory circle. As we anticipated, the symmetry $U(1)_M$ acts as the vector $s_0 - s_1$ on the 3d toric diagram (as subgroup of the $U(1)^4$ toric symmetry it is associated to the vector $(0,0,0,-1)$ in $\bZ^4$). It corresponds to the vertical direction in figure \ref{fig: toric diagram Ypq}. Including the last line in the GLSM (\ref{full GLSM dP0}) yields a toric CY three-fold which is the K\"ahler quotient $\mathrm{CY}_3 = \mathrm{CY}_4 \kq U(1)_M$. Its 2d toric diagram is the vertical projection of the 3d toric diagram of the four-fold.

However the type IIA geometry consists of a simple quotient, as opposed to a K\"ahler quotient, by $U(1)_M$. Therefore we keep the moment map $r_0$ as an unconstrained real field.
The type IIA geometry then involves a fibration of the previous CY$_3$ over the real line, parametrized by the moment map $r_0$ \cite{Aganagic:2009zk}.
To obtain the precise form of the fibration, let us rearrange the charges in (\ref{full GLSM dP0}).
First define
\be
\label{def_zeta_in_t_of_xi}
\zeta_0 = -\infty \;,\qquad \zeta_1 = 0 \;,\qquad \zeta_a= \sum_{b=2}^a \xi_b \qquad (a=2,\cdots,p) \;, \qquad \zeta_{p+1}= +\infty \;.
\ee
Given the constraint $\xi_a \geq 0$, they satisfy $\zeta_a \leq \zeta_{a+1}$.
Then rewrite the GLSM, including the last line, as
\be
\label{GLSM_CY3_redundant}
\begin{array}{c|cccccccccc|c}
\text{CY}_3 & t_1 & t_2 & t_3 & s_0 & s_1 & s_2 & \cdots & s_{p-2} & s_{p-1} & s_p  & \text{FI} \\
\hline
\cC & 1 & 1 & 1 &  q-3 & -q & 0 & \cdots & 0 &  0 & 0    & \xi^c \\
& 0 & 0 & 0 & 1 & -1  & 0 & \cdots & 0 & 0 & 0 & r_0-\zeta_1 \\
& 0 & 0 & 0 & 0 & 1 &  -1 & \cdots & 0 & 0 & 0 & r_0-\zeta_2 \\
& 0 & 0 & 0 & 0 & 0 & 1 & \cdots & 0 & 0 & 0 & r_0-\zeta_3 \\
& \vdots & \vdots & \vdots & \vdots & \vdots & \vdots & \ddots & \vdots & \vdots  & \vdots & \vdots  \\
& 0 & 0 & 0 & 0 & 0 & 0 & \cdots & 1 & -1 & 0 & r_0-\zeta_{p-1}\\
& 0 & 0 & 0 & 0 & 0 & 0 & \cdots & 0 & 1 & -1  & r_0-\zeta_{p}\\
\end{array}
\ee
This is a redundant description of the CY$_3$, as all but one of the $s_{a=0,\cdots,p}$ coordinates can be eliminated. Which one of the $s_a$ is unconstrained depends on the value of $r_0$: given a GLSM with two fields, charges $(1,-1)$ and FI $\xi$, the variable that can be eliminated is the one with charge $\sign(\xi)$. Therefore the unconstrained variable is
\be
\label{s_coordinate}
t_0 = s_a  \quad \mathrm{if} \quad \zeta_a \leq r_0 \leq \zeta_{a+1} \;.
\ee
We can rewrite the CY$_3$ GLSM as
\be
\label{generic_GLSM_of_CY3}
\begin{array}{c|cccc|c}
\text{CY}_3 &  t_1 & t_2 & t_3 & t_0 & \text{FI} \\
\hline
\cC & 1 & 1 & 1 & -3 & \chi(r_0)
\end{array}
\ee
which describes a resolved $\bC^3/\bZ_3$ with a blown-up $\CPt$ of size $\chi(r_0)$. The holomorphic 2-cycle $\cC$ is the hyperplane $\CP^1 \subset \CPt$, and it is exceptional. The FI parameter is
\be
\label{resol_param_CY3}
\chi(r_0) = \xi^c - q \, (r_0 - \zeta_1) +3 \sum_{b=1}^p (r_0 - \zeta_b) \, \Theta(r_0-\zeta_b) \;,
\ee
where $\Theta(x)$ is the Heaviside step function. The K\"ahler parameter $\chi(r_0)$ is continuous in $r_0$, while its first derivative jumps by 3 at $r_0 = \zeta_{a=1,\cdots,p}$ (where the unconstrained coordinate jumps from $s_{a-1}$ to $s_a$). As we will shortly see, this is due to the presence of a $\overline{\text{D6}}$-brane wrapping the toric divisor $\CPt$ in $\bC^3/\bZ_3$.%
\footnote{We explain why they are $\overline{\text{D6}}$-branes rather than D6 in section 4.}

If the CY$_4$ is conical, that is $\xi^c = \zeta_a = 0$ for all $a$, the resolution parameter is
\be
\label{resol_param_CY3_conical}
\chi(r_0) =  \big[ -q + 3p\, \Theta(r_0) \big] \, r_0  \;.
\ee

We can read the intersection numbers of $\bC^3/\bZ_3$ from (\ref{generic_GLSM_of_CY3}). There is a compact toric divisor $D_0$, which is the exceptional blown-up $\CPt$, and noncompact divisors $D_i$ ($i=1,2,3$), subject to linear equivalences $D_1=D_2=D_3$ and $D_0=-3D_1$. Abusing notation we use $D_\alpha$ for the cohomology class Poincar\'e dual to the toric divisor $D_\alpha = \{t_\alpha = 0\}$. The compact 2-cycle $\cC$ is the $\CP^1$ inside $\CPt$. The intersections are:
\be
\label{intersection numbers C3/Z3}
\cC \cdot D_0 = -3 \,\cC \cdot D_i = -3 \;,\qquad\quad D_0 \cdot D_0 \cdot D_0 = 9 \;,\qquad\quad D_0 \cdot D_0 = -3\cC \;.
\ee

\subsection{RR $F_2$ flux and D6-branes}

Since the M-theory circle is non-trivially fibered, the $U(1)_M$ connection gives a RR 2-form field strength in type IIA. In the CY$_4$ the fibration is encoded in a vielbein \cite{Aganagic:2009zk}
\be
\label{11th vielbein Mtheory}
\re \Big( d\theta_0 \,+\, i \sum_{j=1}^3 q_j \frac{dt_j}{t_j} \,+\, i \sum_{a=0}^p b_a \frac{ds_a}{s_a}  \Big) \;,
\ee
from which we read off the $U(1)_M$ connection. Its curvature is a 2-cocycle
\be
[F_2] = \sum_{i=1}^3 q_i D_i \,+\, \sum_{a=0}^p b_a D_{(s_a)}
\ee
where $[\cdot]$ stands for cohomology class. The integers $q_i$, $b_a$ are fixed by gauge invariance of (\ref{11th vielbein Mtheory}): $0 = q_1+q_2+q_3 + (q-3)s_0 - qs_1 = b_a - 2b_{a+1} + b_{a+2}$ for all $a=2, \cdots, p$ and $1 = b_0 - b_1$.
The system is solved if the coefficient in front of each divisor class equals minus the value of the vertical coordinate of the corresponding point in the 3d toric diagram. The solution is unique in cohomology, \ie{} up to linear equivalences:
\be
[F_2] = - q D_3 - \sum_{a=1}^p \, a\,  D_{(s_a)} \;.
\ee
This expression is still in terms of the redundant GLSM for the CY$_3$. In terms of the reduced GLSM in (\ref{generic_GLSM_of_CY3}), in the window $\zeta_a \leq r_0 \leq \zeta_{a+1}$ where the unconstrained coordinate is $t_0 = s_a$, we have%
\footnote{Given the GLSM with two fields, charges $(1,-1)$ and FI $\xi$, the toric divisor corresponding to the variable with charge $\sign(\xi)$ is empty.}
$[F_2] = - q D_3 - a D_0$. Therefore we can write the general expression
\be
\label{F2 on CY3}
[F_2] = - q D_3 - D_0 \sum_{a=1}^p \Theta(r_0 - \zeta_a) \;.
\ee
The flux $[F_2]$ jumps by $-D_0$ at $r_0 = \zeta_{a = 1,\cdots,p}$: such discontinuity is due to a magnetic source for $F_2$---a \aDs-brane wrapping $\CPt$ at $r_0 = \zeta_a$. The $\bC^2/\bZ_p$ K\"ahler parameters $\xi_a$ are thus the separations between $p$ \aDs-branes on $\CPt$ along $r_0$. When the CY$_4$ is conical, that is $\xi^c=\xi_a=0$, the type IIA background has $p$ coincident \aDs-branes wrapping the collapsed $\CPt$ at $r_0 = 0$.

The flux of $F_2$ on the holomorphic 2-cycle $\cC = \CP^1$ can be obtained from its intersection numbers with the toric divisors:
\be
\label{relation_F2_flux-Kaehler_par}
\int_{\cC \text{ at } r_0} F_2 = - q +3 \sum_{a=1}^p \Theta(r_0 - \zeta_a) = \chi'(r_0) \;.
\ee
The equality between 2-form fluxes and derivatives of the K\"ahler parameters is a consequence of supersymmetry.

\subsection{The manifold $M_6$}

The manifold $M_6$ which appears in the type IIA supergravity background $AdS_4 \times_w M_6$, is defined as $Y^{p,q}/U(1)_M$ \cite{Martelli:2008rt}. Recall that $Y^{p,q}$ is an $S^3/\bZ_p$ bundle over $\CPt$. $S^3$ has $SU(2)\times SU(2)_\text{base}$ symmetry, while the $\bZ_p$ quotient breaks the first factor to $U(1)_M$. The circle $U(1)_M$ is the fiber in the Hopf fibration $S^1 \hookrightarrow S^3/\bZ_p \to S^2$ of the lens space. Therefore $M_6$ is an $S^2$ bundle over $\CP^2$,
\be
S^2 \hookrightarrow M_6 \rightarrow \CP^2 \;.
\ee
On the other hand $SU(2)_\text{base}$ acts on the base of the Hopf fibration.
In the fibration over $\CPt$, $S^3/\bZ_p$ is twisted by a $U(1)_\text{twist} \subset SU(2)_\text{base}$. As a result, in $M_6$ the $S^2$ fiber is twisted on $\CPt$ by $U(1)_\text{twist}$. This has two fixed points: the north and the south pole.

The homology of $M_6$ is
\be
\label{homology of M6 for Ypq}
\begin{tabular}{c|ccccccc}
$M_6$ & $H_0$ &$H_1$ & $H_2$ & $H_3$ & $H_4$ & $H_5$ & $H_6$ \\
\hline
  & $\bZ$ & $0$ & $\bZ^{2}$ & $0$ & $\bZ^{2}$ & $0$ &  $\bZ $
\end{tabular}
\ee
Since the poles of $S^2$ are invariant under $U(1)_\text{twist}$, one can construct push forward maps $\sigma_{N,S}: \CPt \to M_6$ which uplift $\CPt$ to global sections of the bundle \cite{Martelli:2008rt}. We define the following representatives of $H_4(M_6)$:
\be
S^2 \hookrightarrow D \rightarrow \CP^1 \;,\qquad\qquad D^+ = \sigma_N \CPt \;,\qquad\qquad D^- = \sigma_S \CP^2 \;.
\ee
Here $D$ is the restriction of the $S^2$ bundle to the hyperplane $\CP^1$ in the base $\CPt$. The 4-cycles $(D, D^-, D^+)$ form an over-complete but convenient basis for $H_4(M_6)$. Similarly we define the 2-cycles
\be
\cC_0 \cong S^2 \;,\qquad\qquad \cC^+ = \sigma_N \CP^1 \;,\qquad \qquad \cC^- = \sigma_S \CP^1 \;,
\ee
where $\cC_0$ is the $S^2$ fiber over any point in $\CPt$. One can show that in homology
\be
D^+ = D^- + 3 D \;, \qquad\qquad \cC^+ = \cC^- + 3 \cC_0 \;.
\ee
The intersections between 2-cycles and 4-cycles are easily worked out from their definitions (recall that $\CP^1 \cdot \CP^1 = 1$ in $\CPt$) and the linear relations above:
\be
\label{intersection numbers}
\begin{array}{c|ccc|ccc}
& \cC_0 & \cC^+ & \cC^- & D & D^+ & D^- \\
\hline
\tabs D & 0 & 1 & 1 & \cC_0 & \cC^+ & \cC^- \\
D^+ & 1 & 3 & 0 & \cC^+ & 3\cC^+ & 0 \\
D^- & 1 & 0 & -3 & \cC^- & 0 & -3\cC^-
\end{array}
\ee

We have seen that if the original CY$_4$ is a cone,
the type IIA manifold $X_7$ can be sliced in two different ways: either as a $\text{CY}_3 = \text{CY}_4 \kq U(1)_M$ fibration along a real line parametrized by $r_0$, or as a real cone over $M_6 = Y_7/U(1)_M$. While the latter is manifest in the type IIA supergravity solution, the former is more useful to identify the dual field theory. In general cycles of $M_6$ are not cycles of the CY$_3$ and viceversa. However---in the conical case---the north pole (south pole) of the $S^2$ fiber of $M_6$ is the resolved tip of $\bC^3/\bZ_3$ at $r_0 > 0$ ($r_0 < 0$). We have:
\be\label{rel btw cycles M6 CY3}
\begin{split}
&\cC^+ = \CP^1 \;,\quad D^+ = \CPt \qquad \text{at } r_0 > 0 \\
&\cC^- = \CP^1 \;,\quad D^- = \CPt \qquad \text{at } r_0 < 0
\end{split}
\ee
where $\CP^1$ and $\CPt$ are at the tip of the resolved $\bC^3/\bZ_3$.
If the CY$_4$ itself is resolved, one has to be careful that $M_6$ degenerates (some of its cycles shrink) at the scale of the resolutions, but the map is still valid for $|r_0|$ big enough. These are the only cycles which are common to the CY$_3$ and $M_6$.
The relation (\ref{rel btw cycles M6 CY3}) can be understood by noting that topologically $M_6$ is the projectivisation of the anti-canonical bundle over $\CPt$ \cite{Gauntlett:2004hh}; see \cite{to:appear} for a more detailed discussion.

\subsection{Page charges and Freed-Witten anomaly}
\label{subsec: Page charges}
\label{subsec: FW anomaly}
\label{subsec:D2_Page}

We move to consider the charges present in the type IIA supergravity solution. We want to study D-brane charges that are conserved and quantized. The most useful notion is that of Page charges \cite{Marolf:2000cb}. In the context of AdS/CFT, they have first been applied to 4d theories in \cite{Evslin:2004vs, Benini:2007gx, Benini:2007kg, Argurio:2008mt, Benini:2008ir} to study duality cascades, and then in \cite{Aharony:2009fc, Hashimoto:2010bq, Hashimoto:2011aj} to 3d theories. Page charges have the advantage that they are sourced only by branes and the worldvolume flux $F$ on them, not by $\hat B_2$ or background fluxes.
On the other hand they are not gauge-invariant: they transform under large gauge transformations of the B-field.

The most interesting charge is the IIA reduction of the torsion flux (\ref{torsion G4 in Ypq}). Since $G_4$ vanishes in real cohomology, the M-theory flat connection $C_3$ descends in type IIA to a flat B-field $B_2$. The gauge invariant RR 4-form flux vanishes ($F_4 = 0$). Nevertheless the Page D4-charge is non-vanishing, and quantized: \be
[G_4] \;\to\; [B_2 \wedge F_2] \neq 0 \;.
\ee
We explicitly show this for $Y^{p,q}$ in appendix \ref{app: torsion flux}.
Gauge transformations of the Page charge are crucial to reproduce the periodicity (\ref{torsion group}) of the torsion group $\Gamma$. The following analysis generalizes the one in \cite{Aharony:2009fc}.

Let us start with the D6-charge. On $M_6$ the RR 2-form is \cite{Martelli:2008rt}
\be\label{F2 flux on M6}
[F_2] = p\, D^+ - q\, D \;.
\ee
It will be convenient to use the homology basis $\{-\cC_0, \cC^+\}$ and its dual basis $\{-D^-, D\}$. Accordingly, from the intersection numbers the D6-charges are:
\be
Q_{6;\,0} \equiv \int_{-\cC_0} F_2 = - p \;,\qquad Q_{6;\,+} \equiv \int_{\cC^+} F_2 = 3p-q \;,\qquad Q_{6;\,-} \equiv \int_{\cC^-} F_2 = -q \;.
\ee
We have included the third linearly dependent charge for completeness.

Next we consider the Page D4-charge. Let us parameterize the flat B-field as
\be
\label{flat B-field}
[B_2] = -b_0 \, D^- + b^+ \, D
\ee
where the coefficients have been chosen in such a way that
\be
b_0 \equiv \int_{-\cC_0} B_2 \;,\qquad\qquad b^+ \equiv \int_{\cC^+} B_2 \;,\qquad\qquad b^- \equiv \int_{\cC^-} B_2  = b^+ + 3 b_0 \;.
\ee
Since $F_4 = F_0 = 0$, the Page D4-charge is simply $\int F_2 \wedge B_2$ which depends on $p,q$ and $b_0, b^+$.
The potentials $b_0, b^+$ are not kinematically quantized in type IIA (although they are periodically identified). However Page D4-charges are quantized. We define them as integrals on the dual basis $\{-D^-, D\}$ and parameterize them by two integers
\be
\label{integer values torsion B}
(n_0 \,,\, n_1) \in \bZ^2 \;.
\ee
We get (see below for an anomalous correction):
\be
\label{def Q4 Page charges for M6}
\begin{array}{llll}
Q_{4;\, -} & = {\displaystyle \int_{-D^- \rule[-.5em]{0pt}{1em}} F_2 \wedge B_2} &=  q(3b_0 + b^+) &\equiv \; n_0 \\
Q_{4;\, D} &= {\displaystyle \int_{D \rule[-.5em]{0pt}{1em}} F_2 \wedge B_2} &=  qb_0 + p b^+ &\equiv\; n_1 \\
Q_{4;\, +} & = {\displaystyle \int_{-D^+} F_2 \wedge B_2} &= (3p-q)b^+ \;.
\end{array}
\ee
These relations can be inverted, to express the B-field in terms of the integers $(n_0, n_1)$:
\be
b_0 = \frac{p\, n_0 - q\, n_1}{q(3p-q)} \;,\qquad\qquad b^+ = \frac{3n_1 - n_0}{3p-q} \;,\qquad\qquad b^- = \frac{n_0}q \;,
\ee
from which the dynamical quantization of the B-field follows. In this discussion we just assumed that Page D4-charges are integers: in fact due to  Freed-Witten anomalies the quantization is shifted by half-integers, as we show below.

The periodicity (\ref{torsion group}) is realized as the following large gauge transformations:
\bea
\label{large gauge transfo and n shift}
\delta (n_0,\, n_1) &= (3q,q) \qquad &&\longleftrightarrow\qquad & \delta(b_0,\, b^+) &= (1,0) \\
\delta (n_0,\, n_1) &= (q,p) \qquad &&\longleftrightarrow\qquad & \delta(b_0,\, b^+) &= (0,1)
\eea
which generate $\bZ^2$ in the B-field space.

\paragraph{Freed-Witten anomaly.} We can engineer the $F_2$ flux (\ref{F2 flux on M6}) on $M_6$ by wrapping $p$ $\overline{D6}$-branes around $D^{+}$ and $q$ D6-branes around $D$, at a finite radius on AdS$_4$, and letting them fall towards the origin. The 4-cycles $D^+$ and $D$ both suffer from the Freed-Witten (FW) anomaly \cite{Freed:1999vc}. The anomaly on a D6-brane is canceled by a half-integrally quantized worldvolume flux on all 2-cycles on which the second Stiefel-Whitney class is non-vanishing (\ie{} the first Chern class is odd).
In fact we find that we can cancel the anomaly on any D6-brane using the pull-back of a single bulk 2-form, that we loosely call $F$.

Consider a D6-brane wrapped on $(- D^+)$. At finite radius it is a domain wall which increases $p$ by one:
\be
[F_2] \rightarrow [F_2] + D^+ \;,\qquad\qquad  \leftrightarrow \qquad\qquad  (p, q) \rightarrow (p+1, q) \;.
\ee
The first Chern class of $D^+$ is $c_1(D^+)|_{D^+ } = 3 D \cdot D^+$, which is odd.
Let us take the bulk 2-form $F$ as $F = \frac12 D$. Its pull-back on the D6-brane is
\be
F\Big|_{ -D^+ } = -\frac12 D \cdot D^+ = -\frac12 \cC^+ \;,
\ee
which cancels the FW anomaly. It also shifts the Page D4-charges according to
\be
\delta Q_{4;\,-} = 0 \;,\qquad\qquad \delta Q_{4;\,+} = \frac32   \;,\qquad\qquad \delta Q_{4;\, D} = - \frac12  \;.
\ee
Similarly, a D6-brane wrapped on $D$ increases $q$ by one:
\be
[F_2] \rightarrow [F_2] - D \;,\qquad\qquad  \leftrightarrow \qquad\qquad  (p, q) \rightarrow (p, q+1) \;.
\ee
The first Chern class is $c_1(D)|_D = (-D+2D^+)\cdot D$, so that the anomaly is canceled by the same $F=\frac12 D$ as before. The pull-back on the D6-brane is $F \big|_D = \frac12 D\cdot D= \frac12 \cC_0$, which gives rise to a shift of Page D4-charges $\delta Q_{4;\,-} = \delta Q_{4;\,+} = - \frac12$ and $\delta Q_{4;\, D}=0$.
We conclude that for the $Y^{p,q}$ geometry the Page D4-charges (\ref{def Q4 Page charges for M6}) are shifted to
\be
\label{D4 Page for Ypq with FW}
Q_{4;\, -}  = n_0 -\frac q2 \;, \qquad\quad  Q_{4;\, D}= n_1 -\frac p2 \;, \qquad\quad Q_{4;\, +} = n_0- 3n_1 -\frac{1}{2}(q-3p) \;.
\ee
The correct B-field periods are therefore
\be
\label{corrected B periods}
b_0 = \frac{p\, n_0 - q\, n_1}{q(3p-q)} \;,\qquad\qquad b^+ = -\frac{1}{2}+ \frac{3n_1 - n_0}{3p-q} \;,\qquad\qquad b^- = -\frac{1}{2} + \frac{n_0}q \;,
\ee
In particular, we have $B_2 = -\frac12 D$  at the torsionless point $n_0=n_1=0$, such that $\cF= \hat{B_2}+F=0$ on the D6-branes.

\paragraph{Page D2-charge.} The Page D2-charge of our background with $F_4= F_0=0$ reads
\be
\label{D2 brane charge}
Q_2 =  \int_{M_6}\left( \ast F_4 - \frac{1}{2} B_2\wedge B_2 \wedge F_2   \right) \;.
\ee
Remark that $Q_2 = Q^{\mathrm{Maxwell}}_2 + \tilde{Q}_2$, where we defined
\be
\tilde{Q}_2 \equiv -\frac{1}{2 }\int_{M_6} B_2\wedge B_2\wedge F_2 \;.
\ee
At the torsionless point ($n_0 = n_1 = 0$) $Q^{\mathrm{Maxwell}}_2 =N$.
Note that there are higher curvature corrections similarly to \cite{Bergman:2009zh}, but we neglect such contributions in this paper since they do not affect our line of argument.
The Page D2-charge is therefore
\be
\label{D2 Page charge in Ypq}
Q_2 = N - \frac p8 \;.
\ee
As we move in the torsion group $\Gamma$, the value of $\tilde{Q}_2$ changes according to (\ref{corrected B periods}):
\be
\tilde Q_2 = -\frac{p}{2} (b^+)^2 - \frac{q}{2} b_0(3 b_0 + 2b^+)
=  -\frac{p}{8} - \frac{p n_0^2 + q n_1(3n_1- 2n_0 -3p +q)}{2q(3p-q)} \;.
\ee
Since Page charges are not sourced by $B_2$, the D2-charge $Q_2$ is invariant as we vary $B_2$ continuously, while the Maxwell D2-charge changes accordingly. On the other hand $Q_2$ shifts by integers under the large gauge transformations (\ref{large gauge transfo and n shift}). We have:
\bea
\label{shifts in D2 Page charge}
\delta(b_0,\, b^+) = (1,0)  \quad :& \qquad  \delta Q_2 \equiv Q_2[n_0+3q,n_1+q]-Q_2[n_0,n_1] &&\!\!= -n_0-q \\
\delta(b_0,\, b^+) = (0,1) \quad :& \qquad  \delta Q_2 \equiv Q_2[n_0 + q,n_1 + p]-Q_2[n_0,n_1] &&\!\!= -n_1 \;.
\eea

\subsection{Remarks on parity and fundamental domain for $\Gamma$}
\label{subsec: parities}

\begin{figure}[t]
\begin{center}
\includegraphics[width=12cm]{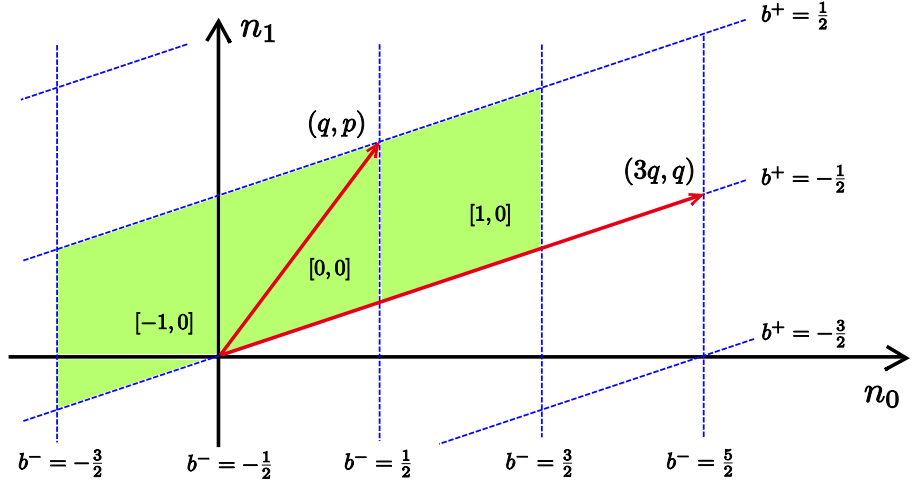}
\caption{\small The $(n_0, n_1)$ plane. The torsion group $\Gamma$ is obtained by quotienting this $\bZ^2$ lattice by the sublattice generated by the periodicity vectors $(q, p)$ and $(3q, q)$ shown in red. The shaded area is a choice
of fundamental domain, where the opposite sides are identified according to the red vectors (for instance the lower boundary of the area $[0,0]$ is identified with the upper boundary of the area $[1,0]$). The three parallelograms denoted  $[-1, 0]$, $[0,0]$ and $[1,0]$ correspond to the three windows needed to cover $\Gamma$ once, as will be explained in Section 5. The dashed blue lines correspond to loci where either $b^-$ or $b^+$ is  half integer, as indicated.  \label{fig: torsion domain}}
\end{center}
\end{figure}

There are two interesting parity transformations in supergravity. First, there is the $U(1)_M$-parity which changes sign to the M-theory circle coordinate. In type IIA it corresponds to
\be
F_2 \rightarrow - F_2 \;,\qquad B_2 \rightarrow -B_2 \;,\qquad \text{with} \quad D^+ \leftrightarrow -D^- \;.
\ee
In the CY$_3 \times\bR$ description, it also corresponds to $r_0 \rightarrow - r_0$. One easily checks that this M-theory parity can be undone by the following change of parameters:
\be
(p, q) \rightarrow (p, 3p-q) \;,\qquad\qquad (n_0, n_1)  \rightarrow (3n_1- n_0, n_1) \;.
\ee
Indeed the space $Y^{p, 3p-q}$ is identical to $Y^{p,q}$ \cite{Martelli:2008si}.

Second, there is the usual parity in M-theory, which changes sign to one spatial coordinate in $\bR^{1,2}$ and to $C_3$. In type IIA it corresponds to parity in $\bR^{1,2}$ and
\be
\label{parity B to -B}
 B_2 \rightarrow -B_2 \;,
\ee
while $F_2$ is invariant. D6-charges $(p,q)$ are left invariant, while the D4-charges (\ref{D4 Page for Ypq with FW}) change sign. This corresponds to the change of parameters
\be
\label{parity acting on n0 n1}
n_0 \rightarrow q- n_0 \;,\qquad\qquad  n_1 \rightarrow p- n_1 \;.
\ee
In the field theory this operation corresponds to parity times charge conjugation (CP).

For each $Y^{p,q}$ there are only few type IIA backgrounds invariant under M-theory parity, among the family indexed by $(n_0, n_1) \in \Gamma$.
One is the torsionless background: $n_0=n_1=0$. The B-field is $b^+ = b^- = -\frac 12$ and the parity transformation (\ref{parity B to -B}) gives an identical background up to a large gauge transformation.%
\footnote{The parity transformed theory is at $(n_0, n_1)=(q, p)$. It is identified with the theory at $(n_0, n_1)=(0,0)$ by a large gauge transformation, with vanishing shift of D2-charge (\ref{shifts in D2 Page charge}).}
Another one, which exists only if $q$ and $p$ are even, is at $(n_0, n_1)=\big( \frac q2, \frac p2 \big)$ and therefore $b^+=b^-=0$.
A third one is at $(n_0,n_1)=\big( 2q,\frac{p+q}2 \big)$ corresponding to $(b_0,b^+)=\big( \frac12,0\big)$.
It is also mapped to itself by \eqref{parity acting on n0 n1} together with a large gauge transformation that does not shift the D2-charge.

The torsion group $\Gamma$ is represented in figure \ref{fig: torsion domain}. The shaded area is a convenient choice of fundamental domain. The full plane is divided into ``windows'' by vertical dashed lines, corresponding to values of $n_0$ for which $b^-$ is half-integral, and diagonal dashed lines, corresponding to $b^+$ being half-integral. As explained below, those are marginal stability walls in the K\"ahler moduli space of the resolved $\bC^3/\bZ_3$ geometry.
In section \ref{sec: field theory} we will call window $[n,m]$ the one where $b^- \in \big[ n-\frac12,\, n + \frac12\big]$ and $b^+ \in \big[ m - \frac12,\, m+ \frac12 \big]$. The fundamental domain is divided into three windows. Under the parity transformation (\ref{parity acting on n0 n1}), window $[0,0]$ is mapped to itself, whilst windows $[1,0]$ and $[-1,0]$ are exchanged.

\subsection{Limiting geometry at $q=0$ or $3p$}\label{subsec: case q eq 0}

When $q=0$ (or equivalently $q=3p$) the $Y^{p,q}$ geometry has additional orbifold singularities. Consider for definiteness the case $q=0$. The CY$_4$ geometry is an orbifold of flat space, $(\bC^3/\bZ_3\times \bC)/\bZ_p$, with
\be
\label{orbifold C4Z3p}
(z_1,\; z_2,\; z_3,\; z_4) \sim  (e^{-\frac{2\pi i}{3p}} z_1,\; e^{-\frac{2\pi i}{3p}}z_2,\; e^{-\frac{2\pi i}{3p}}z_3,\; e^{\frac{2\pi i}{p}}z_4)\, ,
\ee
as one can show from the toric diagram (\ref{toric vectors for Ypq}).
The $M_6$ manifold that appears in type IIA has the topology of the weighted projective space $W\CP^3_{[1,1,1,3]}$ \cite{Martelli:2008rt}. The CY$_3 \times \bR$ background has a $\bC^3/\bZ_3$ singularity along $r_0 \leq 0$, with vanishing RR fluxes on the blown down $\CPt$:
\be
\int_{\CP^1 \text{ at } r_0<0} F_2 = 0 \;,\qquad\qquad \int_{\CPt \text{ at } r_0<0} F_2\wedge B_2 =0 \;,
\ee
and B-field $b^- = -\frac12$.
The manifold  $M_6$ has a single 2-cycle in homology, which we still denote $\cC^+$, and a dual 4-cycle $D$, with $D\cdot \cC^+=3$. $W\CP^3_{[1,1,1,3]}$  has an isolated $\bZ_3$ singularity, where an exceptional 2-cycle $\cC_0$ lives---if we blow up that 2-cycle we recover the same topology as the $M_6$ for $q>0$ \cite{Martelli:2008rt}, hence the terminology.

In the orbifold (\ref{orbifold C4Z3p}) the $\bZ_p$ subgroup acts freely and gives rise to a torsion group $\Gamma= \bZ_p$ in the seven-dimensional base. This is reproduced by the IIA geometry, where the Page D4-charge and B-field periods are
\be
Q_{4;\, D} = \int_D F_2\wedge B_2 = n_1 - \frac p2 \;,\qquad\quad b^+ = \int_{\cC^+} B_2 = \frac{n_1}p  -\frac12 \;.
\ee
In this last formula we included the shift due to the Freed-Witten anomaly, which arises as in the $q>0$ case.

\section{Fractional branes on $\bC^3/\bZ_3$}\label{sec: frac on C3Z3 orbifold}

D-branes on the orbifold $\bC_3/\bZ_3$ have been studied in detail in the literature \cite{Douglas:1996sw, Gukov:1998kn, Diaconescu:1999dt, Douglas:2000qw}. At the singularity, the point-like D-brane (a D2-brane in our context) is a marginal bound state of three fractional branes, which we denote by $E_1^{\vee}$, $E_2^{\vee}$, $E_3^{\vee}$. The dynamics of the fractional branes near the singularity is well described by the quiver of figure \ref{fig: quiver for dP0} below. Each fractional brane corresponds to one node of the quiver.  The $\bC^3/\bZ_3$ orbifold admits a crepant resolution to $\cO_{\CP^2}(-3)$, the canonical bundle over $\CP^2$. As we resolve the singularity, fractional branes are best seen as D-branes wrapping holomorphic cycles (B-branes). In this section we review in some details how to translate between the orbifold basis and the B-brane basis, or in other words between the quiver description and the CY$_3$ geometry.

The non-compact CY$_3$ $\cO_{\CP^2}(-3)$ has a K\"ahler moduli space of complex dimension one, with K\"ahler parameter
\be
t \equiv \int_{H} (B_2 + i J) \, \equiv \, b + i \chi \; ,
\ee
where $H$ is the hyperplane curve $\CP^1 \subset \CP^2$, $J$ is the K\"ahler form and $B_2$ is the B-field. We are interested in D-branes which are wrapped on the compact cycles $\CP^2$, $H$ and on a point, corresponding to D6-, D4- and D2-branes, respectively. The brane charge we consider is the Chern character of the B-brane, which we write as a vector:
\be
ch(E^{\vee}) = (Q_6,\, Q_4,\, Q_2) = (r,\, c_1,\, ch_2) \;,
\ee
where $r$, $c_1$ and $ch_2$ are the rank, the first and the second Chern characters of the vector bundle.
Therefore, a D6-brane with worldvolume flux $F= n H$ would have $ch(\text{D6})= (1, n, \tfrac{1}{2}n^2)$. This is however too naive, because of the Freed-Witten anomaly \cite{Freed:1999vc}. The manifold $\CP^2$ is not spin but only spin$^c$, and in order to define the spin$^c$ bundle we need to turn on half-integral flux $F$ on the D6-brane wrapped on $\CP^2$. With the minimal choice $F= \frac{1}{2} H$, the D6-brane has charge
\be
ch(\text{D6})= \Big( 1 ,\, \frac12 ,\, \frac18 \Big) \;,
\ee
while the naive D6-brane with Chern character $(1,0,0)$ does not exist in the physical spectrum. On the other hand, we have $ch(\text{D4})= (0,1,0)$ and $ch(\text{D2})= (0,0,1)$ by definition. An important quantity characterizing a D-brane state $E^{\vee}$ is its central charge $Z$, which tells us which half of the 8 supercharges of the CY$_3$ background is  preserved by $E^{\vee}$. We have
\be
Z(E^{\vee}) \, = \, ch(E^{\vee}) \cdot \Pi \;, \qquad\quad \text{with} \quad \Pi=(\Pi_6,\, \Pi_4,\, \Pi_2) \;,
\ee
where $\Pi_6$, $\Pi_4$ and $\Pi_2$ are so-called \emph{periods} associated to the states with Chern characters $(1,0,0)$, $(0,1,0)$ and $(0,0,1)$, respectively. At large volume ($ \chi \rightarrow \infty$), the central charge is given explicitly by
\be
\label{central charge Z at LV}
Z(E^{\vee}) = \int_{\CP^2} e^{B_2 + iJ} \, ch(E^{\vee}) \, \sqrt{\frac{\mathrm{\hat A}(T \CP^2)}{\mathrm{\hat A}(N \CP^2)}} \, \; + \; \cO(e^{2\pi i t}) \;,
\ee
where the last term includes $\alpha'$ corrections.
In that limit we have
\be
\label{periods at large vol}
\Pi_6 = \frac12 t^2 + \frac18 + \cO(e^{2\pi i t}) \;,\qquad\qquad \Pi_4 = t \;,\qquad\qquad \Pi_2 = 1 \;.
\ee
Remark that, for $\chi \gg b$, we have $\Pi_6 \sim -\frac{1}{2}\chi^2 < 0$ while $\Pi_2 = 1 >0$. Therefore, the D2- and D6-brane on $\CP^2$ cannot be mutually BPS at large volume;%
\footnote{This corresponds to the fact that mutual BPS-ness requires the worldvolume flux $\cF = \hat B_2 + F$ on the D6 to be anti-self-dual \cite{Marino:1999af}, so that $\int \cF \wedge \cF \leq 0$. Notice also that the flux $F= n H$ as well as $\hat B_2$ can only be self-dual ($H^2=1$), therefore a single $\overline{\text{D6}}$ can be BPS with a D2 only if $\cF=0$. Multiple $\overline{\text{D6}}$-branes can have anti-self-dual non-Abelian (instanton-like) flux.}
instead the D2-brane can be mutually BPS with the $\overline{D6}$-brane (for $b = - \frac12$).

Type II string theory is invariant under the large gauge transformation
\be
\label{large gauge transfo 001}
B_2 \rightarrow B_2 + H \;,\qquad\qquad F \rightarrow F - H \;,
\ee
where $F$ is the worldvolume flux on any D-brane. The action (\ref{large gauge transfo 001}) on the charges of any state $E^{\vee}$ and on the periods $\Pi$ is given by
\be
ch(E^{\vee}) \,\to\, ch(E^{\vee}) \, M_{\infty}^{-1} \;,\qquad \Pi \,\rightarrow\, M_{\infty}\, \Pi  \qquad\quad \text{with} \qquad
M_{\infty} = \mat{1 & 1 & \frac12 \\ 0 & 1 & 1 \\ 0 & 0 & 1} \;,
\ee
so that the central charge $Z(E^{\vee})$ is indeed invariant. While the transformation $M_{\infty}$ corresponds to $t\rightarrow t+1$ in (\ref{periods at large vol}), it should also be true for the exact periods $\Pi$. As one can easily check, this implies that $\Pi_4 = t$ and $\Pi_2=1$ \emph{exactly}; therefore only $\Pi_6$ is subject to $\alpha'$ corrections.

\subsection{Exact periods and fractional branes}

Because of $\alpha'$ corrections, at small K\"ahler volume it is better to consider the mirror geometry where the periods $\Pi$ can be computed exactly from the Picard-Fuchs equation. The quantum moduli space can be visualized as a Riemann sphere with complex coordinate $w$ and $z=1/w$, and with three singularities at $w=0$, $w=1$ and $w=\infty$. The point $w= \infty$ is the large volume limit. Near $z= 1/w =0$, we have \cite{Klemm:1999gm, Diaconescu:1999dt}
\be\label{mirror map LV}
t(z) = \frac{1}{2\pi i} \ln \Big( -\frac{z}{27} \Big) + \, \cO(z) \;.
\ee
The relation $t(z)$ between the parameters $t$ and $z$ is called the mirror map. We see that the transformation $t\rightarrow t+1$ corresponds to the logarithmic monodromy of $t(z)$ near the large volume point $z=0$. Let us define the period
\be
t_6 \equiv \Pi_6 +\frac12 t + \frac18 \;,
\ee
which from the formul\ae{} above is the central charge of the D6-brane: $Z(D6)= t_6$. We have the exact periods
\bea
t_6(w) &= \frac13 + \frac1{4\pi^2} \, \big( \omega^2 G_1(w) + \omega G_2(w) \big) \\
t(w) &=    \frac{\omega^2-\omega}{4\pi^2} \, \big( G_1(w) -G_2(w) \big) \;,
\eea
in terms of the special functions $G_1$, $G_2$ defined in appendix \ref{app: mirror and exact periods}, and $\omega = e^{\frac{2\pi i}{3}}$.
We choose the branch cuts to lie on the positive real axis in the complex plane, from $w=0$ to $w=1$, and from $z=1$ to $z=0$.
The point $w=0$ is called the \emph{orbifold point} in K\"ahler moduli space. One can check that as we circle $w \rightarrow e^{-2\pi i} w$, we have the monodromy:
\be
\label{monodromy periods weq0}
t_6 \rightarrow -2t_6+t+ 1 \;,\qquad\qquad t \rightarrow -3t_6+ t + 1 \;,\qquad\qquad 1 \rightarrow 1 \;.
\ee
The corresponding action on the charges and on $\Pi$ is
\be
ch(E^{\vee}) \,\to\, ch(E^{\vee}) \, M_o^{-1} \;,\qquad \Pi \,\to\, M_o \, \Pi \qquad\quad \text{with} \qquad
M_o = \mat{- \frac12 & \frac14 & \frac5{16} \\ -3 & -\frac12 & \frac58 \\ 0 & 0 & 1} \;.
\ee
This monodromy generates a $\bZ_3$ group, since $(M_o)^3 =1$. This corresponds to the $\bZ_3$ symmetry of the orbifold string theory. At the orbifold point $w=0$ the D2-brane fractionates into three fractional branes $E_i^{\vee}$ ($i=1,2,3$), which are rotated by this $\bZ_3$. We have $Z(D2)= \sum_{i=1}^3 Z(E_i^{\vee})=1$, so we must have $Z(E_i^{\vee})= \frac{1}{3}$ at $w=0$. Since $t_6(0)=\frac13$, the D6-brane is actually one of the fractional branes, and we can obtain the other two by the action of $M_o$ on $t_6$. We have:
\bea
\label{fractional branes at Beq0}
E_1^{\vee} &:& \; t_6\, , &\qquad\qquad ch(E_1^{\vee})= \Big( 1,\, \frac12,\, \frac18 \Big) \\
E_2^{\vee} &:& \; -2 t_6 + t +1\, ,  &\qquad\qquad ch(E_2^{\vee})= \Big( \!-\!2 ,\, 0,\, \frac34 \Big) \\
E_3^{\vee} &:& \; t_6 - t \, , &\qquad\qquad ch(E_3^{\vee})= \Big( 1,\, -\frac12,\, \frac18 \Big) \;.
\eea
This is the map we need. We summarize the relation between the brane charges $\bm{Q}_\text{source}= (Q_6, Q_4, Q_2)$ and the ranks $\bm{N}= (N_1, N_2, N_3)$ in the $\bC^3/\bZ_3$ quiver in a matrix $Q^{\vee}$, which we call a \emph{dictionary}:
\be
\label{relation charges ranks}
\bm{Q}_\text{source} = \bm{N} \, Q^{\vee} \;,\qquad\qquad Q^\vee \equiv Q^{\vee}[0] =
\mat{1 & \frac12 & \frac18 \\ -2 & 0 & \frac34 \\ 1 & -\frac12 & \frac18} \;.
\ee
The reason for the notation $Q^\vee[n]$, with $n\in \bZ$, is explained below.

\subsection{Monodromies in K\"ahler moduli space and dictionaries at any $b$}

In the derivation of the fractional brane states (\ref{fractional branes at Beq0}) we assumed that $b \in \big[ -\tfrac{1}{2}, \tfrac{1}{2} \big]$. This corresponds to a choice of sheet in the mirror map (\ref{mirror map LV}) (with the cut at $z$ real and positive). More generally, let us consider
\be
b \in \Big[ n -\frac{1}{2},\, n + \frac{1}{2} \Big] \qquad \qquad n \in \bZ \;.
\ee
In such window, the relevant fractional branes (objects with $Z= \frac{1}{3}$ at the orbifold point) are not (\ref{fractional branes at Beq0}) anymore, but they can be obtained from (\ref{fractional branes at Beq0}) by a large volume monodromy---in other words, by an active  $M_{\infty}$ transformation.
For later convenience, we accompany this $M_{\infty}$ action by a $\bZ_3$ rotation of the fractional branes. We define the dictionary $Q^{\vee}[n]$ by:
\be
\label{dictionary all windows}
Q^{\vee}[n] \, = \, (M_{o,\,\text{frac}})^n \,  Q^\vee \, (M_{\infty})^{-n} \;,
\ee
with
\be
M_{o,\,\text{frac}} = Q^\vee M_o (Q^{\vee})^{-1} = \mat{ 0 & 1 & 0 \\ 0 & 0 & 1 \\ 1 & 0 & 0} \;.
\ee
In particular, we have
\be
Q^\vee[-1] = \mat{ 1 & \frac12 & \frac18 \\ 1 & \frac32 & \frac98 \\ -2 & -2 & - \frac14} \;,\qquad\qquad Q^\vee[1] = \mat{ -2 & 2 & - \frac14 \\ 1 & - \frac32 & \frac98 \\ 1 & - \frac12 & \frac18} \;,
\ee
corresponding to the dictionaries valid for $b \in \big[-\frac32, -\frac12 \big]$ and $b \in \big[\tfrac{1}{2}, \tfrac{3}{2} \big]$, respectively.

Physically, as we cross some $b= n + \frac{1}{2}$, $n\in \bZ$ locus in K\"ahler moduli space we have some decay and recombination of the fractional branes. Consider for instance the case $b= -\frac{1}{2}$. This occurs on the branch cut that runs from $z=1$ to $z=0$. The point $z=1$ (with $b= -\frac{1}{2}$) is called a \emph{conifold point}: $t_6(1)=0$  and the first fractional brane $E_1^{\vee}$ in (\ref{fractional branes at Beq0}) becomes tensionless. At this point and anywhere on the $b = -\frac{1}{2}$ marginal stability wall, we have a recombination
\be
\label{recomb}
E_2^\vee + E_3^\vee \;\to (E_2^\vee+3E_1^\vee)+(E_3^\vee-3E_1^\vee)\;.
\ee
While at $b>-\frac12$ the states on the left in \eqref{recomb} are the lightest ones, at $b<-\frac12$ the lightest are those on the right. So \eqref{recomb} is a marginal recombination at $b=-\frac12$, but becomes a true decay as we cross the wall.


\section{Derivation of the CS quivers dual to $Y^{p,q}$}
\label{sec: field theory}

\begin{figure}[t]
\begin{center}
\includegraphics[height=4cm]{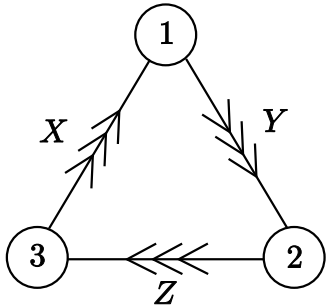}
\caption{\small Quiver diagram of the theory for M2-branes on the real cone over $Y^{p,q}(\CPt)$, the same quiver for D3-branes at $\bC^3/\bZ_3$. \label{fig: quiver for dP0}}
\end{center}
\end{figure}

Our next task is to identify the low energy field theory living on the branes at the singularity. We argued that it is a Chern-Simons quiver gauge theory, whose quiver and superpotential describe point-like D-branes at the tip of $\bC^3/\bZ_3$. The quiver is in figure \ref{fig: quiver for dP0}. The superpotential is
\be
W = \epsilon^{ijk} \Tr (X_i Y_j Z_k) \;,
\ee
and $i,j,k$ are indices of the $SU(3)$ global symmetry.
There are two things to determine: 1) given the Page charges measured on $M_6$, we should find the number of fractional branes of each kind at the singularity, and then the gauge ranks in the quiver making use of the dictionaries (\ref{dictionary all windows}); 2) from the same Page charges we should find the fluxes on cycles at the singularity, and then the CS terms induced by the Wess-Zumino action. Notice that indeed there are 5 parameters on both sides of the duality:%
\footnote{We have set $F_0 = 0$ in IIA, because $F_0$ does not have a M-theory lift \cite{Aharony:2010af}, and correspondingly we will find $k_1 + k_2 + k_3 = 0$.}
in supergravity the Page charges $(Q_{6;\,0},\, Q_{6;\,+},\, Q_{4;\,-},\, Q_{4;\,D},\, Q_2)$; in field theory the ranks and CS levels $(N_1,N_2,N_3,k_1,k_2,-k_1-k_2)$.

The counting of branes is more easily done on $\text{CY}_3 \times \bR$. Let us define Page charges on $\bC^3/\bZ_3$ as functions of $r_0$:
\bea
Q_6(r_0) &= \int_{\CP^1 \text{ at } r_0} F_2 \qquad\quad & Q_4(r_0) &= - \int_{\CPt \text{ at } r_0} F_2 \wedge B_2 \\
Q_2(r_0) &= \int_{\bC^3/\bZ_3 \text{ at } r_0} F_6 - \frac{1}{2}F_2 \wedge B_2 \wedge B_2 \qquad\quad & b(r_0) &= \int_{\CP^1 \text{ at } r_0} B_2
\eea
where the last one is the B-field rather than a Page charge. From the intersection numbers on $\bC^3/\bZ_3$ in (\ref{intersection numbers C3/Z3}) they must satisfy $Q_4(r_0) = -Q_6(r_0)\, b(r_0)$. We detect the presence of branes at $r_0$ by the jump of Page charges: from the intersection numbers we get $\delta Q_6 = -3$ for a D6-brane on $\CPt$ (without worldvolume flux, regardless of its actual existence), $\delta Q_4 = -3$ for a D4-brane on $\CP^1$, and $\delta Q_2 = 1$ for a D2-brane, where $\delta Q_p(r_0) \equiv Q_p(r_0 + \epsilon) - Q_p(r_0 - \epsilon)$.
The relation between branes and gauge ranks is in (\ref{relation charges ranks}): $\bm{N} = \bm{Q}_\text{source} \, Q^{\vee-1}$, where now $\bm{Q}_\text{source} = \big( \!-\!\frac13 \delta Q_6, -\frac13 \delta Q_4, \delta Q_2 \big)$. Exploiting the relation (\ref{rel btw cycles M6 CY3}) between cycles on $\bC^3/\bZ_3$ and on $M_6$, together with $\delta Q_2 = Q_2$, we find:
\be
\label{formula ranks one dictionary}
\bm{N} = \Big( - \frac{Q_{6;\,+} - Q_{6;\,-}}3 \,\Big|\, - \frac{Q_{4;\,+} - Q_{4;\,-}}3 \,\Big|\, Q_2 \Big) \, Q^{\vee-1} \;.
\ee
Notice also that $b(r_0>0) = b^+$ and $b(r_0<0) = b^-$.

Which dictionary should we use in (\ref{formula ranks one dictionary})? The dictionary $Q^\vee[0]$ is applicable only if $b(r_0)$ is within the range $\big[ -\frac12,\, \frac12 \big]$ for any $r_0$. Since $b(r_0)$ jumps when crossing a bunch of branes (as Page charges do), $Q^\vee[0]$ is applicable only for $-\frac12 \leq b^-, b^+ \leq \frac12$. These constraints draw a window in the space of torsion:
\be
\label{window [0,0]}
\text{window [$0,0$]}: \qquad 0 \leq n_0 \leq q \;,\qquad 0 \leq 3n_1 - n_0 \leq 3p-q \qquad\Rightarrow\qquad 0 \leq n_1 \leq p
\ee
where we used (\ref{corrected B periods}).
This window covers only one third of the fundamental domain of torsion, as is clear from figure \ref{fig: torsion domain}.
We define window $[n,m]$ as a subset of the $(n_0,n_1)$-plane where $b^- \in \big[ n -\frac12 ,\, n+ \frac12 \big]$ and $b^+ \in \big[ m -\frac12,\, m+\frac12 \big]$.

In window $[n,n]$, where both $b^+$ and $b^-$ are in the range $\big[ n-\frac12, n+\frac12\big]$, we have a different set of BPS states of minimal tension and therefore a different dictionary $Q^\vee[n]$. The ranks are obtained with the same formula (\ref{formula ranks one dictionary}).
Window $[n,n]$ is the image of window $[0,0]$ under a large gauge transformation of the B-field $\delta(b^+,b_0) = (n,0)$, which indeed shifts the Page charges as $\delta(n_0,n_1) = (nq,np)$ and
\be
\delta N = - n \, n_1 - \frac{n(n-1)}2 p \;.
\ee
The resulting theory is exactly the same theory as before the large gauge transformation, up to a cyclic rotation of the ranks. However the windows $[n,n]$---being images of $[0,0]$---do not help us covering new pieces of the torsion domain.

In windows $[n,m]$ with $n\neq m$, $b^+$ and $b^-$ fall in different ranges. In other words the CY$_3$ is in different K\"ahler domains on the two sides of $r_0 = 0$: the branes sit exactly on top of a K\"ahler wall and no dictionary is straightforwardly applicable. As we cross a bunch of branes moving along $r_0$, Page charges jump and $B_2$ jumps as well, according to $Q_4(r_0) = - Q_6(r_0) \, b(r_0)$.

Let us consider window $[1,0]$, where $\frac12 \leq b^- \leq \frac32$ and $-\frac12 \leq b^+ \leq \frac12$. Since we cannot use neither $Q^\vee[0]$ nor $Q^\vee[1]$, we split the branes along $r_0$ in two groups in such a way that $b(r_0) = \frac12$ in the middle:%
\footnote{In general we cannot do that supersymmetrically, but it is good enough to engineer the field theory: we will release the branes and all will fall at $r_0 = 0$, the tip of the cone.}
this requires $Q_6(r_\text{mid}) = - 2Q_4(r_\text{mid})$. Neglecting the detail of how we actually split the branes, we will simply impose that $Q_4(r_\text{mid}) = \rho$, $Q_6(r_\text{mid}) = -2\rho$ and $Q_2(r_\text{mid}) = \eta$ in terms of undetermined parameters $\rho,\eta$.
We can now use the dictionary $Q^\vee[1]$ with the bunch at $r< r_\text{mid}$ and $Q^\vee[0]$ with the bunch at $r > r_\text{mid}$. We use (\ref{formula ranks one dictionary}) for each bunch of branes separately:
\begin{multline}\nn
\bm{N} = \Big( - \frac{Q_{6;\,+} - (-2\rho)}3 \,\Big|\, - \frac{Q_{4;\,+} - \rho}3 \,\Big|\, Q_2 - \eta \Big) Q^\vee[0]^{-1} \\
+ \Big( - \frac{(-2\rho) - Q_{6;\,-}}3 \,\Big|\, - \frac{\rho - Q_{4;\,-}}3 \,\Big|\, \eta \Big) Q^\vee[1]^{-1} \;.
\end{multline}
In fact the unknowns $\rho,\eta$ cancel out: this just follows from the fact that at $b=\frac12$ the set of three states described by $Q^\vee[0]$ is only marginally unstable to decay to the set of $Q^\vee[1]$, and viceversa. In other words at $b=\frac12$, $Q^\vee[0]$ and $Q^\vee[1]$ produce exactly the same theory.
Running a similar argument with multiple splittings we arrive to a general formula for the ranks in window $[n,m]$:
\be
\bm{N} = \Big( \frac{Q_{6;\,-}}3 \,\Big|\, \frac{Q_{4;\,-}}3 \,\Big|\, Q_2 \Big) Q^\vee[n]^{-1} - \Big( \frac{Q_{6;\,+}}3 \,\Big|\, \frac{Q_{4;\,+}}3 \,\Big|\, 0 \Big) Q^\vee[m]^{-1} \;.
\ee

Chern-Simons levels are induced by the background fluxes \cite{Aganagic:2009zk}. For a D6-brane with Chern character $ch(E^\vee) = (1,F,ch_2)$, the Wess-Zumino action produces CS level
\be
k(E^\vee) = \int_\CPt e^{F+B} \wedge \sum\nolimits_q dC_q = \int_\CPt e^F \wedge \sum\nolimits_p F_p^\text{Page} \;,
\ee
which depends on the Page charges again. Let $\bm{k} = (k_1, k_2, k_3)$ be the vector of CS levels of the quiver. If we place a probe D6-brane bound state at $r_0$, we get on it:
\be
\label{formula CS for probes}
\bm{k} = \big( - Q_4(r_0) \,\big|\, Q_6(r_0) \,\big|\, 0 \big) \, Q^{\vee \trans} \;,
\ee
because the worldvolume flux $F$ and its square are contained in $Q^\vee$. The last entry is taken to be zero because we are not allowing a Romans mass $F_0$ and its D8-charge (the generalization is left for future work).
Remark that, by construction, $\sum_i k_i= 0$.

There are two subtleties, however. First, since the branes source fluxes, we should compute the CS levels induced by the background flux removing the fields produced by the branes themselves. This is easily accomplished by taking the average of the fluxes on the two sides of the branes. Given the linearity of (\ref{formula CS for probes}), this is equivalent to computing the CS levels for probes on the two sides of a bunch of branes, and taking the average. Second, when we are in a window $[n,m]$ with $n\neq m$, we should again split the bunch of branes into groups, with $b(r_0) \in \bZ + \frac12$ between them, and use different dictionaries for each group. All undetermined parameters cancel out, and we arrive at the formula:
\be
\bm{k} = \frac12 \Big[  \big( - Q_{4;\,-} \,\big|\, Q_{6;\,-} \,\big|\, 0 \big) \, Q^\vee[n]^\trans + \big( - Q_{4;\,+} \,\big|\, Q_{6;\,+} \,\big|\, 0 \big) \, Q^\vee[m]^\trans \Big] \;.
\ee
We repeat for convenience the Page charges of the background:
\bea
Q_{6;\,0} &= -p \;,\qquad & Q_{6;\,+} &= 3p-q \;,\qquad & Q_{6;\,-} &= -q \\
Q_{4;\,D} &= n_1 - \frac p2 \;,\qquad & Q_{4;\,+} &= n_0 - 3n_1 + \frac{3p-q}2 \;,\qquad & Q_{4;\,-} &= n_0 - \frac q2 \\
Q_2 &= N - \frac p8 \;.
\eea
We have thus found formul\ae{} for the theories in all windows.

The possible half-integral CS levels that can appear after taking the average are precisely adequate to cancel parity anomalies in the $SU(N_1)\times SU(N_2)\times SU(N_3)$ gauge subgroup. The parity anomaly in $U(1)^3$ is not canceled: we discuss possible solutions in section \ref{sec:_off_diag_CS}.

\begin{table}[t]
\bea\nn
&\qquad \text{Window $[-1,0]$:} \qquad -q \leq n_0 \leq 0 \qquad\qquad 0 \leq 3n_1 - n_0 \leq 3p - q \\
& \boxed{
U(N+n_1-p - n_0)_{-n_0 + \frac32n_1} \times U(N)_{\frac12n_0 - 3n_1 + \frac32p - q} \times U(N - n_1)_{\frac12 n_0 + \frac32n_1 - \frac32p + q} } \\
\\
&\qquad \text{Window $[0,0]$}: \qquad 0 \leq n_0 \leq q \qquad\qquad 0 \leq 3n_1 - n_0 \leq 3p-q \\
& \boxed{
U(N+n_1-p)_{-n_0 + \frac32n_1} \times U(N)_{2n_0 - 3n_1 +\frac32p - q} \times U(N-n_1)_{-n_0 + \frac32n_1 - \frac32p +q} } \\
\\
& \qquad \text{Window $[1,0]$:} \qquad q \leq n_0 \leq 2q \qquad\qquad 0 \leq 3n_1 - n_0 \leq 3p - q \\
& \boxed{
U(N + n_1 - p)_{\frac12n_0 + \frac32 n_1 - \frac32q} \times U(N)_{\frac12n_0 - 3n_1 + \frac32p + \frac12q} \times U(N - n_1 + n_0 - q)_{-n_0 + \frac32n_1 - \frac32p + q} }
\eea
\caption{Theories for $Y^{p,q}$ in a fundamental domain of the torsion group. \label{tab: three theories dP0}}
\end{table}

Each window covers one third of the fundamental domain of torsion, see figure \ref{fig: torsion domain}. We choose windows $[-1,0]$, $[0,0]$ and $[1,0]$ to cover it all. The three set of theories with their respective domains of validity are summarized in Table \ref{tab: three theories dP0}.
Window [0,0] is delimited, in the $(n_0,n_1)$ plane, by $(0,0)$, $(0, p - \frac q3)$, $(q,\frac q3)$, $(q,p)$. The two corners $(n_0,n_1) = (0,0) \cong (q,p)$ are identified, and correspond to the CP-invariant theories of $p$ $\overline{\text{D6}}$'s. The center $(\frac q2, \frac p2)$, present if $p,q$ are even, corresponds to the CP-invariant theory of $\frac p2$ 2$\overline{\text{D6}}$ bound states. As explained in section \ref{subsec: parities}, the parity symmetry of M-theory acts in type IIA as $B_2 \to - B_2$, which means
\be
n_0 \to q - n_0 \;,\qquad\qquad n_1 \to p - n_1 \;,
\ee
together with a parity operation in $\bR^{1,2}$.
In field theory it corresponds to parity times charge conjugation.%
\footnote{This agrees with the sign flip of $B_2$ under a change of orientation of the string worldsheet, which acts as charge conjugation for the open string degrees of freedom.}
This operation leaves window $[0,0]$ invariant, whilst exchanges windows $[1,0]$ and $[-1,0]$.

Let us conclude with a series of remarks.
First, the theories nicely glue on the borders of the torsion domain: On the common edges of the windows, the theories coincide. On edges which are identified (by large gauge transformations), the theories coincide up to a shift of $N$, which agrees with the shift of the Page D2-charge (see section \ref{subsec:D2_Page}):
\bea
Q_2 (n_0+q,\, n_1+p) - Q_2(n_0,n_1) &= -n_1 \\
Q_2 (n_0 + 3q,\, n_1 + q) - Q_2(n_0,n_1) &= - n_0 - q \\
Q_2 (n_0 + 2q,\, n_1 +q-p) - Q_2(n_0,n_1) &= n_1 - n_0 - p \;.
\eea
Second, one can check---although it is not manifest---that the map from $(N,p,q,n_0,n_1)$ to the set of field theories $(N_1,N_2,N_3,k_1,k_2,-k_1-k_2)$ (modulo cyclic permutations of the gauge groups) is injective (up to identifications of $(n_0,n_1)$) and surjective: every possible rank and CS assignment%
\footnote{With the constraint $k_1 + k_2 + k_3 = 0$, which could be relaxed by considering IIA backgrounds with $F_0$ flux \cite{Behrndt:2004mj, Lust:2004ig, Acharya:2006ne, Petrini:2009ur, Lust:2009mb}.}
to the quiver in figure \ref{fig: quiver for dP0} is realized by one and only one background $(N,p,q,n_0,n_1)$.

In particular we do not find the phenomenon of supersymmetry breaking as in the ABJ case \cite{Aharony:2008gk}. In the example of ABJ, the two theories at the boundaries of the torsion domain are identified by a Seiberg-like duality, and theories beyond that are conjectured to break SUSY. The theory on $\bC^3/\bZ_3$ does not enjoy Seiberg-like dualities, so we cannot expect the same mechanism to be at work. Indeed theories at the boundaries of the torsion domain are just trivially identical and, on the other hand, all possible theories arise from some $Y^{p,q}$ with some torsion.

We can further speculate that while ``non-anomalous'' fractional branes (the nomenclature is taken from 4d) enjoy Seiberg-like dualities and suffer SUSY breaking, ``anomalous'' fractional branes are free of both. Whilst $\bC^3/\bZ_3$ has only the latter, in examples with both types of branes---like $S^3/\bZ_p$ bundles over $\CP^1 \times \CP^1$ \cite{to:appear}---we should expect mixed phenomena.

\subsection{Moduli space without resolutions}
\label{subsec: mod space monopoles}

As a first check of the theories in table \ref{tab: three theories dP0}, let us compute the geometric branch of the moduli space of those CFTs. As explained in section \ref{sec:_VMS_monopoles}, it can be done either by considering monopole operators (as in \cite{Benini:2009qs,Jafferis:2009th}) or with a semiclassical computation. Let us adopt the computation with monopoles.

We start with the theory in window $[0,0]$. On the Coulomb branch the gauge group is broken to $\big(U(1)^G\big)\phantom{)\!\!}^{\tilde N} \prod_{i=1}^G U(M_i)$. Let us consider monopoles of the subgroup $U(1)^G$: one copy of the Abelian quiver. The R-, gauge, flavor and topological charges of the two simple monopoles $T$ and $\tilde T$ are:
$$
\begin{array}{c|ccc|c|c|}
& X & Y & Z & T & \tilde T \\
\hline
\tabs R & 2-R_Y - R_Z & R_Y & R_Z & p - \frac32 n_1 R_Y - \frac32 (p-n_1) R_Z & p - \frac32 n_1 R_Y - \frac32 (p-n_1) R_Z \\
g_1 & -1 & 1 & 0 & -n_0 & n_0 - 3n_1 \\
g_2 & 0 & -1 & 1 & 2n_0 - q & -2n_0 + 6n_1 - 3p+q \\
g_3 & 1 & 0 & -1 & -n_0 + q & n_0 - 3n_1 +3p - q \\
SU(3) & \rep{3} & \rep{3} & \rep{3} & \rep{1} & \rep{1} \\
U(1)_M & 0 & 0 & 0 & 1 & -1
\end{array}
$$
Note that we have used an R-symmetry which is not mixed with topological symmetries.
Such charges allow us to write down the following formal gauge invariants:
\be\label{gauge_invariants}
a_{(3)} = XYZ \;,\qquad b_{(q)} = TY^{n_0}Z^{q-n_0} \;,\qquad c_{(3p-q)} = \tilde TY^{3n_1 - n_0}Z^{3p-q-3n_1 + n_0}
\ee
where the subscripts refer to the number of $SU(3)$ indices, and the following formal relations:
\be
T \tilde T Y^{3n_1} Z^{3p-3n_1} = (XYZ)^p \;.
\ee
In appendix \ref{subsec: check of quantum relations dP0} we give some evidence for the relation above from a counting of fermion zero-modes.
The quantum relations in the chiral ring join the classical F-term relations coming from the superpotential:
\be
X_{[i}Y_{j]} = Y_{[i} Z_{j]} = Z_{[i} X_{j]} = 0
\ee
which imply that $SU(3)$ indices are always symmetrized.

The expressions above are only formal because some exponents could be negative: in that case one should multiply the expressions by extra powers of the gauge invariant $(XYZ)$ to eliminate negative powers.
However the boundaries of window $[0,0]$ in the $(n_0,n_1)$-plane are precisely such that all powers are non-negative. The gauge invariants above satisfy the relations \eqref{algebraic_Ypq}
\be
b_{(q)} \, c_{(3p-q)} = a_{(3)}^p\;,
\ee
together with symmetrization of $SU(3)$ indices in products, which provide the algebraic description of $C(Y^{p,q})$. We have found that the moduli space contains a copy of $C(Y^{p,q})$, corresponding to the motion of one M2-brane on the CY$_4$. In fact the theories in window $[0,0]$ contain $\tilde N$ symmetrized copies of the Abelian quiver, where $\tilde N = N - \max(n_1, p-n_1)$. Therefore the Coulomb branch contains the symmetrized product $\text{Sym}_{\tilde N}C(Y^{p,q})$, describing the motion of $\tilde N$ M2-branes on the CY$_4$. The extra piece of the theory will be analyzed in section \ref{subsec: mod space resolutions}, and shown to describe resolutions of the CY$_4$.

The same analysis can be repeated for the other windows, the only changes being in the charges of monopole operators and in the bifundamentals appearing in the gauge invariants $b_{(q)}$ and $c_{(3p-q)}$. The final result is the same: in each window the Coulomb branch contains the symmetric product of a number of copies of $C(Y^{p,q})$. The number of copies equals the smallest gauge rank. The boundaries of the windows are precisely such that the geometric branch is $C(Y^{p,q})$ inside and on the boundaries, but not outside.

This can also be understood in the semiclassical approach by looking at the K\"ahler (FI) parameter space of the quiver theory, which is divided into three chambers characterized by different linear relations between  FI parameters and K\"ahler parameters of $\CP^1\subset\CPt$ (see for instance \cite{Benishti:2011ab}).
As $\sigma$ spans $\bR$, the effective FI parameters of the CS theories draw two rays joined at the origin in the K\"ahler cone. The rays are in the directions of the effective CS levels, \emph{i.e.} the gauge charges of bare monopole operators which determine the dependence of $b_{(q)}$ and $c_{(3p-q)}$ on bifundamentals. The two rays lie either inside the same chamber (for window $[0,0]$ and its images), or inside different chambers (for windows $[0,\pm 1]$ and their images), with a wall crossing at $\sigma=0$. The latter possibility corresponds to the wall in the K\"ahler moduli space of the CY$_3$ crossed at $r_0=0$ by probe D2-branes mobile along $\bR$ in the type IIA setup.
A window thus corresponds to a pair of K\"ahler chambers, both in geometry and in field theory. A dictionary between geometry and field theory is associated to each chamber, and the change of dictionaries as we change windows precisely accounts for the modifications of the field theory needed to reproduce the same $C(Y^{p,q})$ geometric branch.

\subsection{Special cases}
\label{subsec: special cases}

Among the $Y^{p,q}$ theories, there are a few cases deserving special attention.

\paragraph{CP-invariant theories.} Here we discuss the three CP-invariant theories, corresponding to particular values of $(n_0, n_1)$ as described in section \ref{subsec: parities}.

One is the torsionless theory dual to the M-theory background with $(n_0,n_1) = (0,0)$.
In type IIA the B-field is $(b^+,b^-,b_0) = \big( -\frac12,-\frac12, 0\big)$, which under M-theory parity is mapped to itself up to a large gauge transformation. The CP-invariant gauge theory is
\be\label{CP inv quiver 01}
U(N-p)_0 \times U(N)_{\frac32 p - q} \times U(N)_{-\frac32 p + q} \;.
\ee

Since the M-theory background does not have torsion, there are no obstructions to a complete crepant resolution of the CY$_4$ by blowing up $\CPt$ and the $(p-1)$ $\CP^1$'s at the tip of the $\bC^2/\bZ_p$ fiber. In type IIA such resolutions correspond to blowing up $\CPt$ in the CY$_3$ at $r_0 = 0$, and to separate along $r_0$ the $p$ $\overline{\text{D6}}$-branes wrapping $\CPt$. At least when the volume of $\CPt$ is large, we can use the brane action. In our conventions only $\overline{\text{D6}}$-branes can be mutually SUSY with D2-branes (which appear in the background). The SUSY condition for one $\overline{\text{D6}}$-brane is that its worldvolume flux $\cF$ is a $(1,1)$-form and $*_4\cF = -\cF$ \cite{Gomis:2005wc}. The only $(1,1)$-form on $\CPt$ is the K\"ahler form, which however is self-dual, therefore the only solution is $\cF = \hat B_2 + F = 0$. In view of the Freed-Witten anomaly on $\CPt$, this requires $b^+ = b^- = -\frac12$ (up to large gauge transformations). Therefore the B-field has the right value to allow a complete SUSY separation of the $\overline{\text{D6}}$-branes. Comparing with $Q^\vee$ in (\ref{relation charges ranks}) we see that the actual BPS bound-state of D6-charge $-1$ is made of one $\overline{\text{D6}}$ and one D2 (last two rows of $Q^\vee$), and it corresponds to the last two groups in field theory. This suggests that out of the $N$ D2-branes, $p$ are stuck on the $\overline{\text{D6}}$-branes.
The field theory computation of section \ref{subsec: mod space resolutions} confirms this.

We already saw that the field theory moduli space indeed contains $N-p$ symmetrized copies of $C(Y^{p,q})$, describing the $N-p$ free D2-branes. We will see in section \ref{subsec: mod space resolutions} that it also contains all $p$ resolution parameters, related to the adjoint scalars of the last two groups.

The second theory is the one for $(n_0,n_1) = \big( \frac q2, \frac p2 \big)$ (it only exists if $p,q$ are even).  It is also dual to a parity-invariant M-theory background. In type IIA the B-field vanishes: $b^+ = b^- = b_0 = 0$. The CP-invariant gauge theory is
\be
\label{CP inv quiver 02}
U \Big(N - \frac p2 \Big)_{\frac34 p - \frac12 q} \times U(N)_0 \times U \Big( N - \frac p2 \Big)_{- \frac34p + \frac12q } \;.
\ee

The number of geometric resolution parameters allowed by the torsion flux is easily understood in type IIA. First we can resolve $\CPt$. With $B_2 = 0$, the minimal large volume object mutually SUSY with D2-branes is the bound state%
\footnote{On $\CPt$ there is an exceptional stable holomorphic rank 2 bundle $V$ with $c_1 = -1$ and $c_2 = 1$, which does not have moduli and is member of a discrete family \cite{LePotier}. We can put it on two $\overline{\text{D6}}$-branes, forming a bound state, but the Freed-Witten anomaly requires to combine it with $-\frac12$ units of Abelian flux (per brane). The resulting bundle has $c_1 = 0$, $ch_2 = \frac34$, takes values in $SU(2)$ only, is anti-self-dual and does not have moduli.}
of two $\overline{\text{D6}}$-branes with a rigid $SU(2)$ instanton (second row of $Q^\vee$ in (\ref{relation charges ranks})), and it corresponds to the second gauge group. In fact the background is made of $p/2$ of them (a total of $p$ $\overline{\text{D6}}$-branes), which can be separated along $r_0$. We obtain that, due to torsion, only $\frac p2-1$ 2-cycles can be blown-up. This agrees with the field theory moduli space.

The third CP-invariant theory is at $(n_0, n_1)=\big( 2q, \frac{q+p}{2} \big)$ (it exists if $q+p$ is even), on the boundary of window $[1,0]$, corresponding to $(b^-,b^+)=\big( \frac32,0\big)$. It is
\be
\label{CP inv quiver 03}
U\Big(N + \frac{q-p}2 \Big)_{\frac34 p + \frac14 q} \times U(N)_0 \times U\Big( N + \frac{q-p}2 \Big)_{- \frac34 p- \frac14 q } \;.
\ee

\paragraph{Another special case: quiver with three equal ranks.}
Inspection of table \ref{tab: three theories dP0} reveals that this is only possible in window $[1,0]$ with $(n_0,n_1) = (p+q,p)$:
$$
U(N)_{2p-q} \times U(N)_{-p+q} \times U(N)_{-p} \;,
$$
and in window $[-1,0]$ with $(n_0,n_1) = (-p,0)$ which gives the CP-transformed theory. Because of the borders of the windows, they actually contain those points only if $p \leq q \leq 2p$.

The theories above are the ones proposed in \cite{Martelli:2008si} for M2-branes at the tip of the cone over $Y^{p,q}$: we have therefore found a stringy derivation of their proposal. Let us add two comments. First, the authors already noticed that their proposals can only realize $C(Y^{p,q})$ if $p \leq q \leq 2p$: here we understand why. Second, the theories with equal ranks can only describe one resolution parameter of the CY$_4$, the blow-up of $\CPt$, while they cannot describe any resolution of the $\bC^2/\bZ_p$ fiber. The reason is that they are dual to $AdS_4\times Y^{p,q}$ backgrounds with torsion $G_4$ flux, and such torsion obstructs the resolutions. In type IIA, SUSY does not allow the bound state of all D6-branes to be broken in smaller pieces.

\paragraph{Field theory for the limiting case $q=0, 3p$.}
In the case $q=0$, the IIA background has a non-isolated $\bC^3/\bZ_3$ singularity, as discussed in section \ref{subsec: case q eq 0}. The field theory derivation nevertheless runs similarly, giving
$$
U(N+n_1-p)_{\frac{3}{2}n_1} \times U(N)_{-3n_1+\frac{3}{2}p} \times U(N-n_1)_{\frac{3}{2}n_1-\frac{3}{2}p} \;,
$$
where $n_1 \in \bZ_p$ lies in the torsion group of the $(\bC^3/\bZ_3\times \bC)/\bZ_p$ orbifold. This corresponds to the $q=n_0=0$ limit of general theory for $Y^{p,q}$ in window $[0,0]$. Similarly, the $q=3p$ theory is the $U(1)_M$-parity dual to the $q=0$ case.

\subsection{Moduli space with resolutions}
\label{subsec: mod space resolutions}

In section \ref{subsec: mod space monopoles} we computed the chiral ring of our theories using monopoles, and that describes the complex structure of the moduli space. However there can be resolutions, \ie{} K\"ahler deformations, of the CY$_4$, and to see them the semi-classical approach is more powerful.

To begin with, consider the CP-invariant torsionless theory (\ref{CP inv quiver 01}):
\be\label{CP inv quiver 01 bis}
U(N-p)_0 \times U(N)_{\frac32p - q} \times U(N)_{-\frac32p + q} \;.
\ee
Referring to section \ref{subsec:_VMS_from_1loop} and to figure \ref{fig: quiver for dP0}, the classical
D-term equations are%
\footnote{From now on we neglect factors of $2\pi$ in the effective FI parameters.}
\bea
\xi_1 &= YY^\dag - X^\dag X \;,\qquad\qquad &  \xi_2 + \Big( \frac32p - q \Big) \sigma_2 &= ZZ^\dag - Y^\dag Y \\
&& \xi_3 - \Big( \frac32p - q \Big) \sigma_3 &= XX^\dag - Z^\dag Z \;,
\eea
where the $SU(3)$ index has been suppressed and
$\xi_{1,2,3}$ are bare FI terms. Gauge rotations can be used to diagonalize $\sigma_i$, and the gauge group is generically broken to a bunch of $U(1)'s$.
The geometric branch consists of diagonalized configurations:
\be
X = \mat{x \\ 0} \;,\quad Y = \mat{y & 0} \;,\quad Z = \mat{ z & 0 \\ 0 & \tilde z} \;,\quad \sigma_1 = \sigma \;,\quad \sigma_2 = \sigma_3 = \mat{\sigma & 0 \\ 0 & \tilde\sigma} \;,
\ee
where $x,y,z,\sigma$ are diagonal matrices of size $N-p$, while $\tilde z,\tilde\sigma$ are diagonal matrices of size $p$.
The equations admit solutions only for $\xi_1 = 0$ and $\xi_2 = -\xi_3 \equiv \xi$, which we then impose.

The classical moduli space is corrected at one-loop. Following section \ref{subsec:_VMS_from_1loop},
let us focus on one of the untilded directions and its coupling to the tilded directions:
\be
U(1)_1 \times U(1)_2 \times U(1)_3 \times \prod_{a=1}^p U(1)_{2a} \times U(1)_{3a} \;,
\ee
the so-called pseudo-Abelian quiver. We can use the residual Weyl group and a field redefinition in the decoupled totally diagonal $U(1)$ to fix
\be
0 = \tilde\sigma_1 \leq \tilde\sigma_2 \leq \dots \leq \tilde\sigma_p
\ee
and introduce the index $0 \leq I \leq p$ such that $\tilde\sigma_I \leq \sigma \leq \tilde\sigma_{I+1}$ (where formally $\tilde\sigma_0 = -\infty$, $\tilde\sigma_{p+1} = +\infty$). The effective CS terms are
\bea
k^\text{eff}_{11} &= 0 \;,\qquad\qquad k^\text{eff}_{22} = - k^\text{eff}_{33} = 3I-q \;,\qquad\qquad
k^\text{eff}_{2a,2a} = - k^\text{eff}_{3a,3a} = 3a - \frac32 - q \\
k^\text{eff}_{1,2a} &= k^\text{eff}_{2,3a} = - k^\text{eff}_{1,3a} = - k^\text{eff}_{3,2a} = \begin{cases} - \tfrac32 \;, & a\leq I \\ \phantom{-} \tfrac32 \;, & I+1 \leq a \end{cases} \;,\qquad
k_{2a,3b} = \begin{cases} - \frac32 \;, & b < a \\ \phantom{-} \frac32 \;, & a < b \end{cases} \;.
\eea
From these we compute the effective FI terms:
\bea\label{xi eff for torsionless theory}
\xi^\text{eff}_1 &= 0 \\
\xi^\text{eff}_2 &= - \xi^\text{eff}_3 = \xi + (3I-q)\, \sigma + \frac32 \sum_a \tilde\sigma_a - 3 \sum_{a\leq I} \tilde\sigma_a  \\
\xi_{2a}^\text{eff} &= - \xi_{3a}^\text{eff} = \xi + (3a-q) \, \tilde\sigma_a + \frac32 \sum_b \tilde\sigma_b - 3 \sum_{b\leq a} \tilde\sigma_b \;.
\eea
Along the untilded direction, the one-loop corrected D-term equations are (recall that $SU(3)$ indices are implicit):
\be
|y|^2 - |x|^2 = 0 \;,\qquad\quad |z|^2 - |y|^2 = \xi + (3I-q)\,\sigma + \frac32 \sum_a \tilde\sigma_a - 3 \sum_{a\leq I} \tilde\sigma_a \;.
\ee
They describe a resolved $\bC^3/\bZ_3$. For $\sigma \to \pm \infty$, $\xi^\text{eff}_2$ is positive: we are in a chamber where $x=y=0$ is the tip, and there $\CPt$ has size $\xi_2^\text{eff}$. For $\xi$ large enough, the size is always non-negative and the theory is in a geometric phase. We can rewrite the size $\chi(\sigma)$ of $\CPt$ as
\be\label{final result for chi of sigma}
\chi(\sigma) = \Big( \xi + \frac32 \sum_{a=1}^p \tilde\sigma_a\Big) - q\,(\sigma - \tilde\sigma_1) + 3 \sum_{a=1}^p (\sigma - \tilde\sigma_a) \, \Theta(\sigma - \tilde\sigma_a)
\ee
in terms of the Heaviside step function $\Theta(x)$.

Comparing with (\ref{resol_param_CY3}) we see that the semi-classical computation has \emph{exactly} reproduced the type IIA manifold $X_7 = \text{CY}_4/U(1)_M$, upon the identification
\be\label{map FT/Geom}
\sigma = r_0 \;,\qquad\qquad \xi + \frac32 \sum_{a=1}^p \tilde\sigma_a = \xi^c \;,\qquad\qquad \tilde\sigma_a = \zeta_a \;.
\ee
The M-theory geometry is reproduced by gauge fixing the dual photon, as explained in section \ref{subsec:_VMS_from_1loop}. Thus the field theory moduli space contains all $p$ resolutions of the CY$_4$:%
\footnote{More precisely, $\tilde\sigma_a$ are moduli of the theory and should correspond to normalizable deformations of the CY$_4$; $\xi$ is a parameter in field theory and should correspond to a non-normalizable mode in supergravity.}
$\xi$ controls the resolution of $\CPt$, while $\tilde\sigma_{2,\cdots,p}$ control the resolutions of the $\bC^2/\bZ_p$ fiber. From the point of view of type IIA, $\tilde\sigma_a$ are the positions along $r_0$ of $p$ $\overline{\text{D6}}$-branes. The one-loop corrections arise from integrating out massive D2-D6 strings.

Along the $a$-th tilded direction, the quiver is $U(1)_{2a} \times U(1)_{3a}$ with 3 bifundamental fields $z_\alpha$. The one-loop corrected D-term equation is
\be
|\tilde z|^2 = \chi(\tilde\sigma_a) \;.
\ee
This describes a $\CPt$ of size $\chi(\tilde\sigma_a)$. There is also a dual photon, whose gauge fixing provides an extra $U(1)_J$ direction. Indeed $\tilde\sigma_a$ controls the K\"ahler class of a resolved 2-cycle in the CY$_4$: supersymmetry pairs it up with a pseudoscalar $b_a$---acted upon by $U(1)_J$---into a complexified K\"ahler class. We expect $b_a$ to correspond in M-theory to $C_6$ integrated on a 6-cycle, and in type IIA to $C_5$ integrated on the 5-chain between two \aDs-branes.

The natural geometric interpretation of the $\CPt$ is as the moduli of a D2-brane confined on the \aDs-brane at $r_0 = \tilde\sigma_a$. Indeed we observed in section \ref{subsec: special cases} that the bound states at $b=-\frac12$ are D2-$\overline{\text{D6}}$, and supersymmetry does not allow to separated them. This suggests that in M-theory $p$ M2-branes are confined to the tip. We leave a precise understanding of the responsible mechanism for the future.

\

A similar analysis can be performed on the CP-invariant theories (\ref{CP inv quiver 02}),
\be\label{CP inv quiver 02 bis}
U\Big(N - \frac p2 \Big)_{\frac34 p - \frac12 q} \times U(N)_0 \times U\Big( N - \frac p2 \Big)_{- \frac34p + \frac12q } \;.
\ee
(Recall that $p,q$ must be even). The geometric branch is given by diagonalized fields
\be
X = \mat{x} \;,\quad Y = \mat{y & 0} \;,\quad Z = \mat{ z \\ 0} \;,\quad \sigma_1 = \sigma_3 = \sigma \;,\quad \sigma_2 = \mat{\sigma & 0 \\ 0 & \tilde\sigma} \;,
\ee
where $x,y,z,\sigma$ are diagonal matrices of size $N-\frac p2$, while $\tilde\sigma$ is a diagonal matrix of size $\frac p2$. The classical F- and D-term equations admit solutions only for $\xi_1 = -\xi_3 = \xi$ and $\xi_2 = 0$. We use gauge rotations to set $0 = \tilde\sigma_1 \leq \cdots \leq \tilde\sigma_{p/2}$. The one-loop corrected D-term equations along an untilded direction are
\be
|z|^2 - |y|^2 = 0 \;,\qquad\quad |y|^2 - |x|^2 = \xi + \Big(3I- \frac q2 \Big) \,\sigma + \frac32 \sum_{a=1}^{p/2} \tilde\sigma_a - 3 \sum_{a\leq I} \tilde\sigma_a \;.
\ee
We are in a chamber where the tip of the geometry is at $x=0$: the classical F-term equations impose that the vectors $y_\alpha$ and $z_\alpha$ point in the same direction in $\CPt$, which then has size
\be
\chi(\sigma) = \Big( 2\xi - 3 \sum_{a=1}^{p/2} \tilde\sigma_a \Big) -  q \,(\sigma - \tilde\sigma_1) + 3 \sum_{a=1}^{p/2} 2(\sigma - \tilde\sigma_a) \, \Theta(\sigma - \tilde\sigma_a) \;.
\ee
Along the tilded directions we get trivial equations, and the only extra modulus is the dual photon that with $\tilde\sigma_a$ forms a complexified K\"ahler class.

Comparing with (\ref{resol_param_CY3}) we recognize the type IIA manifold $X_7$ with $\frac p2$ resolutions: $\xi$ controls the resolution of $\CPt$, while $\tilde\sigma_{2,\cdots,p/2}$ control half of the resolutions of $\bC^2/\bZ_p$. From the point of view of type IIA, the minimal object which is SUSY in the background is the bound-state of 2$\overline{\text{D6}}$ with non-Abelian gauge bundle, and $\tilde\sigma_a$ are the positions of those $\frac p2$ bound-states along $r_0$. Supersymmetry prevents us to completely break the $p$ $\overline{\text{D6}}$-branes apart. From the point of view of M-theory the mechanism should be similar, even though a better understanding along the lines of \cite{Benishti:2009ky} would be welcome: the torsion flux should not be compatible with a full resolution of the geometry and SUSY. Notice that in this case we do not get extra degrees of freedom on the $\overline{\text{D6}}$-branes.

\

Let us comment on the case of a generic theory in table \ref{tab: three theories dP0}, leaving the details for future work. From the field theory point of view, resolutions are possible only if the one-loop corrected D-term equations admit solutions with non-trivial $\xi_{1,2,3}$ or $\tilde\sigma$. In fact generically these resolutions are obstructed---some resolutions open up only for special values of $(n_0,n_1)$. From the point of view of type IIA, this has a simple interpretation. Partial resolutions are possible only when the B-field is such that either or both of the following conditions are satisfied: 1) the bunch of all $\overline{\text{D6}}$-branes with worldvolume fluxes is SUSY at large volume; 2) the bunch can be broken in smaller bound states, all of which are SUSY at large volume.


\section{Canceling parity anomalies: Off-diagonal CS terms}
\label{sec:_off_diag_CS}

The CS quiver theories we considered in the previous section still suffer from a so-called parity anomaly, making them inconsistent. Cancelation of all the $\bZ_2$ anomalies in a CS quiver requires the conditions \cite{Redlich:1983kn, Redlich:1983dv, Aharony:1997bx}
\bea
\label{parity anomaly SU(N) and U(1)s}
&k_i + \frac12 \sum\nolimits_{j} A_{ij} N_j \, &&\in\, \bZ \\
& \Lambda_{ij} - \frac12 A_{ij} \, &&\in\, \bZ  \;,
\eea
where $A_{ij}$ is the adjacency matrix of the quiver and $\Lambda_{ij}$ are Chern-Simons coupling between $U(1)$ photons of two different gauge groups of the quiver, that we introduce below.
The first line insures the cancelation of anomalies for each $U(N)_i$ factor in the gauge group. Our models satisfy this condition. The second line is a condition for the cancelation of $\bZ_2$ anomalies for the Abelian part of the gauge group. Since we have considered $\Lambda_{ij}=0$ so far, and because $A_{ij}=\pm 3$ for $i\neq j$, the last condition in (\ref{parity anomaly SU(N) and U(1)s}) is \emph{not} satisfied by our $\bC^3/\bZ_3$ quiver; this problem was also noticed in \cite{Cheon:2011th}.

The anomaly can be canceled in different ways. One way is to add some appropriate CS couplings to the Lagrangian. In general this would affect the moduli space, and spoil the nice matching with the M-theory CY$_4$. However, we show below that the anomaly cancelation is possible even without affecting the moduli space. Another way to cancel the anomaly could be to constraint the Abelian field strengths. A further way could be to consider gauge group $SU(N_1)\times SU(N_2) \times SU(N_3) \times U(1)$, the last factor being the diagonal subgroup of the original $U(1)^3$. Unfortunately we were not able to derive what mechanism string theory realizes.

\subsection{Off-diagonal CS terms and moduli space}

Since $U(N) \cong SU(N) \times U(1)/\bZ_N$, the most general CS term we can write for a $U(N)$ gauge field $A$ is
\be
\label{anomaly cancellation}
\cL_\text{CS} = \frac k{4\pi} \int \Tr \Big( A \wedge dA + \frac{2}3 A \wedge A \wedge A \Big) + \frac\Lambda{4\pi} \int \Tr A \wedge d \Tr A \;.
\ee
While the first term is the $U(N)$ CS term we considered so far, the second term is a correction for the $U(1)$ part only. If we have multiple gauge fields $A_i$ of $U(N_i)$, the most general CS interaction is
\be
\label{general CS term}
\cL_\text{CS} = \sum_i \frac{k_i}{4\pi} \int \Tr \Big( A_i \wedge dA_i + \frac{2}3 A_i \wedge A_i \wedge A_i \Big) + \sum_{ij} \frac{\Lambda_{ij}}{4\pi} \int \Tr A_i \wedge d \Tr A_j \;.
\ee
$\Lambda_{ij}$ is a symmetric matrix, that contains off-diagonal CS interactions between the Abelian gauge fields (such mixing is not possible with non-Abelian gauge fields). Consider a vacuum in which the gauge group $\prod_j U(N_j)$ is broken to its maximal torus: let $A_{i,m}$ be the photon in the $U(1)_{i,m} \in U(N_i)$ (with $m=1, \cdots, N_i$). The CS term (\ref{general CS term}) becomes:
\begin{multline}
\cL_\text{CS} = \sum_{i,m} \frac{k_i + \Lambda_{ii}}{4\pi} \int A_{i,m} \wedge dA_{i,m} + \sum_i \sum_{m\neq n} \frac{\Lambda_{ii}}{4\pi} \int A_{i,m} \wedge dA_{i,n} \\
+ \sum_{i\neq j} \sum_{m,l} \frac{\Lambda_{ij}}{4\pi} \int A_{i,m} \wedge dA_{j,l} \;.
\end{multline}
To these Abelian CS couplings the quantization condition applies. Since $k_i$ already satisfy the first condition in (\ref{anomaly cancellation}), the parity anomaly is canceled if
\be
\Lambda_{ij} - \frac12 A_{ij} \;\in\, \bZ \qquad\quad \forall\, i,j \;.
\ee

If the extra CS terms $\Lambda_{ij}$ are generic, they affect the theory in an important way. In particular they modify the gauge charges of the monopole operators $T$, $\tilde{T}$, ruining the match we found with the dual geometry.  In order for that not to happen, the following conditions should be met:
\be
\label{constraint on off diag CS}
\sum\nolimits_{j=1}^3 \Lambda_{ij}= 0\, \qquad  \, i=1,2,3 \;.
\ee
In that case the chiral ring is unmodified. Therefore we can cancel the $\bZ_2$ anomaly by choosing appropriate $\Lambda_{ij}$ CS levels for the $U(1)$ factors in the gauge group, and there is a 3-parameter family of solutions.

The semi-classical computation of section \ref{subsec: mod space resolutions} is only slightly affected. Consider for instance the torsionless theory $n_0=n_1=0$, one can show that the effective FI terms (see (\ref{xi eff for torsionless theory})) are modified to
\bea
\xi^\text{eff}_1 &= - \Lambda_{11} \sum_a \tilde\sigma_a \\
\xi^\text{eff}_2 &=  \xi- \Lambda_{12} \sum_a \tilde\sigma_a + (3I-q)\, \sigma + \frac32 \sum_a \tilde\sigma_a - 3 \sum_{a\leq I} \tilde\sigma_a  \\
\xi^\text{eff}_3 &=  -\xi- \Lambda_{13} \sum_a \tilde\sigma_a - (3I-q)\, \sigma - \frac32 \sum_a \tilde\sigma_a + 3 \sum_{a\leq I} \tilde\sigma_a  \, .
\eea
In the derivation of the above we have used the constraints (\ref{constraint on off diag CS}). Since we need $\xi^\text{eff}_1=0$ and $\xi^\text{eff}_2=-\xi^\text{eff}_3$ in order to have SUSY solutions, we can either introduce a bare $\xi_1^\text{bare}$ or simply take $\Lambda_{11}=0$. Assuming the latter, the final result (\ref{final result for chi of sigma}) becomes
\be
\chi(\sigma) = \Big( \xi + (\frac32-\Lambda_{12}) \sum\nolimits_a \tilde\sigma_a\Big) - (\sigma - \tilde\sigma_1)\, q + 3 \sum\nolimits_a (\sigma - \tilde\sigma_a) \, \Theta(\sigma - \tilde\sigma_a)\, .
\ee
Remark that for $\Lambda_{12}=\frac32$ the map (\ref{map FT/Geom}) between field theory and geometric parameters is diagonal. A convenient choice in this case is
\be
(\Lambda_{ij})= \mat{0 & \frac32 & -\frac32 \\ \frac32 & -3 & \frac32 \\ -\frac32 & \frac32 & 0 } \;.
\ee

While we have shown that from the 3d point of view the $\bZ_2$ anomaly can be canceled at no cost by a small modification of the Chern-Simons interactions, we cannot \emph{a priori} exclude that string theory chooses a more exotic method of anomaly cancelation.


\section{Conclusions}

We have discussed a stringy derivation for the low energy theory of M2-branes at the tip of a class of conical Calabi-Yau four-folds called $C(Y^{p,q}(\CPt))$. We have found that the theories are quiver Chern-Simons theories, with quivers already known in the literature but with shifted ranks and levels. Such theories would be anomalous in 4d. In 3d they are well-defined, and the would-be anomalies appear instead as quantum corrections to the charges of monopole operators. As a result, the chiral rings of the theories receive one-loop corrections. We have shown with localization techniques that in $\cN=2$ superconformal theories the charges of chiral monopoles are one-loop exact; in some sense it is the analogue of the one-loop exactness of 4d anomalies.
Even though what we have discussed is a specific example, we believe that similar considerations apply to more general classes of geometries. In a companion paper \cite{to:appear} we will initiate the systematic study of $\bC^2/\bZ_p$ bundles over del Pezzo surfaces, finding similar results.

There are a number of open questions to be stressed.

We have explicitly shown that a torsion flux $G_4 \in H^4(Y^{p,q},\bZ)$ can be described by a flat $C_3$ connection and descends in the type IIA reduction of \cite{Martelli:2008rt} to a B-field, such that the Page D4-charge $F_2 \wedge B_2$ is quantized.  On the other hand, the Freed-Witten anomaly on D6-branes implies a semi-integral shift of the Page charge. One could then expect a similar mechanism to take place in M-theory, possibly along the lines of \cite{Witten:1996md}.

In the case of the torsionless theory of section \ref{subsec: special cases}, analyzing the moduli space we have found indications that each $\overline{\text{D6}}$-brane has a D2-brane stuck on it. At the orbifold point this follows from the knowledge of the mutually BPS states, described by the quiver. However from the point of view of M-theory the reason is less clear. An observation is that one could expect non-perturbative corrections, both in field theory and M-theory, leading to a runaway behavior and decompactification \cite{Witten:1996bn, deBoer:1997kr, Aharony:1997bx}. In \cite{Benishti:2010jn} it was argued from supergravity that those corrections can be present rather generally when the CY$_4$ has exceptional six-cycles. While EM5-branes wrapped on 6-cycles can make the geometry unstable \cite{Witten:1996bn}, the results of \cite{Benishti:2010jn} suggests that when M2-branes are at the location of an exceptional 6-cycle we can blow it up without instability.
In our field theory analysis we did not include possible non-perturbative corrections. It will be interesting to do so and compare field theory to M-theory expectations.

We have stressed that our field theories still have a $\bZ_2$ anomaly that needs to be cured. One possibility is that extra CS terms are present for the $U(1)^3$ central gauge subgroup, and this would not spoil the moduli space. These extra CS terms, not considered in the literature so far, could be at the origin of the difficulties to reproduce the $N^{3/2}$ behavior of the partition function on $S^3$, but for now this is pure speculation.

Finally, it would be interesting to include a Romans mass (D8-charge) in the analysis \cite{Petrini:2009ur,Lust:2009mb}. From our derivation of the CS levels in section \ref{sec: field theory} it is already obvious that the D8-charge $F_0$ maps to $\sum_i k_i \neq 0$, as already proposed in \cite{Gaiotto:2009mv, Fujita:2009kw, Petrini:2009ur, Aharony:2010af, Tomasiello:2010zz}. A subtlety here is that for $F_0\neq 0$ one has to include gravitational corrections, which conspire to keep the CS levels properly quantized.

\section*{Acknowledgments}
We are grateful to Dario Martelli, Igor Klebanov, Alessandro Tomasiello and Alberto Zaffaroni for their comments on the draft, and to Ofer Aharony, Oren Bergman, Amihay Hanany, Diego Rodr\'i{}guez-G\'o{}mez, Mithat \"Unsal, and Herman Verlinde for interesting discussions and suggestions. FB would like to thank the Aspen Center for Physics, FB and SC are grateful to the Galileo Galilei Institute for Theoretical Physics in Florence, for hospitality during the course of this work. CC would like to thank the Service de PMIF at ULB Brussels for its kind hospitality during the course of this work, and Sophie de Buyl in particular.

The work of FB was supported in part by the US NSF under Grants No. PHY-0844827 and PHY-0756966.
CC is a  Feinberg Postdoctoral Fellow at the Weizmann Institute for Sciences.
SC was supported by the Israel Science Foundation (grant 1468/06), the German-Israel Project Cooperation (grant DIP H52), the German-Israeli Foundation (GIF) and the US-Israel Binational Science Foundation (BSF).


\appendix

\section{One-loop shift of Chern-Simons levels and effective FI terms}
\label{app:_shift_of_CS_levels}

Three-dimensional theories do not suffer from chiral anomalies, however they can suffer from $\bZ_2$ valued \emph{parity anomalies} \cite{Redlich:1983kn, Redlich:1983dv} which depend on the matter content.
Parity anomalies can always be canceled by half-integral Chern-Simons levels (so that the theory eventually breaks parity). One way to understand this is to start with a non-anomalous theory and integrate out some fermions such that the would-be low energy theory is naively anomalous. In fact a one-loop diagram provides a half-integral shift of the Chern-Simons level, which keeps the theory consistent.

Consider an Abelian theory with gauge group $U(1)^G$. It can have generic Chern-Simons interactions
\be
\sum_{ij}^G \frac{k_{ij}}{4\pi} \int A_i \wedge dA_j \;.
\ee
Let the theory have fermions $f$ of electric charges $Q_i(f)$ and real masses $M_f$. Integrating out these fields (for $M_f \neq 0$) shifts the CS levels by \cite{deBoer:1997kr, Aharony:1997bx}
\be
 k_{ij} \, \rightarrow\, k_{ij} + \frac12 \sum_f Q_i(f) Q_j(f) \, \sign(M_f) \;,
\ee
where the sum is over all fermions $f$ in the theory and a regularization of $\sign(x)$ which vanishes if $x=0$ is implied. Invariance of the partition function under large gauge transformations then requires the quantization of bare Chern-Simons levels $k_{ij}$ \cite{Aharony:1997bx}
\be\label{quantization_CS}
 k_{ij} + \frac12 \sum_f Q_i(f) Q_j(f) \,\, \in \,\, \bZ \;.
\ee
Consider in particular an $\cN=2$ theory: real masses have a bare contribution $m_f$ and a contribution from the adjoint real scalars $\sigma_i$ in vector multiplets:
\be
M_f = m_f + \sum_i Q_i(f) \, \sigma_i \;.
\ee
In fact the bare masses $m_f$ can be considered as expectation values of adjoint real scalars in external vector multiplets.

Consider now an $\cN=2$ quiver theory of generic gauge ranks $N_i$ (some gauge fields could be external) with all fields $X_{ij}$ in bifundamental or adjoint representations. When the theory is Higgsed to an Abelian one through generic VEV's for $\sigma_i$,
\be
\sigma_i = (\sigma^{(i)})_{n_i} \qquad n_i = 1,\cdots, N_i \;,
\ee
and recall that they can always be diagonalized by gauge transformations, then chiral fields $X_{ij}$ split into $N_i \times N_j$ components which have real masses
\be
M[ (X_{ij})\ud{n_i}{n_j} ] = \sigma^{(i)}_{n_i} - \sigma^{(j)}_{n_j} \;.
\ee
Integrating them out, the CS levels are shifted by
\bea
\delta k_{n_in_i'} &= \delta_{n_in_i'} \frac12 \sum_j \sum_{n_j} \eta_{ij} \sign(\sigma^{(i)}_{n_i} - \sigma^{(j)}_{n_j}) \\
\delta k_{n_in_j} &= - \frac12 \eta_{ij} \sign(\sigma^{(i)}_{n_i} - \sigma^{(j)}_{n_j} ) \qquad\text{for } i\neq j
\eea
where we denoted by $\eta_{ij}$ the net number of chiral fields between $U(N_i)$ and $U(N_j)$:
\be
\eta_{ij} = \#(X_{ij}) - \#(X_{ji}) \;.
\ee
Notice that $\eta_{ii} = 0$.

The presence of CS terms determines an effective shift of the bare FI parameters along the Coulomb branch \cite{Dorey:1999rb}, and the effective IR shift of the CS levels determines an extra one-loop contribution:
\be
\xi_{n_i}^\text{eff} = \xi_i^\text{bare} + \sum_{j=1}^G \sum_{n_j=1}^{N_j} k_{n_in_j}^\text{eff} \sigma^{(j)}_{n_j} \;.
\ee
Plugging in the actual effective CS levels corrected at one-loop, we obtain:
\be
\label{general formula for xi eff}
\xi_{n_i}^\text{eff} = \xi_i^\text{bare} + \sum_{j=1}^G \sum_{n_j=1}^{N_j} k_{n_in_j}^\text{bare} \sigma^{(j)}_{n_j} + \frac12 \sum_{j=1}^G \sum_{n_j = 1}^{N_j} \eta_{ij} \big| \sigma^{(i)}_{n_i} - \sigma^{(j)}_{n_j} \big| \;.
\ee


\subsection{Monopole operators and dual photon}
\label{app_subsec:_dual_photon}

Consider a free Maxwell theory. In Euclidean signature, the dual photon $\varphi$ is defined by
\be
\label{euclidean rel btw dual photon and F}
\partial_{\mu} \varphi = -\frac{2\pi i}{e^2} \epsilon_{\mu\nu\rho} F^{\mu\nu} \;.
\ee
Due to flux quantization, it is periodic of period $2 \pi$. In a $\cN=2$ supersymmetric theory, $\varphi$ and $\sigma$ pair into a chiral superfield of lowest component
\be
\Phi= \frac{2\pi }{e^2} \sigma \, + i\varphi \;.
\ee
The monopole operator of flux $n$ is defined by
\be
T^{(n)}= \exp( n\Phi ) \;.
\ee
This operator corresponds to the Borokhov-Kapustin-Wu definition of a monopole operator \cite{Borokhov:2002ib}: it is an operator which inserts $n$ units of magnetic flux at $x=0$. We can see it explicitly in the dual photon formulation of the Maxwell theory.
We have
\be
\langle T(0)^{(n)} \rangle = \int \cD\varphi \, \cD\sigma \exp \int d^3x \Big[ -\frac1{2e^2} (\partial\sigma)^2 - \frac{e^2}{8\pi^2} (\partial\varphi)^2 + n \Big( \frac{2\pi}{e^2} \sigma + i\varphi \Big) \delta^3(x) \Big] \;.
\ee
This insertion changes the perturbative saddle point from $\sigma=\varphi=0$ to
\be
\sigma= \frac n{2r} \;,\qquad\qquad \varphi = \frac{2\pi i}{e^2} \, \frac n{2r} \;.
\ee
Such a singularity in the dual photon field corresponds to $n$ units of magnetic flux at $r=0$, as we can see by using the relation (\ref{euclidean rel btw dual photon and F}). This discussion is closely related to the discussion of line operators given by Kapustin in \cite{Kapustin:2005py}.


\section{Flavored ABJM: resolved geometry and field theory}
\label{app:flavored_ABJM}

In \cite{Benini:2009qs} (see also \cite{Jafferis:2009th}) field theories on M2-branes probing a family of toric CY$_4$ cones whose 3d toric diagrams project to that of the conifold were derived reducing M-theory to type IIA. There is a 5-parameter family of such toric CY$_4$ singularities and of dual flavored ABJM quiver gauge theories. These dualities have been thoroughly checked in \cite{Jafferis:2011zi}, where the partition function of the CFTs on $S^3$ and the superconformal R-charges of gauge invariant operators were computed and matched with data of the dual M-theory backgrounds.

Here we explicitly show how real masses in the flavored ABJM models of \cite{Benini:2009qs,Jafferis:2009th} correspond to partial resolutions of the CY$_4$ geometry.
First we sketch the chiral ring of the theories with monopoles, then we compute the geometric moduli space semiclassically, as in section \ref{subsec:_VMS_from_1loop}, in the presence of bare FI parameters and real masses for flavors which map to resolution parameters of the CY$_4$, similarly to section \ref{subsec: mod space resolutions}.
Finally we show how to see the resolutions using the monopole operators of the effective theories.
Some remarks have already appeared in \cite{Cremonesi:2010ae}, which mainly focused on the dual type IIB brane configuration to study fractional M2-branes.

\begin{figure}[t]
\begin{center}
\subfigure[\small Toric diagram.]{
\includegraphics[height=6cm]{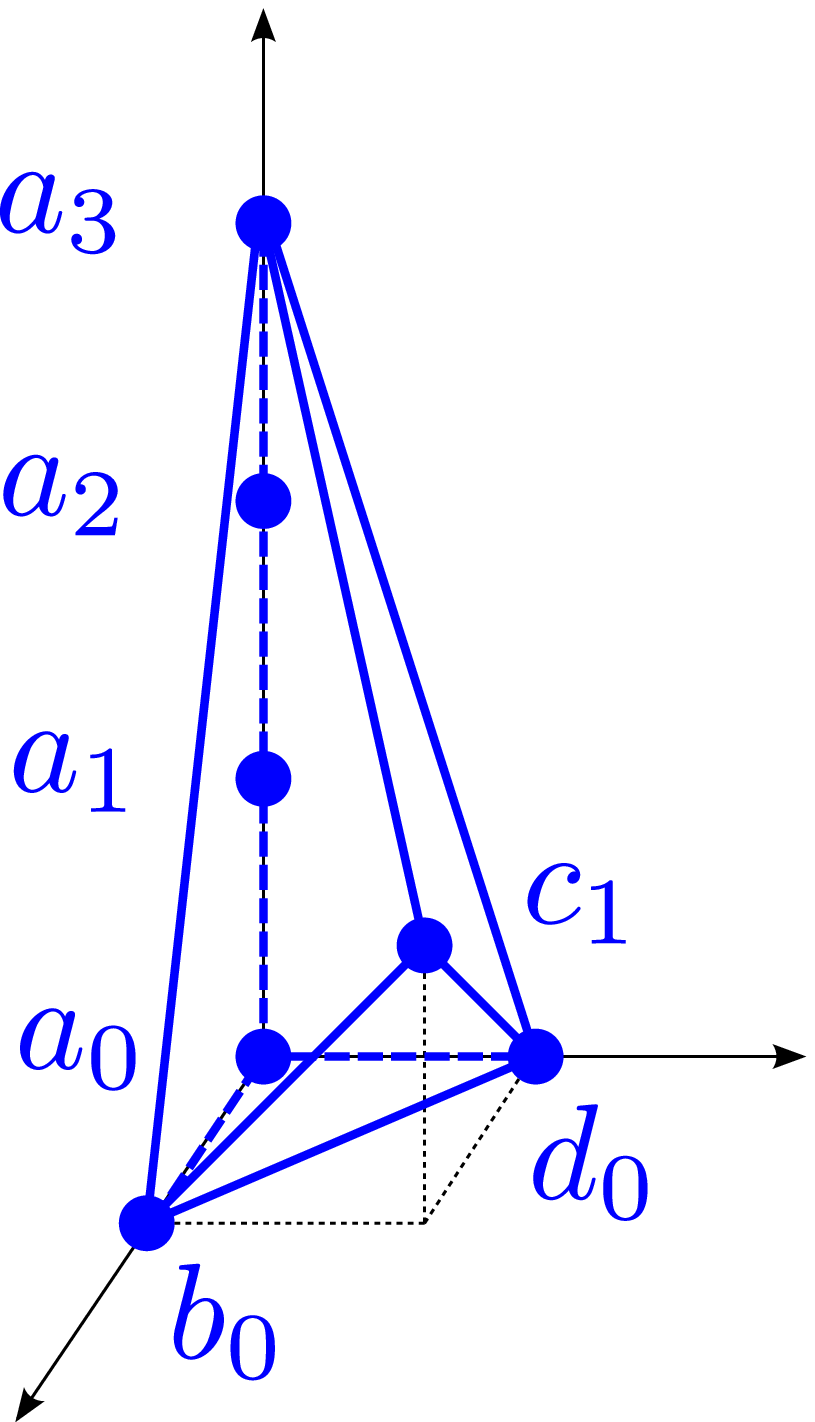}
\label{fig:toric_diag_flav_ABJM}}
\qquad \qquad
\subfigure[\small Quiver diagram.]{
\includegraphics[height=5cm]{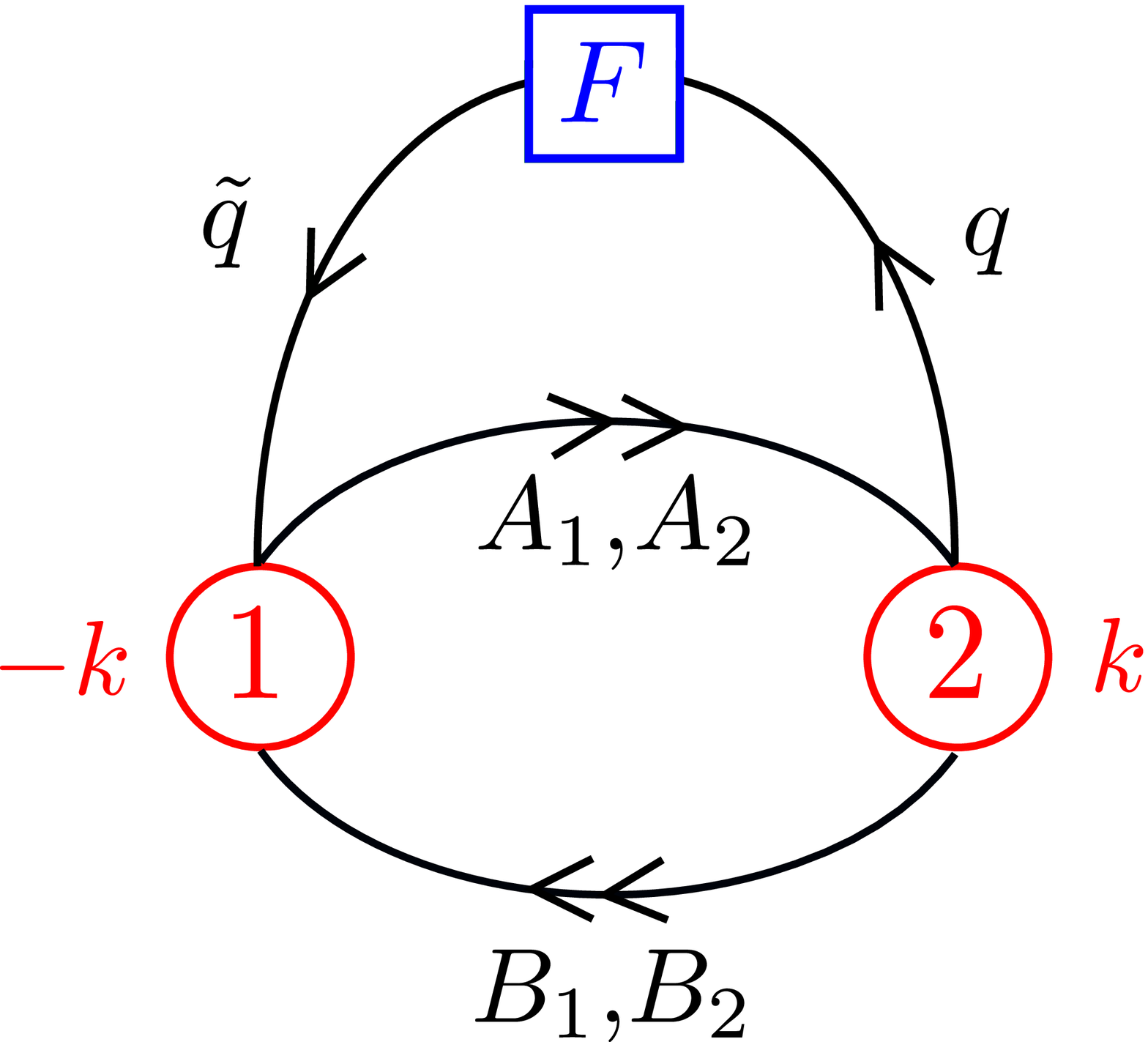}
\label{fig:quiver_flav_ABJM}}
\end{center}
\caption{Toric diagram of a toric CY$_4$ singularity leading to ABJM with $p$ flavors, for $p=3$ and $h=1$, and quiver diagram of the flavored ABJM theory.}
\end{figure}
For comparison with section \ref{sec:M_to_IIA}, we consider a 2-parameter subfamily of CY$_4$ singularities, whose toric diagrams take the form (see fig. \ref{fig:toric_diag_flav_ABJM})
\be\label{toric_diag_flav_ABJM}
a_i=(0,0,i)\;,\qquad b_0=(1,0,0)\;,\qquad c_h=(1,1,h)\;,\qquad d_0=(0,1,0)
\ee
with $i=0,1,\cdots,p$ running over the single vertical column.
The charges of the GLSM for the CY$_4$, including the global $U(1)_M$ charges in the last line, are
\be
\label{GLSM_flav_ABJM}
\begin{array}{c|cccccccccc|c}
\text{CY}_4 &   a_0 & a_1 & a_2 & \cdots & a_{p-2} & a_{p-1} & a_p  & b_0 & c_h & d_0  &\mathrm{FI}\\ \hline
 &  h+1 & -h & 0 & \cdots & 0 & 0 & 0 & -1 & 1 & -1 & \xi^c \\
 & 1 & -2 & 1 & \cdots & 0 & 0 & 0 &  0 & 0 & 0 & \xi_2  \\
 &  0 & 1 & -2 & \cdots  & 0 & 0 & 0  & 0 & 0 & 0 &\xi_3  \\
 &  \vdots & \vdots & \vdots & \ddots  & \vdots & \vdots & \vdots & \vdots &\vdots & \vdots & \vdots \\
 & 0 & 0 & 0 & \cdots & -2 & 1 & 0 & 0 & 0 & 0 & \xi_{p-1}  \\
 & 0 & 0 & 0 & \cdots & 1 & -2 & 1 & 0 & 0 & 0 & \xi_{p}  \\
\hline
U(1)_M &  1 & -1 & 0 &\cdots & 0  & 0 & 0 & 0 & \cdots& 0 & r_0
\end{array}
\ee
with $\xi_a\geq 0$ for all $a=2,\cdots,p$.
The minimal GLSM for the fibered conifold $\cC$ in IIA is
\be\label{GLSM_conif_flav_ABJM}
\begin{array}{c c c  c |c}
  a & b & c & d & FI\\ \hline
  1 & -1 & 1 & -1 & \chi(r_0)
\end{array}
\ee
where
\be\label{resol_param_IIA_flav_ABJM}
\chi(r_0) = \xi^c - h\, (r_0-\zeta_1) - \sum_{a=1}^p (r_0-\zeta_a) \Theta(r_0-\zeta_a)
\ee
and the parameters $\zeta_a$ are defined as in \eqref{def_zeta_in_t_of_xi}. The RR 2-form flux through the 2-cycle of the resolved $\cC$ is $\chi '(r_0)$, and there is a D6-brane along the divisor $D_a$ of the resolved conifold over $r_0=\zeta_a$, for each $a=1,\cdots,p$.

The type IIA background allows us to derive the field theory on $N$ 2-brane probes \cite{Benini:2009qs}: an ABJM $U(N)_{-k}\times U(N)_{k}$ quiver with $p$ pairs of flavors  $\tilde q^a$, $q_a$ as in fig. \ref{fig:quiver_flav_ABJM} and superpotential
\be\label{W_flav_ABJM}
W= \epsilon^{ij} \epsilon^{rs} \Tr (A_i B_r A_j B_s) + \sum_{a=1}^p \tilde{q}^a A_1 q_a\;.
\ee
The CS level $k$ is related to the integers $h$ and $p$ appearing in the CY$_4$ geometry by $k=h+\frac{p}{2}$, in agreement with the quantization law \eqref{quantization_CS}.
The complete map between geometric and field theory data will be provided below.

The flavor symmetry group of the field theory is $SU(p)$ rather than $U(p)$, since the action of the overall $U(1)$ in $U(p)$ coincides with that of the overall diagonal gauge $U(1)$. Its rank $p-1$ equals the number of independent real masses for flavors. However, to compare with \eqref{resol_param_IIA_flav_ABJM}, we find it convenient to work in terms of the flavor group $U(p)$ of the ungauged theory and introduce $p$ real masses $\{m_a\}$ as the eigenvalues of the corresponding background hermitian scalar. One of them can be absorbed by a shift of the real scalar $\sigma$ in the overall diagonal gauge $U(1)$.

Following the approach of section \ref{subsec:_chiral_ring_for_CY4}, the geometric moduli space of the Abelian gauge theory, where the flavor fields do not acquire a VEV, was computed in \cite{Benini:2009qs,Jafferis:2009th} for the superconformal theory with vanishing bare FI parameters and real masses. The gauge charges of bifundamentals and unit flux diagonal monopole operators are
\be\label{gauge_charges_flav_ABJM}
\begin{array}{c | c c c c }
 & A_j & B_r & T & \tilde{T}\\ \hline
U(1)_{-h-\frac{p}{2}} & 1 & -1 & -h & h+p\\
U(1)_{h+\frac{p}{2}} & -1 & 1 & h & -h-p
\end{array}
\ee
and the quantum corrected F-term relation involving monopole operators is $T\tilde{T}=A_1^p$. It is then easy to prove that the geometric moduli space \eqref{geometric_moduli_space} of the Abelian gauge theory reproduces the CY$_4$ cone geometry probed by the M2-brane in M-theory \cite{Benini:2009qs}.

\begin{figure}[t]
\begin{center}
\includegraphics[height=2cm]{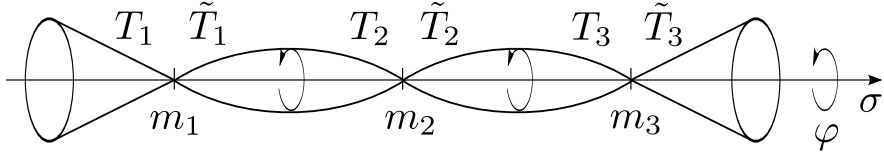}
\end{center}
\caption{Topology of the complex dimension parametrized by bare monopole operators in flavored ABJM for $p=3$ flavors with non-vanishing real masses. \label{fig:monopoles_resolved}}
\end{figure}

We now perform the semiclassical analysis of these flavored ABJM theories, in the general case where real masses $m_a$ and bare FI parameters $\xi_1=-\xi_2$ are allowed. As explained in section \ref{subsec:_VMS_from_1loop}, it suffices to consider the Abelian gauge theory. By a Weyl transformation in the flavor group and a shift of the real scalar $\sigma$ in the gauge vector multiplet, we order the real masses as follows
\be\label{order_masses}
0=m_1\leq m_2\leq \dots\leq m_p\;.
\ee
We first compute the effective CS levels
\be
k_1^\text{eff}(\sigma)=-k_2^\text{eff}(\sigma)= -\left(h+\frac{p}{2}\right) - \frac{1}{2}\sum_{a=1}^p  \sign(\sigma-m_a)    \label{eff_CS_flav_ABJM}
\ee
and then the effective FI parameters of the two gauge groups
\be\label{eff_FI_flav_ABJM}
\xi_1^\text{eff}(\sigma)=-\xi_2^\text{eff}(\sigma)=\xi_1 + \left(\frac{h}{p}+\frac{1}{2}\right)\sum_{a=1}^p m_a - \left(h+\frac{p}{2}\right)\sigma - \frac{1}{2}\sum_{a=1}^p  |\sigma-m_a|\;.
\ee
Rewriting \eqref{eff_FI_flav_ABJM} as
\be\label{eff_FI_flav_ABJM_2}
\xi_1^\text{eff}(\sigma) = \xi_1 + \frac{h}{p}\sum_{a=1}^p m_a - h \,(\sigma -m_1) -  \sum_{a=1}^p (\sigma-m_a) \Theta(\sigma-m_a)\;,
\ee
we compare it to the resolution parameter \eqref{resol_param_IIA_flav_ABJM} of the fibered conifold in IIA to find the precise map between parameters in field theory and geometry:
\be
\sigma = r_0 \;,\qquad m_a = \zeta_a \;,\qquad \xi_1 + \frac{h}{p} \sum_a m_a = \xi^c \;,\qquad \xi_1^\text{eff} = \chi \;.
\ee
Finally, we can introduce the dual photon $\varphi$ in the one-loop field theory analysis to complexify the real scalar $\sigma$. Exponentiating the resulting complex scalars provides BPS diagonal monopole operators of flux one (sec. \ref{app_subsec:_dual_photon}). We are now considering a more general situation compared to section \ref{subsec:_VMS_from_1loop}, as we allow real masses and FI parameters. A crucial insight on the degeneration loci of the $U(1)_M$ circle acting on the dual photon is provided by the quantized effective CS level \eqref{eff_CS_flav_ABJM}, which jumps whenever $\sigma$ equals one of the real masses $m_a$. Such a jump is possible only if the $U(1)_M$ circle shrinks there. As a result, the topology of the complex direction parametrized by bare monopole operators is the one shown in figure \ref{fig:monopoles_resolved}: a chain of $p-1$ $\CP^1$'s is generated, whose sizes equal the differences of adjacent real masses $m_{a+1}-m_a$.

This topology arises from the resolution of the $\bC^2/\bZ_p$ singularity parametrized by the bare monopole operators $T$, $\tilde{T}$ and the flavored bifundamental $A_1$ subject to the quantum relation $T\tilde{T}=A_1^p$, as we now explain (see for instance \cite{Hori:1997ab} for a review of this resolution).
The resolved singularity is a smooth variety covered by $p$ patches $U_a$ parametrized by $(T_a,\tilde{T}_a)$, $a=1,\cdots,p$, with covering maps
\begin{equation}\label{resolved_C2_Zp}
(T_a,\tilde{T}_a) \longmapsto \begin{cases}
T = T_a^a \tilde{T}_a^{a-1}\\
\tilde{T} = T_a^{p-a} \tilde{T}_a^{p+1-a}\\
A_1 = T_a \tilde{T_a}
\end{cases}
\end{equation}
and transitions given by
\begin{equation}\label{transitions_resolved}
\tilde{T}_a T_{a+1}=1\;,\qquad T_a \tilde{T}_a = T_{a+1} \tilde{T}_{a+1}\;.
\end{equation}
In particular the monopole operators of the CFT can be expressed as $T=T_1$ and $\tilde{T}=\tilde{T}_p$, which describe the noncompact cones in figure \ref{fig:monopoles_resolved}. The remaining $T_a$ and $\tilde{T}_a$ describe the $p-1$ blown up $\CP^1$'s, since $\tilde{T}_a T_{a+1}=1$. The gauge charges of $T_a$ and $\tilde{T}_a$ follow from \eqref{resolved_C2_Zp} and \eqref{transitions_resolved},
\begin{equation}\label{gauge_charges_monopoles_resolved}
\begin{split}
Q_i[T_{a}]&=Q_i[T_1]-(a-1) Q_i[A_1]\\
Q_i[\tilde{T}_{a}]&=Q_i[\tilde{T}_p]-(p-a) Q_i[A_1]\;.
\end{split}
\end{equation}
Specializing to the first gauge group we find $Q_1[T_a]=-h-a+1$ and $Q_1[\tilde{T}_a]=h+a$, in agreement with the charges induced on diagonal monopoles of flux one by the effective CS levels \eqref{eff_CS_flav_ABJM} in the corresponding range of $\sigma$.
Finally note that
\begin{equation}\label{quantum_relations_flav_ABJM_resolved}
T_a \tilde{T}_b = A_1^{b-a+1}\;,\qquad b\geq a\;.
\end{equation}
Summarizing, pairs of unit flux monopole operators $(T_a, \tilde{T}_{a+1})$ appear when real mass differences $m_{a+1}-m_a$ are turned on. Pairs $(T_{a},\tilde{T}_a)$ disappear when the corresponding real masses $m_a$ are sent to infinity, and the holomorphic relations between surviving monopole operators take the form \eqref{quantum_relations_flav_ABJM_resolved}.
By this method we can describe, at the level of the geometric moduli space and the chiral ring, real mass deformations of flavored ABJM models which correspond to partial or complete resolutions of the $\bC^2/\bZ_p$ fiber in the CY$_4$ geometry.


\section{Charges of monopoles and localization}
\label{app: charges of monopoles}

In this appendix we show with localization methods that the monopole charges used in section \ref{sec:_VMS_monopoles} are one-loop exact. We will essentially borrow the ideas in \cite{Borokhov:2002ib, Borokhov:2002cg, Bashkirov:2010kz}, and integrate them with the results in \cite{Kim:2009wb, Imamura:2011su}, whose conventions we follow.

In any 3d conformal theory there is a one-to-one correspondence between local operators on $\bR^3$ and states on $S^2 \times \bR$ (we will consider the Euclidean theory). Therefore to determine the charges of a monopole operator at the origin of $\bR^3$, we compute the charges on $S^2\times \bR$ of a state carrying some magnetic flux on $S^2$.

We are interested in half-BPS monopole operators (described by chiral multiplets) in an $\cN=2$ superconformal theory, because to them we can apply localization methods. Localization allows us to exactly compute the path-integral of a supersymmetric theory, with the insertion of $\cQ$-closed operators and with $\cQ$-invariant boundary conditions ($\cQ$ being a supercharge). The operators we insert are the charges we want to measure, the boundary conditions are those of the monopole state.

Flavor charges commute with the supercharges and do not pose any problem. On the contrary the R-charge does not commute with the supercharges, and cannot be directly computed with localization by inserting it in the path-integral. Likewise on $S^2 \times \bR$ the supercharges do not commute with the Hamiltonian $H$ that generates translations on $\bR$ (on $\bR^3$, $H$ generates dilations and the supercharges have dimension $1/2$):
\be
[R,Q_\alpha] = Q_\alpha \;,\quad [R, \bar Q_\alpha] = - \bar Q_\alpha \;,\quad
[H, Q_\alpha] = \frac12 Q_\alpha \;,\quad [H, \bar Q_\alpha] = \frac12 \bar Q_\alpha \;.
\ee
Here both $H$ and $R$ are Hermitian operators.
Notice that in radial quantization $Q_\alpha$ and $\bar Q_\alpha$ are no longer Hermitian conjugate to each other: their Hermitian conjugates are the special supercharges $S_\alpha = Q_\alpha^\dag$ and $\bar S_\alpha = \bar Q_\alpha^\dag$, of dimension $-1/2$. Therefore
\be
[R, S_\alpha] = - S_\alpha \;,\quad [R, \bar S_\alpha] = \bar S_\alpha \;,\quad
[H, S_\alpha] = - \frac12 S_\alpha \;,\quad [H, \bar S_\alpha] = - \frac12 \bar S_\alpha \;.
\ee

To find the eigenvalues of $H$ and $R$ on the monopole background, we select a particular supercharge $\cQ$ (specified below) and its conjugate $\cQ^\dag$, which preserve the monopole background and satisfy the algebra \cite{Bhattacharya:2008zy}
\be
\{\cQ,\cQ\} = \{\cQ^\dag, \cQ^\dag\} = 0 \;,\qquad [\cQ,H+J_3] = 0 \;,\qquad \{\cQ, \cQ^\dag\} = H - R - J_3
\ee
where $J_i$ are the generators of rotations on $S^2$. Notice that the last relation implies the BPS energy bound $E \geq r + j_3$ (for the eigenvalues of $H$, $R$, $J_3$), while it also implies that $(H - R - J_3)$ commutes with $\cQ$ and $\cQ^\dag$. Therefore we can compute them localizing the path integral with respect to $\cQ$. We are interested in rotationally invariant monopole states, $J_3 = 0$, so the computation of $(H+J_3)$ gives us the dimension $E$ of the monopole operator. The charge $(H-R-J_3)$ simply vanishes, because on scalars in short representations $E = r$.

To compute the energy $E$ of a monopole state we will need the dimensions $\Delta_\Phi$ of fundamental fields. As a consequence, we can only compute the R-charge of a monopole operator as a function of the R-charges of fundamental fields. The values of the charges at the IR fixed point could be computed with an extremization method \cite{Jafferis:2010un}. However having them as functions will be good enough to our purposes of determining quantum relations in the chiral ring.

\

Let us review the computation in \cite{Imamura:2011su}.
The $\cN=2$ superconformal algebra has 8 supercharges; in Euclidean signature they are parametrized by the independent spinors $\epsilon$ and $\bar\epsilon$. On a conformally flat background the spinor $\epsilon$ satisfies the Killing equation $D_\mu \epsilon = \gamma_\mu \kappa$, where $\kappa$ is an arbitrary spinor, and it is parameter for the global supercharges $Q$ and $S$.
Similarly $\bar\epsilon$ satisfies $D_\mu \bar\epsilon = \gamma_\mu \bar\kappa$ and parametrizes $\bar Q$ and $\bar S$.

The supercharge $\cQ$ we will use for localization is constructed out of $\bar Q$ and $\bar S$. Therefore let us look at the anti-holomorphic transformations of fields. For a vector multiplet $V = (A_\mu, \sigma, D, \lambda)$:
\bea
\delta \sigma &= \bar\epsilon \lambda \;,\qquad\qquad \delta A_\mu = - i \, \bar\epsilon \gamma_\mu \lambda \;,\qquad\qquad \delta \lambda = 0 \\
\delta D &= i \, \bar\epsilon \gamma^\mu D_\mu \lambda + i \, \bar\epsilon [\sigma,\lambda] + \frac i3 \, D_\mu \bar\epsilon \gamma^\mu \lambda \\
\delta \bar\lambda &= - \frac i2\, \gamma^{\mu\nu} \bar\epsilon F_{\mu\nu} - \gamma^\mu \bar\epsilon \, D_\mu \sigma + i \, D\bar\epsilon - \frac23 \, \gamma^\mu D_\mu \bar\epsilon \, \sigma \;.
\eea
The transformations for a chiral multiplet $\Phi = (\phi, \psi, F)$ with Weyl weight $\Delta_\Phi$ are:
\bea
\delta \phi^\dag &= \sqrt2\, \bar\epsilon \bar\psi \;,\qquad\qquad \delta \phi = 0 \;,\qquad\qquad \delta \bar\psi = \sqrt2\, i \, \bar\epsilon F^\dag \;,\qquad\qquad \delta F^\dag = 0  \\
\delta \psi &= \sqrt2 \, \bar\epsilon \sigma \phi - \sqrt2 \, \gamma^\mu \bar\epsilon \, D_\mu\phi - \frac{2\sqrt2}3 \, \Delta_\Phi \phi \, \gamma^\mu D_\mu \bar\epsilon \\
\delta F &= \sqrt2 \, i \, \bar\epsilon \gamma^\mu D_\mu \psi + \sqrt2\, i \, \bar\epsilon \sigma \psi + 2i\, \bar\epsilon \bar\lambda \, \phi + \frac{2\sqrt2 \, i}3 \Big( \Delta_\Phi - \frac12 \Big) D_\mu \bar\epsilon \gamma^\mu \psi \;.
\eea
$\cN=2$ superconformal symmetry forces the superconformal R-charge of a chiral multiplet to be equal to $\Delta_\Phi$.

We deform the theory by some $\delta_1$-exact piece:
\bea
S_\text{def}^\text{vector} &= \delta_1 \delta_2 \int d^3x\, \sqrt g \, \tr \Big( - \frac t2 \bar\lambda \bar\lambda \Big) \\
S_\text{def}^\text{chiral} &= \delta_1 \delta_2 \int d^3x\, \sqrt g \, \Big( - \frac{it}2 \, \phi^\dag F \Big)
\eea
where $\delta_i$ are SUSY variations with respect to two linearly independent anti-holomorphic spinors. On $S^2 \times \bR$ we will take as conformally Killing spinors the two solutions $\bar\epsilon_{1,2}$ of
\be
D_\mu \bar\epsilon = \frac1{2r} \gamma_\mu \gamma_3 \bar\epsilon \;.
\ee
These form a doublet of $SO(3)$ rotations of $S^2$, and we take the third component $j_3$ such that $\bar\epsilon_1$ and $\bar\epsilon_2$ have $j_3$ eigenvalues $+\frac12$ and $-\frac12$ respectively. Therefore our charge $\cQ$ corresponds to $\bar\epsilon_1$.
With that choice of $\bar\epsilon_{1,2}$, the deformation action becomes
\bea
S_\text{def}^\text{vector} &= t \int d^3x\, \sqrt g\, \tr \Big[ V_\mu V^\mu + D^2 - 2 \bar\lambda \gamma^\mu D_\mu \lambda - 2 \bar\lambda [\sigma,\lambda] - \frac 1r\, \bar\lambda \gamma_3 \lambda \Big] \\
S_\text{def}^\text{chiral} &= t \int d^3x\, \sqrt g\, \Big[ \frac{1-2\Delta_\Phi}r \Big( \phi^\dag D_3 \phi + \frac12 \, \bar\psi \gamma_3 \psi \Big) + \frac{\Delta_\Phi (1-\Delta_\Phi)}r^2 \, \phi^\dag\phi \Big]
\eea
where the vector $V_\mu$ is defined as
\be
V_1 = F_{23} - D_1\sigma \;,\qquad V_2 = F_{31} - D_2\sigma \;,\qquad V_3 = F_{12} - D_3 \sigma - \frac1r \sigma \;.
\ee
The path-integral localizes around $V_\mu = 0$, $\phi = 0$. Among the solutions, we find BPS monopole states:
\be
A_\mu dx^\mu = H\, B_i dx^i \;,\qquad\qquad \sigma = \frac H{2r}
\ee
with all other fields vanishing,
where $B_i$ is the Dirac monopole configuration with unit magnetic charge, and up to gauge fixing $H$ takes value in the Cartan part $\fh$ of the gauge Lie algebra $\fg$. Notice that due to the boundary conditions on $\bR$ the classical background has no zero-modes.

The localization is carried out in \cite{Kim:2009wb} for the vector multiplet, and in \cite{Imamura:2011su} for the chiral multiplet. We do not need their full result, only the charges,
because everything else cancels out in the normalization of the monopole state. After a suitable regularization, their result is
\bea
E &= \frac12 \sum_\Phi (1- \Delta_\Phi) \sum_{\rho \in R_\Phi} |\rho(H)| - \frac12 \sum_{\alpha \in \cG} |\alpha(H)| \\
q_i &= - \frac12 \sum_\Phi F_i(\Phi) \sum_{\rho \in R_\Phi} |\rho(H)| \\
b_0(a) &= - \frac12 \sum_\Phi \sum_{\rho \in R_\Phi} |\rho(H)| \, \rho(a) \;,
\eea
where $\rho$ are weights of the representation $R_\Phi$, $\alpha$ are roots of the gauge group $\cG$, $F_i$ are the flavor charges of the fields $\Phi_i$, $E$ is the dimension (equal to the superconformal R-charge) and $q_i$ are global charges. The linear function $b_0(a)$ on $\fh$ appears in the path-integral as a phase $e^{ib_0(a)}$, function of the Cartan gauge field $a \in \fh$. In fact it is an element of $\fh^*$.
The topological charges are not quantum corrected because no fundamental fields are charged under them.

Flavor and gauge charges are the result we were looking for. The dimension $E = r$, where $r$ is the superconformal R-symmetry, still depends on the unknowns $\Delta_\Phi = r_\Phi$. However we can rewrite the formulae{} above as a sum over fermions (gaugini and in chiral multiplets):
\be
\delta q = -\frac12 \sum_\text{fermions $f$ \;} \sum_{\rho \in R_f} |\rho(H)| \, q_f \;.
\ee
This is true for both superconformal R-charge, flavor charges and gauge charges ($\delta$ means quantum correction: classically monopoles can have gauge charges). Since the formula is linear in the charges, it is correct for any R-charge, even those which are not in the stress-tensor multiplet (they are combinations of the superconformal R-symmetry and of flavor symmetries).

\subsection{Quantum F-term relations}
\label{subsec: check of quantum relations dP0}

In section \ref{sec:_VMS_monopoles} we showed how to obtain the chiral ring of $\cN=2$ CS-matter theories: first compute the (one-loop exact) charges of monopole operators, then construct quantum F-term relations compatible with those charges. We constructed the shortest gauge-invariant relations compatible with the charges. However it is also crucial that the coefficient in those relations is not zero, otherwise the chiral ring is different. As partial evidence, we show here that the counting of fermionic zero-modes is compatible with a non-vanishing coefficient, following an argument in \cite{Borokhov:2002cg}.

Let us consider, for simplicity, window $[0,0]$ of the $Y^{p,q}(\CPt)$ quiver with the extra restriction $n_1 \leq p/2$ (similar considerations apply in all other windows). The quantum F-term relation reads schematically
\be
T\tilde T Y^{3n_1} Z^{3p-3n_1} \sim (XYZ)^p
\ee
and we want to check whether the coefficient is non-vanishing. To do that, we should compute the amplitude
\be
\langle \tilde T^\dag | Z^{\dag p} Y^{\dag p} X^{\dag p} Y^{3n_1} Z^{3p-3n_1} |T \rangle
\ee
in radial quantization. To be precise, such amplitude is not gauge invariant, because we are using bare monopoles. To make it gauge-invariant we should either connect the bare monopoles with the undaggered operators by Wilson lines, or move the undaggered operators to $\pm\infty$ in Euclidean time (essentially using the gauge invariant monopoles $b_{(q)}$ and $c_{(3p-q)}$ in place of the bare monopoles $T$ and $\tilde T$). We can think of adding the Wilson lines, and take the limit at the end.
The amplitude has a chance to be non-vanishing if the operators inserted absorb all fermionic zero-modes on the monopole background.

We essentially already computed the number of zero-modes when computing the mo\-no\-pole charges.
We have the following number of fermionic zero-modes:
\bea\nn
3(p-2n_1) &\text{ for } \psi_X \qquad\qquad & 2(p-n1) &\text{ for } \lambda_2 \\
3(p-n_1) &\text{ for } \psi_Y & 2(p-2n_1) &\text{ for } \lambda_3 \\
3(2p-3n_1) &\text{ for } \psi_Z \;.
\eea
These very zero-modes are responsible for the charges of bare monopoles.

The operators we have inserted are only scalars, however they are coupled to matter fermions by terms from the superpotential:
\be
V \supset \partial_i \partial_j W \, \psi_i \psi_j \qquad\Rightarrow\qquad X^\dag \sim \psi_Y \psi_Z \;,\qquad Y^\dag \sim \psi_Z \psi_X \;,\qquad Z^\dag \sim \psi_X \psi_Y
\ee
and to gaugini by terms from the D-term kinetic terms:
\be
\int d^4\theta \, \Phi^\dag e^V \Phi \;\to\; \phi^\dag \, \lambda \, \psi_\phi
\ee
so that we get
\be
X \sim \lambda_3 \psi_X \sim \lambda_2 \psi_X \;,\qquad Y \sim \lambda_1 \psi_Y \sim \lambda_2 \psi_Y \;,\qquad Z \sim \lambda_2\psi_Z \sim \lambda_3\psi_Z \;.
\ee
If we just substitute these relations, we can soak up the following zero-modes:
$$
\psi_X^{2p} \psi_Y^{2p+3n_1} \psi_Z^{5p-3n_1} \lambda_2^{p+3n_1} \lambda_3^{2p-3n_1}
$$
which is not quite enough. However there are other terms in the Lagrangian that couple scalar fields: for instance the scalar potential contains
$$
V \subset XYY^\dag X^\dag \qquad\Rightarrow\qquad X^\dag \sim X^\dag YY^\dag \sim \psi_Y \psi_Z (\lambda_3 \psi_X\psi_Y\psi_Z)
$$
and so on. Using these other relations all the fermionic zero-modes can be soaked up.

\section{Mirror geometry and exact periods}
\label{app: mirror and exact periods}

For completeness, we write down in this appendix the exact central charges for the B-branes on $\cO_{\CP^2}(-3)$, as was found in \cite{Diaconescu:1999dt}.%
\footnote{See also \cite{Aspinwall:2004jr} for an interesting discussion.}
The periods are solutions of Picard-Fuchs equations in the mirror CY$_3$.
The mirror of $\cO_{\CP^2}(-3)$ can be described as a double fibration over a plane $\bC \cong \{W \}$:
\be\label{SigmaW}
W= P(x,y) = x+ y + \frac{1}{xy} -\psi \,  \, , \qquad W = uv\, .
\ee
The parameter $\psi$ is a complex structure modulus of the mirror geometry, corresponding to the single K\"ahler modulus of $\cO_{\CP^2}(-3)$.
The first equation in (\ref{SigmaW}) describes a Riemann surface of genus one, fibered over the plane, which degenerates at 3 critical points on the $\{W \}$ plane, namely at
\be\label{3 critical points of mirror}
W + \psi = 2 \omega^n + \omega^{-2n}\, , \qquad\quad \mathrm{where} \qquad x=y= \omega^{n}\, ; \quad \quad n=0,1,2\, .
\ee
$\omega$ is the third rood of unity. A good coordinate on the complex moduli space of the curve is
\be
z =  \frac{27}{\psi^3}
\ee
The factor of $27$ is there for convenience,  such that $z=1$ is the conifold point in moduli space (where one of the fractional branes becomes tensionless).
The Picard-Fuchs equation is \cite{Klemm:1999gm}
\be\label{PF equ for dP0}
\left( \theta_z^3 - z \left( \theta_z +\frac{2}{3}\right) \left( \theta_z +\frac{1}{3}\right)\theta_z \right) \Pi\, = \, 0\, ,
\ee
with $\theta_z \equiv z \partial_z$. Obviously, the constant period $\Pi=1$ is a solution. This equation is a Meijer equation; a basis for the non-trivial solutions is given by two Meijer G-functions; see \textit{e.g.} \cite{Meijer:wiki}. Instead of the large volume point $z=0$, we are rather interested in the form of $\Pi$ near the orbifold point $w=1/z=0$. We can write the two relevant Meijer G-functions in terms of generalized hypergeometric functions:
\bea
G_1(w)  \, &=& \,
\frac{9}{2} \frac{\Gamma(\frac{2}{3})^2}{\Gamma(\frac{1}{3})} (-w)^{\frac{2}{3}} \, {}_3F_2\left(\frac{2}{3}, \frac{2}{3}, \frac{2}{3}; \frac{4}{3}, \frac{5}{3}; w\right) \, , \label{def g1 g2 periods} \\
G_2(w)  \, &=& \,
3 \frac{\Gamma(\frac{1}{3})^2}{\Gamma(\frac{2}{3})}(-w)^{\frac{1}{3}} \, {}_3F_2\left(\frac{2}{3}, \frac{2}{3}, \frac{2}{3}; \frac{4}{3}, \frac{5}{3}; w\right)\, .
\eea
These functions are defined on the whole $\{w\}$-plane by analytic continuation; we choose the cuts to lie on the real positive axis, from $w=0$ to $w=1$, and from $w=1$ to $w= \infty$.
One can prove that the exact periods for the D4-brane on $\CP^1 \subset \CP^2$ and for the D6-brane on $\CP^2$ are:
\bea
Z(D4) &=& t(w) &=&    \frac{\omega^2-\omega}{4\pi^2}\left(G_1(w) -G_2(w)\right)\, ,\\
Z(D6) &=& t_6(w) &=& \frac{1}{3} + \frac{1}{4\pi^2} \left(\omega^2 G_1(w) +\omega G_2(w)\right)\, ,
\eea
where $\omega= e^{\frac{2\pi i}{3}}$.

\section{Torsion flux, B-field and Page charges}
\label{app: torsion flux}

In this appendix we want to explicitly show, at least for the case of $Y^{p,q}(\CPt)$, that an M-theory torsion flux $G_4 \in H^4(Y^{p,q},\bZ) = \Gamma$ is in fact represented by a flat connection $C_3$ (well-defined only patch by patch), which descends in type IIA to a vanishing $F_4$ and a non-trivial B-field $B_2$, constrained in such a way that the Page D4-charge $F_2\wedge B_2$ is quantized. Moreover $C_3$ in M-theory and $B_2$ in type IIA are pure gauge precisely on the kernel of the projection from $\bZ^2$ to $\Gamma$. The discussion here is classical, so it is not corrected by the Freed-Witten anomaly which requires a deeper analysis.

The metric of $Y^{p,q}(\CPt)$ is \cite{Gauntlett:2004hh, Martelli:2008rt}:
\bea
ds^2(Y^{p,q}) &= \rho^2 d\tilde s^2(\CPt) + \frac{d\rho^2}{U(\rho)} + q(\rho) (d\psi + A)^2 + w(\rho) \ell^2 \big[ d\gamma + \ell^{-1} f(\rho)(d\psi + A) \big]^2 \\
d\tilde s^2(\CPt) &= \frac34 \Big[ d\beta^2 + \sin^2\beta\, \sigma_3^2 + \sin^2 \frac\beta2 \, (d\theta^2 + \sin^2\theta\, d\varphi^2) \Big] \\
A &= 3 \sin^2\frac\beta2\, \sigma_3 \qquad\qquad\qquad \sigma_3 = \frac{d\chi - \cos\theta\, d\varphi}2 \\
\tilde J &= \frac34 \sin\beta\, d\beta \wedge \sigma_3 + \frac34 \sin^2\frac\beta2\, \sin\theta\, d\theta \wedge d\varphi \;.
\eea
The range of coordinates is $\theta \in [0,\pi]$, $\varphi \in [0,2\pi)$, $\chi \in [0,4\pi)$, $\beta \in [0,\pi]$, $\rho \in [\rho_1,\rho_2]$, $\psi \in [0,2\pi)$, $\gamma \in [0,2\pi)$ and intervals with an open end are compactified. The functions $U(\rho)$, $q(\rho)$, $w(\rho)$ and $f(\rho)$ are defined in \cite{Gauntlett:2004hh, Martelli:2008rt} and depend on a parameter: $U$ and $q$ have simple roots at $\rho_{1,2}$ with $0<\rho_1<\rho_2$, while $w$ is positive in $[\rho_1,\rho_2]$. $\tilde J$ is the K\"ahler form of $\CPt$ and satisfies $2\tilde J = dA$. $(\rho,\psi)$ describe an $S^2$ bundle over $\CPt$, that is $M_6$, and $\gamma$ describes a $U(1)$ bundle over $M_6$.
The metric is normalized such that $\widetilde{\text{Ric}} = 2\tilde g$ and $\text{Ric} = 6g$. The constant $\ell$ is chosen in such a way that
\be
C_1 \equiv \ell^{-1} f(\rho)(d\psi + A) \;,\quad \int_{\cC_0} \frac{dC_1}{2\pi} = f(\rho_2) - f(\rho_1) \equiv p \;,\quad \int_{\cC^+} \frac{dC_1}{2\pi} = f(\rho_2) \equiv 3p-q
\ee
that is the $U(1)$ bundle over $M_6$ is well-defined. $\cC_0$ is the $S^2$ fiber of $M_6$ and $\cC^+$ is $\CP^1 \subset \CPt$ at $\rho = \rho_2$. In fact $F_2 = dC_1$, living on $M_6$, becomes the RR 2-form in type IIA.

We add torsion flux $G_4 \in \Gamma$ on $Y^{p,q}$ in M-theory, where $\Gamma = \bZ^2 / \langle (3q,q),\, (q,p)\rangle$. We represent it with a non-trivial flat connection $C_3$, which will be expressed in terms of a closed 2-form $B_2$ on $M_6$: in fact $H^2(M_6, \bR) = \bR^2$ is generated by $dC_1$ and $\tilde J$. We cannot use $[d\gamma + C_1]\wedge B_2$ because it is not closed, nor can we use $d\gamma \wedge B_2$ because $d\gamma$ is singular at 4 points on $M_6$: $\beta = \pi$, $\rho = \rho_{2,1}$, $\theta = 0,\pi$. Thus we cover $Y^{p,q}$ with four patches, each including one singular point and not the other three, and write:
\bea
\label{C3 ansatz}
C_3\big|_{NN} &= \Big[ d\gamma + \ell^{-1} f(\rho_2)\Big( d\psi + 3 \frac{d\chi - d\varphi}2 \Big) \Big] \wedge B_2 \\
C_3\big|_{NS} &= \Big[ d\gamma + \ell^{-1} f(\rho_2)\Big( d\psi + 3 \frac{d\chi + d\varphi}2 \Big) \Big] \wedge B_2 \\
C_3\big|_{SN} &= \Big[ d\gamma + \ell^{-1} f(\rho_1)\Big( d\psi + 3 \frac{d\chi - d\varphi}2 \Big) \Big] \wedge B_2 \\
C_3\big|_{SS} &= \Big[ d\gamma + \ell^{-1} f(\rho_1)\Big( d\psi + 3 \frac{d\chi + d\varphi}2 \Big) \Big] \wedge B_2
\eea
where the first N/S refers to $\rho = \rho_{2,1}$, the second N/S to $\theta = 0,\pi$, respectively. These are constructed in such a way to coincide with $[d\gamma + C_1]\wedge B_2$ at the singular points, yet being closed.

The connections smoothly glue together if and only if the transition functions are well-defined. For instance, $C_3|_{NN} - C_3|_{NS} = -3\ell^{-1} f(\rho_2) d\varphi \wedge B_2 \equiv d\lambda_2$, and the transition function $\lambda_2$ is well-defined only if $\lambda_2(\varphi = 2\pi) - \lambda_2(\varphi = 0) \in 2\pi H^2(M_6,\bZ)$. We can rewrite the condition as
\be
H^2(M_6,\bZ) \ni \int_\varphi \frac{d\lambda_2}{2\pi} = \int_\varphi \frac{C_3|_{NN} - C_3|_{NS}}{2\pi} \;.
\ee
By construction, $C_3|_{NN} - C_3|_{NS} = [d\gamma + C_1]\wedge B_2|_{NN} - [d\gamma + C_1]\wedge B_2|_{NS} = \int_\theta dC_1 \wedge B_2$ where the integral goes from one singular point to another one. Substituting in the previous formula we get that $F_2\wedge B_2$ must be quantized on $D^+$. The analysis of the other transition functions work similarly, and we get the quantization condition
\be
F_2 \wedge B_2 \in H^4(M_6,\bZ) \;.
\ee
So the M-theory flat connection $C_3$ is well-defined whenever $F_2 \wedge B_2$ is quantized on $M_6$. On the other hand, the reduction of $C_3$ gives a B-field $B_2$ in type IIA, and $F_4 = 0$. Thus the condition is that the Page D4-charge is quantized in type IIA.

Finally $C_3$ could be pure gauge. This happens if all its periods are trivial: $\int C_3 \in 2\pi \bZ$. Now $H_3(Y^{p,q},\bZ)$ is generated by $U(1)$ bundles over $H_2(M_6,\bZ)$, so $C_3$ is trivial whenever
\be
H^2(M_6,\bZ) \ni \int_\gamma \frac{C_3|_{**}}{2\pi} = \int_\gamma \frac{[d\gamma + C_1]\wedge B_2}{2\pi} = B_2 \;.
\ee
So $C_3$ is pure gauge in M-theory whenever $B_2$ is pure gauge in type IIA. The resulting torsion group is $\Gamma$, as shown in section \ref{subsec: Page charges}.

Summarizing, we showed that, given the ansatz (\ref{C3 ansatz}) for a flat $C_3$ connection, the set of bundles we obtain is precisely $\Gamma$; moreover the quantization of $F_2 \wedge B_2$ in IIA comes from the quantization of $G_4$ and gauge transformations of $B_2$ come from transformations of $C_3$.


\bibliographystyle{JHEP}
\bibliography{bib2}{}

\providecommand{\href}[2]{#2}\begingroup\raggedright\begin{thebibliography}{10}

\bibitem{Bagger:2006sk}
J.~Bagger and N.~Lambert, {\it {Modeling multiple M2's}},  {\em Phys. Rev.}
  {\bf D75} (2007) 045020, [\href{http://xxx.lanl.gov/abs/hep-th/0611108}{{\tt
  hep-th/0611108}}].

\bibitem{Gustavsson:2007vu}
A.~Gustavsson, {\it {Algebraic structures on parallel M2-branes}},  {\em Nucl.
  Phys.} {\bf B811} (2009) 66--76,
  [\href{http://xxx.lanl.gov/abs/0709.1260}{{\tt arXiv:0709.1260}}].

\bibitem{Bagger:2007jr}
J.~Bagger and N.~Lambert, {\it {Gauge Symmetry and Supersymmetry of Multiple
  M2-Branes}},  {\em Phys. Rev.} {\bf D77} (2008) 065008,
  [\href{http://xxx.lanl.gov/abs/0711.0955}{{\tt arXiv:0711.0955}}].

\bibitem{Aharony:2008ug}
O.~Aharony, O.~Bergman, D.~L. Jafferis, and J.~Maldacena, {\it {N=6
  superconformal Chern-Simons-matter theories, M2-branes and their gravity
  duals}},  {\em JHEP} {\bf 10} (2008) 091,
  [\href{http://xxx.lanl.gov/abs/0806.1218}{{\tt arXiv:0806.1218}}].

\bibitem{Benna:2008zy}
M.~Benna, I.~Klebanov, T.~Klose, and M.~Smedback, {\it {Superconformal
  Chern-Simons Theories and AdS$_4$/CFT$_3$ Correspondence}},  {\em JHEP} {\bf
  09} (2008) 072, [\href{http://xxx.lanl.gov/abs/0806.1519}{{\tt
  arXiv:0806.1519}}].

\bibitem{Imamura:2008nn}
Y.~Imamura and K.~Kimura, {\it {On the moduli space of elliptic
  Maxwell-Chern-Simons theories}},  {\em Prog. Theor. Phys.} {\bf 120} (2008)
  509--523, [\href{http://xxx.lanl.gov/abs/0806.3727}{{\tt arXiv:0806.3727}}].

\bibitem{Aharony:2008gk}
O.~Aharony, O.~Bergman, and D.~L. Jafferis, {\it {Fractional M2-branes}},  {\em
  JHEP} {\bf 11} (2008) 043, [\href{http://xxx.lanl.gov/abs/0807.4924}{{\tt
  arXiv:0807.4924}}].

\bibitem{Jafferis:2008qz}
D.~L. Jafferis and A.~Tomasiello, {\it {A simple class of N=3 gauge/gravity
  duals}},  {\em JHEP} {\bf 10} (2008) 101,
  [\href{http://xxx.lanl.gov/abs/0808.0864}{{\tt arXiv:0808.0864}}].

\bibitem{Hanany:2008cd}
A.~Hanany and A.~Zaffaroni, {\it {Tilings, Chern-Simons Theories and M2
  Branes}},  {\em JHEP} {\bf 10} (2008) 111,
  [\href{http://xxx.lanl.gov/abs/0808.1244}{{\tt arXiv:0808.1244}}].

\bibitem{Hanany:2008fj}
A.~Hanany, D.~Vegh, and A.~Zaffaroni, {\it {Brane Tilings and M2 Branes}},
  {\em JHEP} {\bf 03} (2009) 012,
  [\href{http://xxx.lanl.gov/abs/0809.1440}{{\tt arXiv:0809.1440}}].

\bibitem{Klebanov:2008vq}
I.~Klebanov, T.~Klose, and A.~Murugan, {\it {AdS$_4$/CFT$_3$ -- Squashed,
  Stretched and Warped}},  {\em JHEP} {\bf 03} (2009) 140,
  [\href{http://xxx.lanl.gov/abs/0809.3773}{{\tt arXiv:0809.3773}}].

\bibitem{Aganagic:2009zk}
M.~Aganagic, {\it {A Stringy Origin of M2 Brane Chern-Simons Theories}},
  \href{http://xxx.lanl.gov/abs/0905.3415}{{\tt arXiv:0905.3415}}.

\bibitem{Martelli:2009ga}
D.~Martelli and J.~Sparks, {\it {AdS$_4$/CFT$_3$ duals from M2-branes at
  hypersurface singularities and their deformations}},
  \href{http://xxx.lanl.gov/abs/0909.2036}{{\tt arXiv:0909.2036}}.

\bibitem{Hanany:1996ie}
A.~Hanany and E.~Witten, {\it {Type IIB superstrings, BPS monopoles, and
  three-dimensional gauge dynamics}},  {\em Nucl.Phys.} {\bf B492} (1997)
  152--190, [\href{http://xxx.lanl.gov/abs/hep-th/9611230}{{\tt
  hep-th/9611230}}].

\bibitem{Imamura:2008ji}
Y.~Imamura and S.~Yokoyama, {\it {N=4 Chern-Simons theories and wrapped
  M-branes in their gravity duals}},  {\em Prog. Theor. Phys.} {\bf 121} (2009)
  915--940, [\href{http://xxx.lanl.gov/abs/0812.1331}{{\tt arXiv:0812.1331}}].

\bibitem{Gaiotto:2009tk}
D.~Gaiotto and D.~L. Jafferis, {\it {Notes on adding D6 branes wrapping RP$^3$
  in AdS(4) $\times$ CP$^3$}},  \href{http://xxx.lanl.gov/abs/0903.2175}{{\tt
  arXiv:0903.2175}}.

\bibitem{Hohenegger:2009as}
S.~Hohenegger and I.~Kirsch, {\it {A note on the holography of Chern-Simons
  matter theories with flavour}},  {\em JHEP} {\bf 04} (2009) 129,
  [\href{http://xxx.lanl.gov/abs/0903.1730}{{\tt arXiv:0903.1730}}].

\bibitem{Hikida:2009tp}
Y.~Hikida, W.~Li, and T.~Takayanagi, {\it {ABJM with Flavors and FQHE}},  {\em
  JHEP} {\bf 0907} (2009) 065, [\href{http://xxx.lanl.gov/abs/0903.2194}{{\tt
  arXiv:0903.2194}}].

\bibitem{Benini:2009qs}
F.~Benini, C.~Closset, and S.~Cremonesi, {\it {Chiral flavors and M2-branes at
  toric CY4 singularities}},  {\em JHEP} {\bf 1002} (2010) 036,
  [\href{http://xxx.lanl.gov/abs/0911.4127}{{\tt arXiv:0911.4127}}].

\bibitem{Jafferis:2009th}
D.~L. Jafferis, {\it {Quantum corrections to N=2 Chern-Simons theories with
  flavor and their AdS(4) duals}},
  \href{http://xxx.lanl.gov/abs/0911.4324}{{\tt arXiv:0911.4324}}.

\bibitem{Gauntlett:2004hh}
J.~P. Gauntlett, D.~Martelli, J.~F. Sparks, and D.~Waldram, {\it {A New
  infinite class of Sasaki-Einstein manifolds}},  {\em Adv.Theor.Math.Phys.}
  {\bf 8} (2006) 987--1000, [\href{http://xxx.lanl.gov/abs/hep-th/0403038}{{\tt
  hep-th/0403038}}].

\bibitem{Martelli:2008rt}
D.~Martelli and J.~Sparks, {\it {Notes on toric Sasaki-Einstein seven-manifolds
  and AdS(4) / CFT(3)}},  {\em JHEP} {\bf 0811} (2008) 016,
  [\href{http://xxx.lanl.gov/abs/0808.0904}{{\tt arXiv:0808.0904}}].

\bibitem{to:appear}
F.~Benini, C.~Closset, and S.~Cremonesi.
\newblock To appear.

\bibitem{Martelli:2008si}
D.~Martelli and J.~Sparks, {\it {Moduli spaces of Chern-Simons quiver gauge
  theories and AdS(4)/CFT(3)}},  {\em Phys. Rev.} {\bf D78} (2008) 126005,
  [\href{http://xxx.lanl.gov/abs/0808.0912}{{\tt arXiv:0808.0912}}].

\bibitem{Benishti:2009ky}
N.~Benishti, Y.-H. He, and J.~Sparks, {\it {(Un)Higgsing the M2-brane}},  {\em
  JHEP} {\bf 1001} (2010) 067, [\href{http://xxx.lanl.gov/abs/0909.4557}{{\tt
  arXiv:0909.4557}}].

\bibitem{Kapustin:2005py}
A.~Kapustin, {\it {Wilson-'t Hooft operators in four-dimensional gauge theories
  and S-duality}},  {\em Phys. Rev.} {\bf D74} (2006) 025005,
  [\href{http://xxx.lanl.gov/abs/hep-th/0501015}{{\tt hep-th/0501015}}].

\bibitem{Kapustin:2006pk}
A.~Kapustin and E.~Witten, {\it {Electric-magnetic duality and the geometric
  Langlands program}},  \href{http://xxx.lanl.gov/abs/hep-th/0604151}{{\tt
  hep-th/0604151}}.

\bibitem{Borokhov:2002ib}
V.~Borokhov, A.~Kapustin, and X.-k. Wu, {\it {Topological disorder operators in
  three-dimensional conformal field theory}},  {\em JHEP} {\bf 11} (2002) 049,
  [\href{http://xxx.lanl.gov/abs/hep-th/0206054}{{\tt hep-th/0206054}}].

\bibitem{Borokhov:2002cg}
V.~Borokhov, A.~Kapustin, and X.-k. Wu, {\it {Monopole operators and mirror
  symmetry in three dimensions}},  {\em JHEP} {\bf 12} (2002) 044,
  [\href{http://xxx.lanl.gov/abs/hep-th/0207074}{{\tt hep-th/0207074}}].

\bibitem{Borokhov:2003yu}
V.~Borokhov, {\it {Monopole operators in three-dimensional N = 4 SYM and mirror
  symmetry}},  {\em JHEP} {\bf 03} (2004) 008,
  [\href{http://xxx.lanl.gov/abs/hep-th/0310254}{{\tt hep-th/0310254}}].

\bibitem{Kim:2009wb}
S.~Kim, {\it {The Complete superconformal index for N=6 Chern-Simons theory}},
  {\em Nucl.Phys.} {\bf B821} (2009) 241--284,
  [\href{http://xxx.lanl.gov/abs/0903.4172}{{\tt arXiv:0903.4172}}].

\bibitem{Benna:2009xd}
M.~K. Benna, I.~R. Klebanov, and T.~Klose, {\it {Charges of Monopole Operators
  in Chern-Simons Yang-Mills Theory}},
  \href{http://xxx.lanl.gov/abs/0906.3008}{{\tt arXiv:0906.3008}}.

\bibitem{Bashkirov:2010kz}
D.~Bashkirov and A.~Kapustin, {\it {Supersymmetry enhancement by monopole
  operators}},  \href{http://xxx.lanl.gov/abs/1007.4861}{{\tt
  arXiv:1007.4861}}.

\bibitem{Samtleben:2010eu}
H.~Samtleben and R.~Wimmer, {\it {N=6 Superspace Constraints, SUSY Enhancement
  and Monopole Operators}},  {\em JHEP} {\bf 1010} (2010) 080,
  [\href{http://xxx.lanl.gov/abs/1008.2739}{{\tt arXiv:1008.2739}}].

\bibitem{Jafferis:2010un}
D.~L. Jafferis, {\it {The Exact Superconformal R-Symmetry Extremizes Z}},
  \href{http://xxx.lanl.gov/abs/1012.3210}{{\tt arXiv:1012.3210}}.

\bibitem{Gaiotto:2008ak}
D.~Gaiotto and E.~Witten, {\it {S-Duality of Boundary Conditions In N=4 Super
  Yang-Mills Theory}},  \href{http://xxx.lanl.gov/abs/0807.3720}{{\tt
  arXiv:0807.3720}}.

\bibitem{Goddard:1976qe}
P.~Goddard, J.~Nuyts, and D.~I. Olive, {\it {Gauge Theories and Magnetic
  Charge}},  {\em Nucl. Phys.} {\bf B125} (1977) 1.

\bibitem{Imamura:2011su}
Y.~Imamura and S.~Yokoyama, {\it {Index for three dimensional superconformal
  field theories with general R-charge assignments}},  {\em JHEP} {\bf 1104}
  (2011) 007, [\href{http://xxx.lanl.gov/abs/1101.0557}{{\tt
  arXiv:1101.0557}}].

\bibitem{Niemi:1983rq}
A.~Niemi and G.~Semenoff, {\it {Axial Anomaly Induced Fermion Fractionization
  and Effective Gauge Theory Actions in Odd Dimensional Space-Times}},  {\em
  Phys.Rev.Lett.} {\bf 51} (1983) 2077.

\bibitem{Redlich:1983kn}
A.~N. Redlich, {\it {Gauge Noninvariance and Parity Nonconservation of
  Three-Dimensional Fermions}},  {\em Phys. Rev. Lett.} {\bf 52} (1984) 18.

\bibitem{Redlich:1983dv}
A.~N. Redlich, {\it {Parity Violation and Gauge Noninvariance of the Effective
  Gauge Field Action in Three-Dimensions}},  {\em Phys. Rev.} {\bf D29} (1984)
  2366--2374.

\bibitem{Benvenuti:2006qr}
S.~Benvenuti, B.~Feng, A.~Hanany, and Y.-H. He, {\it {Counting BPS operators in
  gauge theories: Quivers, syzygies and plethystics}},  {\em JHEP} {\bf 11}
  (2007) 050, [\href{http://xxx.lanl.gov/abs/hep-th/0608050}{{\tt
  hep-th/0608050}}].

\bibitem{Kennaway:2007tq}
K.~D. Kennaway, {\it {Brane Tilings}},  {\em Int. J. Mod. Phys.} {\bf A22}
  (2007) 2977--3038, [\href{http://xxx.lanl.gov/abs/0706.1660}{{\tt
  arXiv:0706.1660}}].

\bibitem{Davey:2009sr}
J.~Davey, A.~Hanany, N.~Mekareeya, and G.~Torri, {\it {Phases of M2-brane
  Theories}},  {\em JHEP} {\bf 06} (2009) 025,
  [\href{http://xxx.lanl.gov/abs/0903.3234}{{\tt arXiv:0903.3234}}].

\bibitem{Imamura:2008qs}
Y.~Imamura and K.~Kimura, {\it {Quiver Chern-Simons theories and crystals}},
  {\em JHEP} {\bf 10} (2008) 114,
  [\href{http://xxx.lanl.gov/abs/0808.4155}{{\tt arXiv:0808.4155}}].

\bibitem{deBoer:1997kr}
J.~de~Boer, K.~Hori, and Y.~Oz, {\it {Dynamics of N = 2 supersymmetric gauge
  theories in three dimensions}},  {\em Nucl. Phys.} {\bf B500} (1997)
  163--191, [\href{http://xxx.lanl.gov/abs/hep-th/9703100}{{\tt
  hep-th/9703100}}].

\bibitem{Aharony:1997bx}
O.~Aharony, A.~Hanany, K.~A. Intriligator, N.~Seiberg, and M.~J. Strassler,
  {\it {Aspects of N = 2 supersymmetric gauge theories in three dimensions}},
  {\em Nucl. Phys.} {\bf B499} (1997) 67--99,
  [\href{http://xxx.lanl.gov/abs/hep-th/9703110}{{\tt hep-th/9703110}}].

\bibitem{Dorey:1999rb}
N.~Dorey and D.~Tong, {\it {Mirror symmetry and toric geometry in three
  dimensional gauge theories}},  {\em JHEP} {\bf 05} (2000) 018,
  [\href{http://xxx.lanl.gov/abs/hep-th/9911094}{{\tt hep-th/9911094}}].

\bibitem{Tong:2000ky}
D.~Tong, {\it {Dynamics of N = 2 supersymmetric Chern-Simons theories}},  {\em
  JHEP} {\bf 07} (2000) 019,
  [\href{http://xxx.lanl.gov/abs/hep-th/0005186}{{\tt hep-th/0005186}}].

\bibitem{Hanany:2005ss}
A.~Hanany and D.~Vegh, {\it {Quivers, tilings, branes and rhombi}},  {\em JHEP}
  {\bf 10} (2007) 029, [\href{http://xxx.lanl.gov/abs/hep-th/0511063}{{\tt
  hep-th/0511063}}].

\bibitem{Marolf:2000cb}
D.~Marolf, {\it {Chern-Simons terms and the three notions of charge}},
  \href{http://xxx.lanl.gov/abs/hep-th/0006117}{{\tt hep-th/0006117}}.

\bibitem{Evslin:2004vs}
J.~Evslin, {\it {The Cascade is a MMS instanton}},
  \href{http://xxx.lanl.gov/abs/hep-th/0405210}{{\tt hep-th/0405210}}.
  Published in 'Advances in Soliton Research'. Edited by L.V. Chen. Nova
  Science Publishers, 2006. pp. 153-187.

\bibitem{Benini:2007gx}
F.~Benini, F.~Canoura, S.~Cremonesi, C.~Nunez, and A.~V. Ramallo, {\it
  {Backreacting Flavors in the Klebanov-Strassler Background}},  {\em JHEP}
  {\bf 09} (2007) 109, [\href{http://xxx.lanl.gov/abs/0706.1238}{{\tt
  arXiv:0706.1238}}].

\bibitem{Benini:2007kg}
F.~Benini, {\it {A chiral cascade via backreacting D7-branes with flux}},  {\em
  JHEP} {\bf 10} (2008) 051, [\href{http://xxx.lanl.gov/abs/0710.0374}{{\tt
  arXiv:0710.0374}}].

\bibitem{Argurio:2008mt}
R.~Argurio, F.~Benini, M.~Bertolini, C.~Closset, and S.~Cremonesi, {\it
  {Gauge/gravity duality and the interplay of various fractional branes}},
  {\em Phys. Rev.} {\bf D78} (2008) 046008,
  [\href{http://xxx.lanl.gov/abs/0804.4470}{{\tt arXiv:0804.4470}}].

\bibitem{Benini:2008ir}
F.~Benini, M.~Bertolini, C.~Closset, and S.~Cremonesi, {\it {The N=2 cascade
  revisited and the enhancon bearings}},  {\em Phys. Rev.} {\bf D79} (2009)
  066012, [\href{http://xxx.lanl.gov/abs/0811.2207}{{\tt arXiv:0811.2207}}].

\bibitem{Aharony:2009fc}
O.~Aharony, A.~Hashimoto, S.~Hirano, and P.~Ouyang, {\it {D-brane Charges in
  Gravitational Duals of 2+1 Dimensional Gauge Theories and Duality Cascades}},
   {\em JHEP} {\bf 01} (2010) 072,
  [\href{http://xxx.lanl.gov/abs/0906.2390}{{\tt arXiv:0906.2390}}].

\bibitem{Hashimoto:2010bq}
A.~Hashimoto, S.~Hirano, and P.~Ouyang, {\it {Branes and fluxes in special
  holonomy manifolds and cascading field theories}},
  \href{http://xxx.lanl.gov/abs/1004.0903}{{\tt arXiv:1004.0903}}.

\bibitem{Hashimoto:2011aj}
A.~Hashimoto and P.~Ouyang, {\it {Quantization of charges and fluxes in warped
  Stenzel geometry}},  \href{http://xxx.lanl.gov/abs/1104.3517}{{\tt
  arXiv:1104.3517}}.

\bibitem{Freed:1999vc}
D.~S. Freed and E.~Witten, {\it {Anomalies in string theory with D-branes}},
  \href{http://xxx.lanl.gov/abs/hep-th/9907189}{{\tt hep-th/9907189}}.

\bibitem{Bergman:2009zh}
O.~Bergman and S.~Hirano, {\it {Anomalous radius shift in AdS(4)/CFT(3)}},
  {\em JHEP} {\bf 0907} (2009) 016,
  [\href{http://xxx.lanl.gov/abs/0902.1743}{{\tt arXiv:0902.1743}}].

\bibitem{Douglas:1996sw}
M.~R. Douglas and G.~W. Moore, {\it {D-branes, quivers, and ALE instantons}},
  \href{http://xxx.lanl.gov/abs/hep-th/9603167}{{\tt hep-th/9603167}}.

\bibitem{Gukov:1998kn}
S.~Gukov, M.~Rangamani, and E.~Witten, {\it {Dibaryons, strings, and branes in
  AdS orbifold models}},  {\em JHEP} {\bf 12} (1998) 025,
  [\href{http://xxx.lanl.gov/abs/hep-th/9811048}{{\tt hep-th/9811048}}].

\bibitem{Diaconescu:1999dt}
D.-E. Diaconescu and J.~Gomis, {\it {Fractional branes and boundary states in
  orbifold theories}},  {\em JHEP} {\bf 0010} (2000) 001,
  [\href{http://xxx.lanl.gov/abs/hep-th/9906242}{{\tt hep-th/9906242}}].

\bibitem{Douglas:2000qw}
M.~R. Douglas, B.~Fiol, and C.~Romelsberger, {\it {The Spectrum of BPS branes
  on a noncompact Calabi-Yau}},  {\em JHEP} {\bf 0509} (2005) 057,
  [\href{http://xxx.lanl.gov/abs/hep-th/0003263}{{\tt hep-th/0003263}}].

\bibitem{Marino:1999af}
M.~Marino, R.~Minasian, G.~W. Moore, and A.~Strominger, {\it {Nonlinear
  instantons from supersymmetric p-branes}},  {\em JHEP} {\bf 01} (2000) 005,
  [\href{http://xxx.lanl.gov/abs/hep-th/9911206}{{\tt hep-th/9911206}}].

\bibitem{Klemm:1999gm}
A.~Klemm and E.~Zaslow, {\it {Local mirror symmetry at higher genus}},
  \href{http://xxx.lanl.gov/abs/hep-th/9906046}{{\tt hep-th/9906046}}.

\bibitem{Aharony:2010af}
O.~Aharony, D.~Jafferis, A.~Tomasiello, and A.~Zaffaroni, {\it {Massive type
  IIA string theory cannot be strongly coupled}},  {\em JHEP} {\bf 1011} (2010)
  047, [\href{http://xxx.lanl.gov/abs/1007.2451}{{\tt arXiv:1007.2451}}].

\bibitem{Behrndt:2004mj}
K.~Behrndt and M.~Cvetic, {\it {General N=1 supersymmetric fluxes in massive
  type IIA string theory}},  {\em Nucl.Phys.} {\bf B708} (2005) 45--71,
  [\href{http://xxx.lanl.gov/abs/hep-th/0407263}{{\tt hep-th/0407263}}].

\bibitem{Lust:2004ig}
D.~Lust and D.~Tsimpis, {\it {Supersymmetric AdS(4) compactifications of IIA
  supergravity}},  {\em JHEP} {\bf 02} (2005) 027,
  [\href{http://xxx.lanl.gov/abs/hep-th/0412250}{{\tt hep-th/0412250}}].

\bibitem{Acharya:2006ne}
B.~S. Acharya, F.~Benini, and R.~Valandro, {\it {Fixing moduli in exact type
  IIA flux vacua}},  {\em JHEP} {\bf 0702} (2007) 018,
  [\href{http://xxx.lanl.gov/abs/hep-th/0607223}{{\tt hep-th/0607223}}].

\bibitem{Petrini:2009ur}
M.~Petrini and A.~Zaffaroni, {\it {N=2 solutions of massive type IIA and their
  Chern-Simons duals}},  {\em JHEP} {\bf 09} (2009) 107,
  [\href{http://xxx.lanl.gov/abs/0904.4915}{{\tt arXiv:0904.4915}}].

\bibitem{Lust:2009mb}
D.~Lust and D.~Tsimpis, {\it {New supersymmetric AdS(4) type II vacua}},  {\em
  JHEP} {\bf 0909} (2009) 098, [\href{http://xxx.lanl.gov/abs/0906.2561}{{\tt
  arXiv:0906.2561}}].

\bibitem{Benishti:2011ab}
N.~Benishti, {\it {Emerging Non-Anomalous Baryonic Symmetries in the
  $AdS_5$/CFT$_4$ Correspondence}},
  \href{http://xxx.lanl.gov/abs/1102.1979}{{\tt arXiv:1102.1979}}.

\bibitem{Gomis:2005wc}
J.~Gomis, F.~Marchesano, and D.~Mateos, {\it {An open string landscape}},  {\em
  JHEP} {\bf 11} (2005) 021,
  [\href{http://xxx.lanl.gov/abs/hep-th/0506179}{{\tt hep-th/0506179}}].

\bibitem{LePotier}
J.~Le~Potier, {\em {Lectures on Vector Bundles}}.
\newblock Cambridge University Press, 1997.

\bibitem{Cheon:2011th}
S.~Cheon, D.~Gang, S.~Kim, and J.~Park, {\it {Refined test of AdS4/CFT3
  correspondence for N=2,3 theories}},
  \href{http://xxx.lanl.gov/abs/1102.4273}{{\tt arXiv:1102.4273}}.

\bibitem{Witten:1996md}
E.~Witten, {\it {On flux quantization in M-theory and the effective action}},
  {\em J. Geom. Phys.} {\bf 22} (1997) 1--13,
  [\href{http://xxx.lanl.gov/abs/hep-th/9609122}{{\tt hep-th/9609122}}].

\bibitem{Witten:1996bn}
E.~Witten, {\it {Non-Perturbative Superpotentials In String Theory}},  {\em
  Nucl. Phys.} {\bf B474} (1996) 343--360,
  [\href{http://xxx.lanl.gov/abs/hep-th/9604030}{{\tt hep-th/9604030}}].

\bibitem{Benishti:2010jn}
N.~Benishti, D.~Rodriguez-Gomez, and J.~Sparks, {\it {Baryonic symmetries and
  M5 branes in the $AdS_4$/CFT$_3$ correspondence}},  {\em JHEP} {\bf 07}
  (2010) 024, [\href{http://xxx.lanl.gov/abs/1004.2045}{{\tt
  arXiv:1004.2045}}].

\bibitem{Gaiotto:2009mv}
D.~Gaiotto and A.~Tomasiello, {\it {The gauge dual of Romans mass}},  {\em
  JHEP} {\bf 1001} (2010) 015, [\href{http://xxx.lanl.gov/abs/0901.0969}{{\tt
  arXiv:0901.0969}}].

\bibitem{Fujita:2009kw}
M.~Fujita, W.~Li, S.~Ryu, and T.~Takayanagi, {\it {Fractional Quantum Hall
  Effect via Holography: Chern- Simons, Edge States, and Hierarchy}},  {\em
  JHEP} {\bf 06} (2009) 066, [\href{http://xxx.lanl.gov/abs/0901.0924}{{\tt
  arXiv:0901.0924}}].

\bibitem{Tomasiello:2010zz}
A.~Tomasiello and A.~Zaffaroni, {\it {Parameter spaces of massive IIA
  solutions}},  \href{http://xxx.lanl.gov/abs/1010.4648}{{\tt
  arXiv:1010.4648}}.

\bibitem{Jafferis:2011zi}
D.~L. Jafferis, I.~R. Klebanov, S.~S. Pufu, and B.~R. Safdi, {\it {Towards the
  F-Theorem: N=2 Field Theories on the Three-Sphere}},
  \href{http://xxx.lanl.gov/abs/1103.1181}{{\tt arXiv:1103.1181}}.

\bibitem{Cremonesi:2010ae}
S.~Cremonesi, {\it {Type IIB construction of flavoured ABJ(M) and fractional M2
  branes}},  {\em JHEP} {\bf 1101} (2011) 076,
  [\href{http://xxx.lanl.gov/abs/1007.4562}{{\tt arXiv:1007.4562}}].

\bibitem{Hori:1997ab}
K.~Hori, H.~Ooguri, and Y.~Oz, {\it {Strong coupling dynamics of
  four-dimensional N=1 gauge theories from M theory five-brane}},  {\em
  Adv.Theor.Math.Phys.} {\bf 1} (1998) 1--52,
  [\href{http://xxx.lanl.gov/abs/hep-th/9706082}{{\tt hep-th/9706082}}].

\bibitem{Bhattacharya:2008zy}
J.~Bhattacharya, S.~Bhattacharyya, S.~Minwalla, and S.~Raju, {\it {Indices for
  Superconformal Field Theories in 3,5 and 6 Dimensions}},  {\em JHEP} {\bf 02}
  (2008) 064, [\href{http://xxx.lanl.gov/abs/0801.1435}{{\tt
  arXiv:0801.1435}}].

\bibitem{Aspinwall:2004jr}
P.~S. Aspinwall, {\it {D-branes on Calabi-Yau manifolds}},
  \href{http://xxx.lanl.gov/abs/hep-th/0403166}{{\tt hep-th/0403166}}.

\bibitem{Meijer:wiki}
{\url{http://en.wikipedia.org/wiki/Meijer_G-function}}.

\end{thebibliography}\endgroup

\end{document}